\documentclass[10pt,a4paper,twoside]{book}
\usepackage{amsmath}
\usepackage{amssymb}
\usepackage{graphicx}
\usepackage{float}
\usepackage{pictex}
\usepackage[dvipsnames,usenames]{color}
\usepackage{fancyhdr}
\usepackage{cite}
\usepackage{longtable}
\usepackage{epsfig}
\usepackage{psfrag}

\definecolor{Conf}{named}{Black}
\definecolor{GBE}{named}{Black}
\definecolor{OGE}{named}{Black}
\definecolor{Text-normal}{named}{Black}
\definecolor{Bonn}{named}{Black}

\begin{document}
\begin{titlepage}
\centering
{\large Elmar P. Biernat}\\[4cm]
{\LARGE  \bfseries Quantization on\\[0.4cm] Space-Time Hyperboloids  }\\[8cm]
{\large \bfseries Diplomarbeit}\\[2pt]
zur Erlangung des akademischen Grades eines\\
 Magisters\\
 an der Naturwissenschaften Fakult\"at der\\
Karl-Franzens-Universit\"at Graz\\[2cm]
Betreuer:\\
Ao.~Univ.-Prof.~Dr.~Wolfgang Schweiger\\
Institut f\"ur Physik, Fachbereich Theoretische Physik\\[1cm]
September 2007
\thispagestyle{empty}
\end{titlepage}\thispagestyle{empty}
\cleardoublepage
\thispagestyle{empty}\newpage
\begin{titlepage}
\centering \vspace*{8cm}
{\large {\textit{F\"ur Justine}} }
\thispagestyle{empty}
\end{titlepage}\thispagestyle{empty}
\begin{titlepage} \cleardoublepage
\subsubsection*{Abstract} We quantize a relativistic massive
complex spin-$0$ field and a relativistic massive spin-$\frac12$
field on a space-time hyperboloid. We call this procedure
\emph{point-form canonical quantization}. Lorentz invariance of
the hyperboloid implies that the 4 generators for translations
become dynamic and interaction dependent, whereas the 6 generators
for Lorentz transformations remain kinematic and interaction free.
We expand the fields in terms of usual plane waves and prove the
equivalence to equal-time quantization by representing the
Poincar\'e generators in a momentum basis. We formulate a
generalized scattering theory for interacting fields by
considering evolution of the system generated by the interaction
dependent four-momentum operator. Finally we expand our
generalized scattering operator in powers of the interaction and
show its equivalence to the Dyson expansion of usual time-ordered
perturbation theory.
\end{titlepage}\thispagestyle{empty}
\tableofcontents
\chapter{Introduction and Overview}
With his theory of relativity Einstein replaced the absolute
character of space and time in Newtonian mechanics by giving them
a relative meaning. In Newtonian physics the three-dimensional
physical space and the one-dimensional time are represented
separately by the three-dimensional Euclidean space and the real
numbers, respectively. In special relativity space and time are
treated equally, forming the combined notion of \emph{space-time}
represented by the four-dimensional \emph{Minkowski space}. On the
other hand, classical mechanics was replaced by the quantum theory
of Schr\"odinger and Heisenberg, treating the space observable as
operator and leaving time as a c-number
parameter.\\
In order to make quantum theory consistent with the theory of
special relativity Dirac and others initiated what is known as
\emph{relativistic quantum mechanics}. In his famous paper
\emph{Forms of Relativistic Dynamics}~\cite{Dirac:1949cp} Dirac
found a way to make the Poincar\'e group applicable to quantum
theory. Furthermore, he pointed out the possibility of formulating
Poincar\'e invariant relativistic dynamics in different ways,
depending on the foliation of Minkowski space. He found three
forms: the \emph{instant, front} and \emph{point form}. Each form
corresponds to a different choice of a spacelike hypersurface
defining an instant in the time parameter. This hypersurface is
invariant under the action of certain Poincar\'e generators
(\emph{kinematic generators}) which span the, so called,
\emph{stability group}. The others, the \emph{dynamic generators},
generate evolution of the system and contain interactions (if
present) whereas the kinematic generators stay interaction free.
Therefore each form describes a different way of
including interactions into the free theory.  \\
Inconsistencies of relativistic quantum mechanics with the
existence of anti-particles and relativistic causality can be
resolved by going from a finite number of degrees of freedom to
infinitely many degrees of freedom. This corresponds to setting up
a local quantum field theory. The arbitrariness of choosing a time
parameter will still be present in a quantum field theory and is
reflected in the arbitrariness of the choice of hypersurface on
which canonical (anti)commutation relations are imposed. This
issue was taken up by Tomonaga~\cite{Tomonaga:1946} and
Schwinger~\cite{Schwinger:1948yk} by formulating generalized
canonical (anti)commutation relations on arbitrary spacelike
hypersurfaces.
\\ Field quantization on the Lorentz-invariant forward hyperboloid
$x_{\lambda}\,x^{\lambda}=\tau^2$, with $\tau$ arbitrary but
fixed, provides a simple example of field quantization on a curved
hypersurface. Following Dirac's nomenclature~\cite{Dirac:1949cp}
we speak in this context of \emph{point-form quantum field
theory}. Due to the curvilinear
 nature of the hyperboloid field quantization is not straightforward. Therefore only a few papers exist about point-form quantum field
 theory~\cite{Fubini:1972mf,Sommerfield:1974fa,Gromes:1974yu,Disessa:1974xd,Disessa:1974ve} and
 they mainly don't go beyond free fields.
 Most of these attempts to quantize field theories on the hyperboloid made
 use of
 hyperbolic coordinates. The \lq\lq Hamiltonian\rq\rq\,in these coordinates is identified with the generator for dilatation
 transformations, which is explicitly \lq\lq hyperbolic-time\rq\rq\,dependent and does not belong to the Poincar\'e
 group. Furthermore, it is first only defined in the forward light
 cone. This restriction
can be overcome by analytical continuation, although it does not
seem very convenient to consider development in hyperbolic time,
especially if one wants to describe scattering. The field
equations in these coordinates are solved in terms of Hankel
functions. The associated field quanta are characterized by the
eigenvalues of the generators for Lorentz boosts, which become
diagonal in the corresponding Fock representation (\emph{Lorentz
basis}). However, the definition of the translation generator as a
self-adjoint operator acting on square integrable functions of
these boost eigenvalues is not completely
straightforward~\cite{Macdowell:1972ef}. Altogether, these
approaches do not seem to be very useful for massive theories and
lead to difficulties in describing scattering.\\
\\
In this thesis we argue that it is more convenient to work with
the usual Cartesian coordinates and to expand the fields in terms
of usual plane waves. The associated field quanta are
characterized by the eigenvalues of the three-momentum operator
and a spin number. The four-momentum operator represented in this
\emph{Wigner basis} becomes diagonal. Moreover, a
Lorentz-invariant formulation of scattering can be easily achieved
by considering evolution generated by the
four-momentum operator.\\
\\
 In point-form quantum mechanics, many
ideas for the construction of interaction potentials and current
operators are motivated by quantum field theoretical
considerations. This emphasizes the necessity of formulating an
interacting point-form quantum field theory. Furthermore,
point-form quantum field theory can be viewed as a special case of
field quantization in curved space-time, that is, with a classical
gravitational background~\cite{Wald:1995yp,Birrell:1984}.
Altogether, this should provide enough motivation for setting up a
point-form quantum field theory.\\
\\
In Chapter~\ref{cha:sqft} some basics of quantum field theory and
an overview of its symmetries are given, which will be frequently
used and referred to in the following chapters. The topic of
Chapter~\ref{cha:TP} is the problem of time parameterization.
Invariance under reparameterization is a typical feature of
parameterized Hamiltonian systems. This is first discussed for the
classical free relativistic particle and then Dirac's forms of
relativistic dynamics are introduced. In Chapter~\ref{cha:ccfq} a
free complex massive scalar field and a free massive spinor field
are quantized on the hyperboloid by means of Lorentz-invariant
(anti)commutation relations. Furthermore, the equivalence between
instant- and point-form quantization of free fields is proved by
using the Wigner representation of the Poincar\'e group. Finally,
in Chapter~\ref{cha:st}, a manifest covariant formulation of
scattering is presented. This leads to the same series expansion
of a corresponding scattering operator as usual time-ordered
perturbation theory does. All the longer calculations are put into
five
Appendices~\ref{eq:app:hse},~\ref{app:WWmu},~\ref{app:cs0f},~\ref{app:s12f}
and~\ref{app:st}.
\chapter{Fundamentals of Quantum Field Theory}
\label{cha:sqft} In this chapter we give an overview of the basic
theorems and definitions of quantum field theory, which will be
frequently used and referred to throughout the thesis.
\section{Poincar\'e Group}

In his paper~\cite{Wigner:1939cj} Wigner realized that the
fundamental symmetry group of a relativistic quantum theory is the Poincar\'e
group $\mathcal{P}$ of special relativity. It is the group of all
transformations of Minkowski space-time\footnote{Minkowski
space-time is the $\mathbb{R}^4$ together with a flat Lorentz
metric $g$ of signature $(+,-,-,-)$.} that leave the distance
between 2 points invariant. It can be written as the semi-direct
product of $\mathbb{R}^4$ with the Lorentz group $\mathcal{L}$. We will restrict ourselves to the restricted
Poincar\'e group $\mathcal{P}_{+}^{\uparrow}$ being the group of space-time translations together with
rotations and boosts. Its
elements, denoted by a pair $(a,\Lambda)$ with $a\in\mathbb{R}^4$
 and $\Lambda\in\mathcal{L}_{+}^{\uparrow}$,
form a ten-dimensional Lie group\footnote{An n-dimensional Lie
group is a continuous group which has the properties of an
n-dimensional manifold~\cite{Isham:1999}. }. Its parameters are
the four-vector $a^{\mu}$ and the skew-symmetric, real
$\omega^{\mu\nu}$.\footnote{Minkowski-vector indices are denoted
by $\rho,\,\mu,\,\nu,\,\lambda,\,\sigma=0,\ldots 3$, three-vector
indices by $i,\,j,\,k=1,\ldots 3$. The Dirac spin indices are also
denoted by $\rho,\, \lambda,\, \sigma=\pm\frac12$, but it should
be clear from the context which ones are meant.} An operator
representation of infinitesimal $\mathcal{P}_{+}^{\uparrow}$
transformations which act on scalar functions of Minkowski
space-time is given by\cite{Bogolubov:1990}
\begin{eqnarray}\label{eq:D}
\Omega\left(a,\Lambda\right)=\mathbf{1}+ia^{\lambda}\,P_{\lambda}-\frac{i}{2}\omega^{\lambda\sigma}\,M_{\lambda\sigma}+\mathcal{O}\left(a^2,\omega^2\right),\end{eqnarray}
where $M^{\mu\nu}=-M^{\nu\mu}$. The operators $P^{\mu}$ and
$M^{\mu\nu}$ are then given by
\begin{eqnarray}\label{eq:P}
P^{\mu}&=&i\frac{\partial }{\partial x_{\mu}},\\
\label{eq:M}
M^{\mu\nu}&=&x^{\mu}P^{\nu}-x^{\nu}P^{\mu},\quad\text{with}\quad
x\in \mathbb{R}^4.
\end{eqnarray}
They satisfy the following commutation relations of the Lie
algebra of $\mathcal{P}$:
\begin{eqnarray}\label{eq:la}
\left[ P^{\mu},P^{\nu}\right]& =&0,\\
\left[M^{\mu\nu}, P^{\rho}\right]&=&i\left(g^{\nu\rho}P^{\mu}-g^{\mu\rho}P^{\nu}
\right),\\
\left[M^{\mu\nu}, M^{\lambda\sigma}\right]&=&
-i\left(g^{\mu\lambda}M^{\nu\sigma}-g^{\nu\lambda}M^{\mu\sigma}+g^{\nu\sigma}M^{\mu\lambda}-g^{\mu\sigma}M^{\nu\lambda}
\right).
\end{eqnarray}
Only these commutation relations are essential for the definition
of the Lie algebra, they are satisfied for any arbitrary
representation of $\mathcal{P}_{+}^{\uparrow}$. $P^{\mu}$ and
$M^{\mu\nu}$ are called generators of
$\mathcal{P}_{+}^{\uparrow}$, they generate space-time
translations and Lorentz transformations parameterized by $a$ and
$\omega$, respectively. Since the proper Lorentz group has
covering group $SL\left(2,\mathbb{C}\right)$, we shall call the
covering group of the $\mathcal{P}_{+}^{\uparrow}$
\emph{inhomogeneous} $SL\left(2,\mathbb{C}\right)$.
\section{Fields}
All relativistic theories should be invariant under
$\mathcal{P}_{+}^{\uparrow}$. For relativistic quantum theories there is a physical Hilbert
space $\mathcal{H}$ in which a unitary representation
$\hat{U}\left(a,\Lambda\right)$ of the inhomogeneous
$SL\left(2,\mathbb{C}\right)$
 acts\footnote{Operators acting on a Hilbert space are denoted by \lq\lq\,\,$\widehat{\,}$\,\,\rq\rq.}, giving the relativistic transformation law of the states.
 $\hat{U}\left(a,\mathbf{1}\right)$ can be written as
 $\hat{U}\left(a,\mathbf{1}\right)=\mathrm{e}^{ia_{\lambda}\,\hat{P}^{\lambda}}$
 with $\hat{P}^{\lambda}$ unbounded and hermitian. The operator
 $\hat{P}_{\lambda}\,\hat{P}^{\lambda}=\hat{M}^2$ is interpreted as the
 square of the mass and the eigenvalues of $\hat{P}^{\mu}$ lie in or
 on the forward light cone\footnote{The light cone is the region of all timelike and lightlike four-vectors $v$ ($v_{\lambda}\,v^{\lambda}\geq0$) of Minkowski
 space.}.
\subsection{Transformation Laws}
Let us consider classical fields\footnote{We label the fields by
greek letters $\alpha,\,\beta,\,\gamma,\,\delta\ldots$, which have
not to be confused with Lorentz indices.}
$\chi_{\alpha}\left(x\right)$ that transform under a Poincar\'e
transformation $(a,\Lambda)$ as
\begin{eqnarray}\label{eq:clf}
\chi_{\alpha}\left(x\right)\rightarrow \chi_{\alpha}'\left(x'\right)=S\left(\Lambda\right)_{\alpha}^{~\beta}\chi_{\beta}\left(x\right),\quad x'=\Lambda x+a.
\end{eqnarray}
After quantization the classical fields $\chi_{\alpha}\left(x\right)$ are replaced by field operators $\hat{\chi}_{\alpha}\left(x\right)$ that act on a Hilbert space $\mathcal{H}$. Quantum states $\vert\Phi\rangle$, which are elements of $\mathcal{H}$, behave under Poincar\'e transformations like
\begin{eqnarray}\label{eq:state}
\mathcal{H}\ni\vert \Phi\rangle \rightarrow
\vert\Phi'\rangle=\hat{U}\left(a,\Lambda\right)\vert\Phi\rangle\in\mathcal{H},
\end{eqnarray}
with $\hat{U}\left(a,\Lambda\right)$ being a unitary operator. The
classical fields $\chi_{\alpha}\left(x\right)$ correspond to
expectation values of the field operators
$\hat{\chi}_{\alpha}\left(x\right)$ $\langle\Phi\vert
\hat{\chi}_{\alpha}\left(x\right)\vert\Phi\rangle$ and the
transformed fields $\chi'_{\alpha}\left(x'\right)$ correspond to a
transformed matrix element $\left\langle\Phi'\left\vert
\hat{\chi}_{\alpha}\left(x'\right)\right\vert\Phi'\right\rangle$.
From~(\ref{eq:clf}) and~(\ref{eq:state}) we find
\begin{eqnarray}\label{eq:matrixe}
S\left(\Lambda\right)_{\alpha}^{~\beta}\langle\Phi\vert
\hat{\chi}_{\beta}\left(x\right)\vert\Phi\rangle=\langle\Phi\vert\,\hat{U}\left(a,\Lambda\right)^{-1}
\hat{\chi}_{\alpha}\left(x'\right)\hat{U}\left(a,\Lambda\right)\vert\Phi\rangle.
\end{eqnarray}
This equation is valid for arbitrary states, therefore a field operator
transforms under
$\hat{U}\left(a,\Lambda\right)$ as
\begin{eqnarray}\label{eq:sl2c}
\hat{U}\left(a,\Lambda\right)\hat{\chi}_{\alpha}\left(x\right)\hat{U}\left(a,\Lambda\right)^{-1}
=S\left(\Lambda^{-1}\right)_{\alpha}^{~\beta}\hat{\chi}_{\beta}\left(\Lambda x +a\right).
\end{eqnarray}
For a spin-0 field operator $\hat{\phi}$ (Lorentz scalar) we
have then
\begin{eqnarray}\label{eq:scalarfield}
\hat{U}\left(a,\Lambda\right)\hat{\phi}\left(x\right)\hat{U}\left(a,\Lambda\right)^{-1}
=\hat{\phi}\left(\Lambda x+a\right).
\end{eqnarray}
For a spin-$\frac12$ field operator $\hat{\psi}$ (Lorentz
four-spinor) we have for the components
$\hat{\psi}_{\alpha},\,\alpha=1,\ldots 4$
\begin{eqnarray}
\hat{U}\left(a,\Lambda\right)\hat{\psi}_{\alpha}\left(x\right)\hat{U}\left(a,\Lambda\right)^{-1}
=S\left(\Lambda^{-1}\right)_{\alpha}^{~\beta}\hat{\psi}_{\beta}\left(\Lambda
x+a\right),
\end{eqnarray}
where $S\left(\Lambda^{-1}\right)$ is a $4\times4$-matrix
representation of the $SL\left(2,\mathbb{C}\right)$\cite{Kugo:1997fs}.
\\
\subsection{Noether Theorem}
\label{sec:Nt} A symmetry of a theory is equivalent with the
invariance of the action under a certain transformation. According
to Neother's theorem, every symmetry of the action corresponds to
an integral of motion of the theory. The classical action is given
by
\begin{eqnarray}
\label{eq:action}
S\left[\chi\right]:=\int_{\mathbb{R}^4} \mathrm{d}^4 x\,\mathcal{L}\left(\chi_{\alpha}\left(x\right),\partial_{\mu}\chi_{\alpha}\left(x\right)\right),
\end{eqnarray}
with the Lagrangian density
$\mathcal{L}\left(\chi_{\alpha}\left(x\right),\partial_{\mu}\chi_{\alpha}\left(x\right)\right)$ being
a function of the fields and their first derivatives.\footnote{In
order to simplify notation, we will write in the following $\mathcal{L}\left(x\right)$ instead of
$\mathcal{L}\left(\chi_{\alpha}\left(x\right),\partial_{\mu}\chi_{\alpha}\left(x\right)\right)$. } The Hamiltonian principle of
making the action stationary gives the Euler-Lagrange equations as
\begin{eqnarray}
\label{eq:elegen} \frac{\delta
S}{\delta\chi_{\alpha}\left(x\right)}=\frac{\partial
\mathcal{L}}{\partial\chi_{\alpha}\left(x\right)}-\partial_{\mu}\left(\frac{\partial
\mathcal{L}}{\partial\left(\partial_{\mu}\chi_{\alpha}\left(x\right)\right)}
\right)\stackrel{!}{=}0,
\end{eqnarray}where $\frac{\delta}{\delta\chi\left(x\right)}$ denotes the functional differentiation.
\\
Let us now consider an infinitesimal symmetry transformation of the form
\begin{eqnarray}
\chi_{\alpha}\left(x\right)\rightarrow\chi'_{\alpha}\left(x\right)=\chi_{\alpha}\left(x\right)+\epsilon
T_{\alpha}\left(\chi_{1}\left(x\right),\chi_{2}\left(x\right),\ldots
\right).
\end{eqnarray}
Remarkably, the integral of motion $G$, associated with this
symmetry transformation, is its infinitesimal generator in the
sense that
 \begin{eqnarray}\label{eq:pbnt}
T_{\alpha}\left(\chi_{1}\left(x\right),\chi_{2}\left(x\right),\ldots
\right)=\left\lbrace \chi_{\alpha}\left(x\right),G\right\rbrace_P,
\end{eqnarray} where $\left\lbrace \ldots\right\rbrace_P$ denotes the Poisson bracket.
For field operators $\hat{\chi}_{\alpha}$ the Poisson bracket is
replaced by the commutator
\begin{eqnarray}T_{\alpha}\left(\hat{\chi}_{1}\left(x\right),\hat{\chi}_{2}\left(x\right),\ldots \right)
=\left[i\hat{G},\hat{\chi}_{\alpha}\left(x\right)\right].
\end{eqnarray}
\subsubsection{Global Gauge Transformations}
We assume that the action and even the Lagrangian density of a complex field is
invariant under a global $U\left(1\right)$ phase transformation, i.e.
\begin{eqnarray}\label{eq:deltalagrangian}
\delta \mathcal{L}\left(x\right):=
\mathcal{L}\left(\chi'\left(x\right),\partial_{\mu}\chi'\left(x\right)\right)-\mathcal{L}\left(\chi\left(x\right),\partial_{\mu}\chi\left(x\right)\right)
\stackrel{!}{=}0,\\
\text{if}\quad
\chi'\left(x\right)=\mathrm{e}^{-i\epsilon}\chi\left(x\right)\quad\text{and}\quad\chi'^{\ast}\left(x\right)=\mathrm{e}^{i\epsilon}\chi^{\ast}\left(x\right),\quad
\epsilon=\mathrm{const.}\quad.
\end{eqnarray}
With the help of $\delta\left(\partial_{\mu}\chi\right)=\partial_{\mu}\left(\delta\chi\right)$
and~(\ref{eq:elegen}) we find for infinitesimal transformations that
\begin{eqnarray}\label{eq:nc}
\delta \mathcal{L}\left(x\right)=\partial_{\mu}\left[
\frac{\partial\mathcal{L}\left(x\right)}{\partial\left(\partial_{\mu}\chi\left(x\right)\right)}\delta\chi\left(x\right)\right]+\partial_{\mu}\left[
\frac{\partial\mathcal{L}\left(x\right)}{\partial\left(\partial_{\mu}\chi^{\ast}\left(x\right)\right)}\delta\chi^{\ast}\left(x\right)\right]
=0,\end{eqnarray} where
\begin{eqnarray}
\label{eq:itr}
\delta\chi\left(x\right):=\chi'\left(x\right)-\chi\left(x\right)\approx-i\epsilon\chi\left(x\right)\quad\text{and}\quad
\delta\chi^{\ast}\left(x\right)\approx i\epsilon\chi^{\ast}\left(x\right)
\end{eqnarray}
are the variations of $\chi\left(x\right)$ and $\chi^{\ast}\left(x\right)$ at point $x$.
Then the quantity in the square brackets in~(\ref{eq:nc}) is a
conserved symmetry current
\begin{eqnarray}\label{eq:cc}\partial_{\mu}\,\mathcal{J}^{\mu}\left(x\right)=0,\quad\text{with}\quad
\mathcal{J}^{\mu}\left(x\right):=i
\frac{\partial\mathcal{L}\left(x\right)}{\partial\left(\partial_{\mu}\chi\left(x\right)\right)}\chi\left(x\right)-i
\frac{\partial\mathcal{L}\left(x\right)}{\partial\left(\partial_{\mu}\chi^{\ast}\left(x\right)\right)}\chi^{\ast}\left(x\right).\nonumber\\
\end{eqnarray} This current integrated over a spacelike
hypersurface\footnote{A hypersurface is a three-dimensional
submanifold embedded in a four-dimensional manifold. A
hypersurface is spacelike, if its normal vector
$n^{\mu}\left(x\right)$ is timelike, i.e.
$n_{\lambda}\left(x\right)\,n^{\lambda}\left(x\right)>0$.} gives a
conserved charge
\begin{eqnarray}\label{eq:cch}
Q=\int_{\Sigma}\mathrm{d}\Sigma^{\mu}\left(x\right)\mathcal{J}_{\mu}\left(x\right),
\end{eqnarray}
with $\mathrm{d}\Sigma^{\mu}\left(x\right)$ denoting the oriented
hypersurface element. In a quantum field theory, $Q$ becomes an
operator $\hat{Q}$ generating global gauge transformations
~(\ref{eq:itr}) in the sense that
\begin{eqnarray}\label{eq:gfggt}
\hat{\chi}\left(x\right)=\left[\hat{\chi}\left(x\right),\hat{Q}\right],\quad\hat{\chi}^{\dag}\left(x\right)=-\left[\hat{\chi}^{\dag}\left(x\right),\hat{Q}\right].
\end{eqnarray}
\subsubsection{Translations}
Let the Lagrangian density $\mathcal{L}\left(x\right)$ be form
invariant, i.e.
$\mathcal{L}'\left(x'\right)=\mathcal{L}\left(x\right)$, under a
translation $x'=x+a$ that transforms the fields as
\begin{eqnarray}
\chi_{\alpha}\left(x\right)\rightarrow\chi'_{\alpha}\left(x'\right)=\chi_{\alpha}\left(x\right).
\end{eqnarray}
Again it is sufficient to consider infinitesimal displacements.
The variation of the fields and their derivatives at the point $x$
is then
\begin{eqnarray}\label{eq:deltachialpha}
\delta\chi_{\alpha}\left(x\right)&:=&\chi_{\alpha}'\left(x\right)-\chi_{\alpha}\left(x\right)=a^{\nu}\partial_{\nu}\chi_{\alpha}\left(x\right),\\
\label{eq:deltapartialchialpha}\delta
\partial_{\mu}\chi_{\alpha}\left(x\right)&=&\partial_{\mu}\delta\chi_{\alpha}\left(x\right).
\end{eqnarray}
Expansion in the small parameters $a^{\nu}$ gives for the change
in the Lagrangian density (at point $x$)
\begin{eqnarray}\label{eq:deltaL}
\delta
\mathcal{L}\left(x\right)=a^{\nu}\partial_{\nu}\mathcal{L}\left(x\right).
\end{eqnarray}
On the other hand we get from~(\ref{eq:deltalagrangian}) with the
help of~(\ref{eq:deltachialpha}),~(\ref{eq:deltapartialchialpha})
and~(\ref{eq:elegen})
\begin{eqnarray}\label{eq:deltaL2}
\delta
\mathcal{L}\left(x\right)=
\partial_{\mu}
\left(\frac{\partial\mathcal{L}\left(x\right)}{\partial\left(\partial_{\mu}\chi_{\alpha}\left(x\right)\right)}a^{\nu}\partial_{\nu}\chi_{\alpha}\left(x\right)\right).
\end{eqnarray}Since $a$ is arbitrary, we get from comparison
of~(\ref{eq:deltaL}) and~(\ref{eq:deltaL2})
\begin{eqnarray}
\label{eq:DT}
\partial_{\mu}\left[\frac{\partial\mathcal{L}\left(x\right)}{\partial\left(\partial_{\mu}\chi_{\alpha}\left(x\right)\right)}\partial_{\nu}\chi_{\alpha}\left(x\right)-g^{\mu}_{\nu}\mathcal{L}\left(x\right)\right]=0
.
\end{eqnarray}
The quantity in the square brackets is a conserved Noether current
called energy-momentum tensor
\begin{eqnarray}
\label{eq:emt}
\mathcal{T}^{\mu}_{\nu}\left(x\right):=\frac{\partial\mathcal{L}\left(x\right)}{\partial\left(\partial_{\mu}\chi_{\alpha}\left(x\right)\right)}\partial_{\nu}\chi_{\alpha}\left(x\right)-g^{\mu}_{\nu}\mathcal{L}\left(x\right)
.
\end{eqnarray}
At this point it is important to note for later purposes, that
this expression is valid for both interacting theories and free
theories. If the interaction terms in $\mathcal{L}$ do not contain
derivatives of the fields, all interaction terms are included in the second part only.\\
Integration of~(\ref{eq:emt}) over a spacelike hypersurface gives
the four-momentum
\begin{eqnarray}
\label{eq:Pmu}P^{\mu}=\int_{\Sigma}\mathrm{d}\Sigma^{\nu}\left(x\right)\mathcal{T}^{\mu}_{\nu}\left(x\right).
\end{eqnarray}
In a quantum field theory $P^{\mu}$ becomes an operator
$\hat{P}^{\mu}$ generating space-time translations of the field
operator in the sense that
\begin{eqnarray}
\label{eq:commP}
\partial^{\mu}\hat{\chi}\left(x\right)=i\left[\hat{
P}^{\mu},\hat{\chi}\left(x\right)\right].
\end{eqnarray}
\subsubsection{Lorentz Transformations}
We consider an infinitesimal Lorentz transformation
\begin{eqnarray}
\Lambda^{\mu}_{~\nu}=\delta^{\mu}_{~\nu}+\omega^{\mu}_{~\nu},
\end{eqnarray} with a corresponding matrix
representation of the
$SL\left(2,\mathbb{C}\right)$~(cf.~(\ref{eq:sl2c}))
\begin{eqnarray}
S\left(\Lambda\right)^{\alpha}_{~\beta}=\delta^{\alpha}_{~\beta}-\frac{i}{2}\omega_{\mu\nu}\left(S^{\mu\nu}\right)^{\alpha}_{~\beta}.\end{eqnarray}
The fields transform according to~(\ref{eq:clf}) as
\begin{eqnarray}
\chi'^{\alpha}\left(x'\right)=S\left(\Lambda\right)^{\alpha}_{~\beta}\chi^{\beta}\left(x\right).\end{eqnarray}
 A Taylor expansion of the left hand side yields
\begin{eqnarray}\chi'^{\alpha}\left(x'\right)=\left[1+\frac12 \omega^{\lambda\sigma}\left(x_{\lambda}\partial_{\sigma}-x_{\sigma}\partial_{\lambda}
\right)\right]\chi'^{\alpha}\left(x\right)+\mathcal{O}\left(\omega^2\right)
.\end{eqnarray} The difference
$\delta\chi'^{\alpha}\left(x\right)=\chi'^{\alpha}\left(x\right)-\chi^{\alpha}\left(x\right)$
is then
\begin{eqnarray}
\label{eq:deltachi} \delta\chi^{\alpha}\left(x\right)=
-\frac{i}{2}\omega^{\lambda\sigma}\left[\left[S_{\lambda\sigma}\right]^{\alpha}_{~\beta}+L_{\lambda\sigma}\delta^{\alpha}_{~\beta}\right]\chi^{\beta}\left(x\right),\end{eqnarray}
where
$L_{\lambda\sigma}:=i\left(x_{\lambda}\partial_{\sigma}-x_{\sigma}\partial_{\lambda}\right)$.
Similarly, we can write the variation of the derivative of the
field as
\begin{eqnarray}\label{eq:dmudeltachi}\delta\partial_{\mu}\chi^{\alpha}\left(x\right)=
-\frac{i}{2}\omega^{\lambda\sigma}\left[\left[S_{\lambda\sigma}\right]^{\alpha}_{~\beta}+L_{\lambda\sigma}\delta^{\alpha}_{~\beta}\right]\partial_{\mu}\chi^{\beta}\left(x\right)-\omega_{\mu}^{~\sigma}\partial_{\sigma}\chi^{\beta}\left(x\right).
\end{eqnarray}
We assume $\mathcal{L}$ to be form invariant under the Lorentz
transformations~(\ref{eq:deltachi}) and~(\ref{eq:dmudeltachi}). On
using~(\ref{eq:elegen}),~(\ref{eq:emt}) and integration by parts
we get
\begin{eqnarray}
i\partial_{\mu}\left[\frac{\partial\mathcal{L}}{\partial\left(\partial_{\mu}\chi^{\alpha}\right)}\left[S_{\lambda\sigma}\right]_{~\beta}^{\alpha}\chi^{\beta}\right]-\mathcal{T}_{\lambda\sigma}+\mathcal{T}_{\sigma\lambda}=0.
\end{eqnarray}
Since the conservation law~(\ref{eq:DT}) does not define the
current in a unique way, we can always add a divergence of a total
antisymmetric tensor $\mathcal{A}^{\lambda\mu\nu}$ satisfying the
same conservation law.\footnote{This suggests to construct a new,
symmetric energy-momentum tensor $\tilde{\mathcal{T}}$, known as
Belinfante tensor \cite{Weinberg:1995},
\begin{eqnarray}
\tilde{\mathcal{T}}^{\mu\nu}=\mathcal{T}^{\mu\nu}+\partial_{\lambda}\mathcal{A}^{\lambda\mu\nu}.\end{eqnarray}
It can be shown that
\begin{eqnarray}
\partial_{\mu}\tilde{\mathcal{T}}^{\mu\nu}\left(x\right)=0,\quad P^{\mu}=\int_{\Sigma}\mathrm{d}\Sigma^{\nu}\left(x\right)\tilde{\mathcal{T}}^{\mu}_{\nu}\left(x\right).
\end{eqnarray}
From now on we will leave the tilde away and assume that
$\mathcal{T}^{\mu\nu}$ is symmetric,
$\mathcal{T}^{\mu\nu}=\mathcal{T}^{\nu\mu}$.} For symmetric
$\mathcal{T}^{\mu\nu}$ we may construct the angular-momentum
density as~\cite{Weinberg:1995}
\begin{eqnarray}\label{eq:Mmunusigma}
\mathcal{M}^{\mu\nu\sigma}\left(x\right):=x^{\nu}\mathcal{T}^{\mu\sigma}\left(x\right)-x^{\sigma}\mathcal{T}^{\mu\nu}\left(x\right)
,\end{eqnarray} conserved in the sense that
\begin{eqnarray}
\partial_{\mu}\mathcal{M}^{\mu\nu\sigma}\left(x\right)=0.
\end{eqnarray}
Then the corresponding conserved charge is the integral over a spacelike hypersurface
\begin{eqnarray}
\label{eq:Mmunu}M^{\mu\nu}=\int_{\Sigma}\mathrm{d}\Sigma_{\lambda}\left(x\right)\mathcal{M}^{\lambda\mu\nu}
\left(x\right).
\end{eqnarray}
The antisymmetric tensor $M^{\mu\nu}$ becomes an operator
$\hat{M}^{\mu\nu}$ in a quantum field theory generating
Lorentz transformations of the field operator, i.e. Lorentz boosts and spatial rotations in
the sense that
\begin{eqnarray}\label{eq:commM}
\left[\left(x^{\mu}\partial^{\nu}-x^{\nu}\partial^{\mu}\right)\delta^{\alpha}_{~\beta}-i\left[S^{\mu\nu}\right]^{\alpha}_{~\beta}
\right]\hat{\chi}^{\beta}\left(x\right)&=&
-i\left[L^{\mu\nu}\delta^{\alpha}_{~\beta}+\left[S^{\mu\nu}\right]^{\alpha}_{~\beta}\right]\hat{\chi}^{\beta}\left(x\right)\nonumber\\&=&i\left[\hat{\chi}^{\alpha}\left(x\right),\hat{M}^{\mu\nu}\right]
.\end{eqnarray}
\subsection{Microscopic Causality}
\label{sec:mc} Next we want to mention what is known as
microscopic causality. Two operators that describe integer spin
fields should commute, if they are spacelike separated, i.e.
\begin{eqnarray}\label{eq:mc}
\left[\hat{\phi}_{\alpha}\left(x\right),\hat{\phi}_{\beta}\left(y\right)\right]=0,\quad
\forall\left(x-y\right)_{\lambda}\,\left(x-y\right)^{\lambda}<0.
\end{eqnarray}
Similarly two operators that describe half-integer spin fields
should anticommute, if they are spacelike separated, i.e.
\begin{eqnarray}\left\lbrace\hat{\psi}_{\alpha}\left(x\right),\hat{\psi}_{\beta}\left(y\right)\right\rbrace=0,\quad
\forall\left(x-y\right)_{\lambda}\,\left(x-y\right)^{\lambda}<0.
\end{eqnarray}
\subsection{Fock Space}
\label{sec:Focksp} In all free field theories the total number of
particles $N$ is a constant in time. The Hilbert space of states
can be written as a direct infinite sum over all $N$ of tensor
products of $N$ single-particle Hilbert spaces
\begin{eqnarray}\label{eq:focksp}
\mathcal{F}=\bigoplus_{N=0}^{\infty}\left(\mathcal{H}_1\right)^{\otimes
N}=\mathbb{C}\oplus\mathcal{H}_{1}\oplus\left(\mathcal{H}_{1}\otimes\mathcal{H}_{1}\right)\oplus\left(
\mathcal{H}_{1}\otimes\mathcal{H}_{1}\otimes\mathcal{H}_{1}\right)\oplus\ldots
\end{eqnarray}
where the single-particle space $\mathcal{H}_{1}$ is a
representation space for a unitary irreducible representation of
the $\mathcal{P}_{+}^{\uparrow}$~\cite{Streater:2000ff}. The
linear Hilbert space $\mathcal{F}$ is called \emph{Fock space}. On
$\mathcal{F}$ we can define a complete set of (anti)commuting
self-adjoint operators that create or annihilate field quanta.
Thus every multi-particle state can be constructed by the action
of these creation operators on the vacuum. A complete set of these
multi-particle states form a basis that span $\mathcal{F}$. The
most common choice of a basis is the, so called, \emph{Wigner
basis}, which consists of simultaneous eigenstates of the
three-momentum operator $\hat{P}^{i}$ and an additional operator
describing the spin orientation $\sigma$. Therefore, a general
field operator representing particles with a certain mass and spin
can be written as an expansion of these creation and annihilation
operators. Furthermore, the generators for space-time translations
$\hat{P}^{\mu}$
expanded in the Wigner basis become diagonal. This Fock-space representation of the $\mathcal{P}_{+}^{\uparrow}$ is called \emph{Wigner representation}. \\
Another representation is the, so called, \emph{Lorentz
representation}. In the Lorentz basis, the Casimir operator of
$\mathcal{L}_{+}^{\uparrow}$,
$\propto\hat{M}_{\mu\nu}\,\hat{M}^{\mu\nu}$ together with the
square of the operator for total angular momentum
$\hat{J}^i=\epsilon_{ijk}\hat{M}^{jk}$ and one of its components
become diagonal \cite{Gromes:1974yu}. A problem of the Lorentz
basis is the definition of the four-momentum operator
$\hat{P}^{\mu}$ as self-adjoint operator acting on the Hilbert
space of square-integrable functions~\cite{Macdowell:1972ef}.
Therefore we will rather use the Wigner representation in the
following.
\subsection{Scattering Operator}
\label{sec:sop}
 The asymptotic incoming (outgoing) multi-particle
states labelled as $\vert \Phi_{\mathrm{in}}\rangle$
$(\vert\Phi_{\mathrm{out}}\rangle)$ span
 the Hilbert space
 $\mathcal{F}_{\mathrm{in}}$ $(\mathcal{F}_{\mathrm{out}})$ with a
 Fock-space structure as~(\ref{eq:focksp}).
 If
 \emph{asymptotic completeness} holds, namely that
 $\mathcal{F}_{\mathrm{in}}=\mathcal{F}_{\mathrm{out}}=\mathcal{F}$
 where $\mathcal{F}$ is the Fock space of the full interacting
 theory, then a unitary operator $\hat{S}:\,\mathcal{F}_{\mathrm{out}}\rightarrow\mathcal{F}_{\mathrm{in}}$ can be
 defined.
 $\hat{S}$ maps $\vert\Phi_{\mathrm{out}}\rangle$ of given momenta and spins to $\vert\Phi_{\mathrm{in}}\rangle$
 of the same momenta and spins~\cite{Streater:2000ff},
 \begin{eqnarray}\label{eq:scattopdef}
 \hat{S}:\,\vert\Phi_{\mathrm{out}}\rangle\mapsto\hat{S}\vert\Phi_{\mathrm{out}}\rangle=\vert\Phi_{\mathrm{in}}\rangle.\end{eqnarray}
 This operator is called \emph{scattering operator} (S operator). The
 S-matrix between the two states, $\vert\Phi\rangle$ and $\vert\Psi\rangle$, is then given by~\cite{Streater:2000ff}
\begin{eqnarray}
\langle\Phi_{\mathrm{out}}\vert
 \Psi_{\mathrm{in}}\rangle=\langle\Phi_{\mathrm{out}}\vert\hat{S}\,\Psi_{\mathrm{out}}\rangle=\langle\Phi_{\mathrm{in}}\vert\hat{S}\,\Psi_{\mathrm{in}}\rangle.
 \end{eqnarray}
 %
%
\chapter{Time Parameterization}
\label{cha:TP} Hamiltonian mechanics is the usual starting point
for canonical quantization of a non-relativistic theory. For a
relativistic theory the Hamiltonian formalism has to be adapted in
such a way that it is consistent with the requirements of
relativity, namely treating space and time equally. This
generalization results in a freedom of time choice, due to the
fact that we have to deal with a singular system. To illustrate
this, it is sufficient to consider a free classical relativistic
particle~\cite{Heinzl:1998kz,Heinzl:2000ht,Scheck:1999fg,Marnelius:1973yt}.
\section{Free Relativistic Particle}
\subsection{Singular System}
The state of motion of a free particle is characterized by the
relativistic energy-momentum vector lying on the mass shell,
\begin{eqnarray}\label{eq:p^2}
p_{\lambda}\,p^{\lambda}=m^2.
\end{eqnarray}
Since we have free relativistic motion and a flat space-time, the
solutions of Hamiltonian's variational principle will be straight
lines joining two points $y_1$ and $y_2$. This results in the
Lorentz-invariant ansatz for the action as integral over the path
between the timelike separated points $y_1$ and $y_2$
\begin{eqnarray}\label{eq:S}
S=-m\int_{s_1}^{s_2}\mathrm{d}s,\quad
\left(y_1-y_2\right)_{\lambda}\,\left(y_1-y_2\right)^{\lambda}>0.
\end{eqnarray}
On choosing an arbitrary parameterization, $\tau\mapsto
x^{\mu}\left(\tau\right)$, the invariant infinitesimal distance
becomes
\begin{eqnarray} \label{eq:ds}\mathrm{d}s=\sqrt{g_{\mu\nu}\mathrm{d}x^{\mu}\mathrm{d}x^{\nu}}=
\sqrt{\frac{\mathrm{d}x^{\mu}}{\mathrm{d}\tau}\frac{\mathrm{d}x_{\mu}}{\mathrm{d}\tau}}\mathrm{d}\tau=\sqrt{\eta\left(\tau\right)}\mathrm{d}\tau.
\end{eqnarray}
Here we have introduced the \emph{word-line metric}
$\eta\left(\tau\right)=\dot{x}_{\mu}\,\dot{x}^{\mu}=\left(\mathrm{d}s/\mathrm{d}\tau\right)^2$
with the four-velocity
$\frac{\mathrm{d}x^{\mu}}{\mathrm{d}\tau}=:\dot{x}^{\mu}$.
$\eta\left(\tau\right)$ can be viewed as an auxiliary
parameter.\footnote{We see that for $\eta\left(\tau\right)=1$ the
infinitesimal distance (since we have chosen the velocity of light
$c=1$ this coincides with the proper time) provides a natural
parameterization~\cite{Green:1998}.}
 Inserting~(\ref{eq:ds}) into~(\ref{eq:S}), we introduce the Lagrangian as
\begin{eqnarray}\label{eq:S2}
S=-m\int_{\tau_1}^{\tau_2}\mathrm{d}\tau\sqrt{\frac{\mathrm{d}x^{\mu}}{\mathrm{d}\tau}\frac{\mathrm{d}x_{\mu}}{\mathrm{d}\tau}}
=\int_{\tau_1}^{\tau_2}\mathrm{d}\tau L\left(\tau\right).
\end{eqnarray}
We see that the four velocity $\dot{x}^{\mu}$ is a timelike (or
lightlike) vector as long as the world-line metric is positive
(or zero) in order to preserve relativity. The Euler-Lagrange
equations corresponding to~(\ref{eq:S2}), resulting from
Hamiltonian's variational principle, namely that the action of the
chosen path becomes stationary, are
\begin{eqnarray}\label{eq:ele2}\frac{\partial L}{\partial x_{\mu}}-\frac{\mathrm{d}}{\mathrm{d}\tau}\frac{\partial
L}{\partial\dot{x}_{\mu}}=0 \quad \Rightarrow
\frac{\mathrm{d}}{\mathrm{d}\tau} \frac{m\dot{x}^{\mu}
}{\sqrt{\eta\left(\tau\right)}}=0.
\end{eqnarray}
The momentum canonically conjugate to $x$ is
\begin{eqnarray}\label{eq:cm}
\pi_{x}^{\mu}:=- \frac{\partial L}{\partial
\dot{x}_{\mu}}=m\frac{\dot{x}^{\mu}}{\sqrt{\eta\left(\tau\right)}}=\frac{m}{\sqrt{\eta\left(\tau\right)}}\frac{\mathrm{d}x^{\mu}}
{\mathrm{d}s} \frac{\mathrm{d}s}
{\mathrm{d}\tau}=m\frac{\mathrm{d}x^{\mu}} {\mathrm{d}s}\equiv
p^{\mu}.
\end{eqnarray}
From now on we will use $p$ for the momentum conjugate to
$x$.\footnote{Note that $\mathrm{d}x^{\mu}/\mathrm{d}s=:v^{\mu}$
is the invariant velocity by choosing the natural parameterization
$s$, i.e. $v_{\mu}\,v^{\mu}=1$. We see that for the choice of the
natural parameterization $s$ the length scale is fixed.} We see
that the canonical momenta are independent of
$\eta\left(\tau\right)$ and thus independent of the chosen
parameterization since squaring gives the mass shell constraint~(\ref{eq:p^2}). Unlike for the momenta~(\ref{eq:p^2}) the length
scale $\eta\left(\tau\right)$ for the velocities is not fixed in
general. The canonical Hamiltonian is given by the Legendre
transformation of the Lagrangian
\begin{eqnarray}\label{eq:H_c}
H_c=\frac{\partial L}{\partial\dot{x}^{\lambda}}\dot{x}^{\lambda}-L=0.
\end{eqnarray}
The canonical Hamiltonian vanishes and it seems that there is no
generator for time evolution. This is due to the fact that this
description of motion contains a redundant degree of freedom,
namely $\dot{x}^{0}$. Thus the dynamics of the system is hidden
in the constraint~(\ref{eq:p^2}). In fact, the Legendre
transformation~(\ref{eq:H_c}) from $L$ to $H_c$ cannot be
performed.\footnote{ The Legendre transformation~(\ref{eq:H_c})
cannot be performed since the condition
\begin{eqnarray}\label{eq:detA}
\mathrm{det}\left(\frac{\partial^2L}{\partial\dot{x}^{\mu}\partial\dot{x}^{\nu}}\right)\neq
0\nonumber
\end{eqnarray}
is not satisfied. Indeed, we calculate
\begin{eqnarray}\label{eq:detA2}
\frac{\partial^2L}{\partial\dot{x}^{\mu}\partial\dot{x}^{\nu}}=-\frac{m}{\left(\dot{x}_{\lambda}\,\dot{x}^{\lambda}\right)^{\frac{3}{2}}}\left(\dot{x}_{\sigma}\,\dot{x}^{\sigma}g_{\mu\nu}-\dot{x}_{\mu}\,\dot{x}_{\nu}\right)=
-\frac{m}{\left(\dot{x}_{\lambda}\,\dot{x}^{\lambda}\right)^{\frac{3}{2}}}A_{\mu\nu}.\nonumber
\end{eqnarray}
The determinant of the Hessian matrix $A_{\mu\nu}$ vanishes as
follows.
\\
The linear, homogeneous system
\begin{eqnarray}\label{eq:Av}
A_{\mu\nu}u^{\nu}=0\nonumber
\end{eqnarray}
has a non-trivial solution, if and only if
$\mathrm{det}\mathbf{A}=0$. Such a non-trivial solution may be
given by $u^{\nu}=c\dot{x}^{\nu}$ with $c=\mathrm{const.}$ .}
 Such a classical system is called \emph{singular}. If
it is not uniquely soluble for the $\dot{x}^{\mu}$, then the
momenta are not completely independent from each other, but they
have to satisfy constraints. These are called \emph{primary
constraints}~\cite{Dirac:1964aa} and their number is the
number of equations~(\ref{eq:cm}) minus the rank of the
determinant of the Hessian
$A^{\mu\nu}$~\cite{Kugo:1997fs}. Thus we have one
primary constraint given by~(\ref{eq:p^2}).
\subsection{Reparameterization Invariance}
\label{sec:ri}
To proceed we consider a reparameterization of the world line,
\begin{eqnarray}\label{eq:rwl}
\tau\mapsto\tau',\quad x^{\mu}\left(\tau\right)\mapsto
x^{\mu}\left(\tau'\left(\tau\right)\right),
\end{eqnarray} where the mapping $\tau\mapsto\tau'$ is injective
and $\mathrm{d}\tau'/\mathrm{d}\tau>0$ to conserve the
orientation. Since the Lagrangian is homogeneous of first degree
in
$\dot{x}^{\mu}$,\begin{eqnarray}\label{eq:lfd}L\left(c\dot{x}^{\mu}\right)=cL\left(\dot{x}^{\mu}\right),
\end{eqnarray}
it changes under reparameterization to
\begin{eqnarray}\label{eq:lrp}
L\left(\mathrm{d}x^{\mu}/\mathrm{d}\tau\right)=L\left(\left(\mathrm{d}x^{\mu}/\mathrm{d}\tau'\right)\left(\mathrm{d}\tau'/\mathrm{d}\tau\right)\right)=
\frac{\mathrm{d}\tau'}{\mathrm{d}\tau}L\left(\mathrm{d}x^{\mu}/\mathrm{d}\tau'\right).
\end{eqnarray}
This makes the action invariant under reparameterization, i.e.
\begin{eqnarray}\label{eq:arp}
S=\int_{\tau_1}^{\tau_2}\mathrm{d}\tau
L\left(\mathrm{d}x^{\mu}/\mathrm{d}\tau\right)=\int_{\tau'_1}^{\tau'_2}\mathrm{d}\tau'
\frac{\mathrm{d}\tau}{\mathrm{d}\tau'}
\frac{\mathrm{d}\tau'}{\mathrm{d}\tau}L\left(\mathrm{d}x^{\mu}/\mathrm{d}\tau'\right)=S',
\end{eqnarray} leaving the endpoint fixed,
$\tau_{1,2}=\tau'_{1,2}$. Thus the choice of the time parameter
is arbitrary and there is no absolute time.
Therefore~(\ref{eq:S2}) really characterizes the world line of
the particle independent of a particular choice of
coordinates~\cite{Green:1998}. According to Euler's theorem for
homogeneous functions we have
\begin{eqnarray}
L=\frac{\partial L}{\partial \dot{x}^{\mu}}\dot{x}^{\mu}=-p_{\mu}\,\dot{x}^{\mu},
\end{eqnarray}
which is equivalent to the vanishing canonical
Hamiltonian~(\ref{eq:H_c}). Clearly, in this case, the momenta are
homogeneous of degree zero. Thus reparameterization invariance of
the action implies that the Lagrangian is homogeneous of degree
one in velocities.\\
\\
The primary constraint
\begin{eqnarray}\label{eq:Theta}
\Theta\left(p\right):=p_{\lambda}\,p^{\lambda}-m^2=0
\end{eqnarray}
has zero Poisson bracket with itself. Any dynamical variable that
has vanishing Poisson bracket with the primary constraint is
called \emph{first class}~\cite{Dirac:1964aa}. Therefore $\Theta$
is called first class. It is important to note that we must not
use the constraint~(\ref{eq:Theta}) before working out a Poisson
bracket, therefore~(\ref{eq:Theta}) is called a weak equation.
That this first class primary constraint generates the
reparameterization invariance can be seen with the help
of~(\ref{eq:pbnt}).\\ On using the Poisson bracket $\left\lbrace
x^{\mu},p^{\nu}\right\rbrace_{P}=-g^{\mu\nu}$ and~(\ref{eq:cm})
 we calculate the change of coordinates induced by an infinitesimal reparameterization $\tau\mapsto\tau'=\tau+\delta\tau$,
\begin{eqnarray}
\delta x^{\mu}&:=&x^{\mu}\left(\tau+\delta\tau\right)-x^{\mu}\left(\tau\right)=\left\lbrace x^{\mu},
\Theta\left(p\right)\delta\epsilon\right\rbrace_{P}=-2p^{\mu}\delta\epsilon\nonumber\\&=&
-2m\frac{\dot{x}^{\mu}}{\sqrt{\eta\left(\tau\right)}}\delta\epsilon\equiv\dot{x}^{\mu}\delta\tau,
\end{eqnarray}
where we have identified $\delta\tau=-2m\frac{\delta\epsilon}{\sqrt{\eta\left(\tau\right)}}$ to account for the different dimensionalities. Thus reparameterization of the world line is indeed generated by
the constraint~(\ref{eq:Theta}).\\
\\
Invariance under reparameterization can be viewed as a redundancy
symmetry. There is a freedom of time choice similar to a freedom
of choice of gauge.
 A single world line (trajectory) can be
described by an infinite number of different parameterizations. We
are free to parameterize a world line by any parameter which can
be expressed by a monotonic increasing function of the particle's
proper time $s$. The trajectories are therefore equivalence
classes obtained by identifying all reparameterizations. The
choice of a particular time $\tau$ corresponds to the choice of a
particular foliation of Minkowski space-time in space and time. An
instant in the chosen time is described by a
three-dimensional hypersurface of equal $\tau$. 
Then time development is a continuous
evolution from one hypersurface $\Sigma_{\tau_0}:\tau=\tau_0$ to
another $\Sigma_{\tau_1}:\tau=\tau_1>\tau_0$. Consequently,
Minkowski space is decomposed into hypersurfaces of equal time
$\tau$.
\subsection{Space-Time Foliation}
\label{sec:STF} To find a particular foliation of Minkowski space,
we introduce a general coordinate transformation from the
Cartesian chart to a new chart
\begin{eqnarray}\label{eq:xi}
x^{\mu}\mapsto \xi^{\mu}\left(x\right)=\frac{\partial
\xi^{\mu}\left(x\right)}{\partial x^{\lambda}}x^{\lambda},
\end{eqnarray}
where the new coordinates $\xi^{\mu}\left(x\right)$ may be
curvilinear. General relativity demands invariance of the
infinitesimal line element in Riemann space\footnote{Riemann
space-time is a four-dimensional, connected, smooth manifold
$\mathcal{M}$ together with a Lorentz metric
$\eta\left(\xi\right)$ with signature $(+,-,-,-)$ defined on
$\mathcal{M}$.} under arbitrary coordinate transformations. Thus,
we have
\begin{eqnarray}\label{eq:ds^2}
\mathrm{d}s^2=g_{\mu\nu}\mathrm{d}x^{\mu}
\mathrm{d}x^{\nu}=g_{\mu\nu}\frac{\partial x^{\nu}}{\partial
\xi^{\sigma}} \frac{\partial x^{\mu}}{\partial
\xi^{\lambda}}\mathrm{d}\xi^{\lambda}\mathrm{d}\xi^{\sigma}=\eta_{\lambda\sigma}\left(\xi\right)\mathrm{d}\xi^{\lambda}\mathrm{d}\xi^{\sigma},
\end{eqnarray}
where
$\eta_{\lambda\sigma}\left(\xi\right)=g_{\mu\nu}\frac{\partial
x^{\nu}}{\partial \xi^{\sigma}} \frac{\partial x^{\mu}}{\partial
\xi^{\lambda}}$ denotes the coordinate dependent metric defined in
the new, non-inertial reference system. If we now choose
$\xi^0\left(x\right)$ to represent our time variable, i.e.
\begin{eqnarray}
 \tau\equiv\xi^0\left(x\right)=\frac{\partial \xi^{0}\left(x\right)}{\partial x^{\lambda}}x^{\lambda},
\end{eqnarray}
 then the remaining spatial coordinates $\xi^i\left(x\right), i=1,\ldots 3$ parameterize the three-dimensional hypersurface $\Sigma_{\tau}\left(x\right)$, which is curved in general.
The normal vector $n\left(x\right)$ on $\Sigma_{\tau}$ is defined
by
\begin{eqnarray}\label{eq:nv}
n_{\mu}\left(x\right)=\left.\frac{\partial
\xi^{0}\left(x\right)}{\partial
x^{\mu}}\right\vert_{\Sigma_{\tau}} .\end{eqnarray} The vector in
$\xi^0$-direction, i.e. the new velocity is
\begin{eqnarray}\label{eq:vel}
\frac{\partial x^{\mu}\left(\xi\right)}{\partial \xi^{0}}\equiv\dot{x}^{\mu}.
\end{eqnarray}
The relation between $n$ and $\dot{x}$ is
\begin{eqnarray}\label{eq:rvn}
n_{\lambda}\,\dot{x}^{\lambda}=\frac{\partial \xi^{0}}{\partial x^{\lambda}}\frac{\partial x^{\lambda}}{\partial \xi^{0}}=1.
\end{eqnarray}
From (\ref{eq:ds}) we find for the world-line metric

\begin{eqnarray}\label{eq:ds^22}
\mathrm{d}s^2&=&\eta\left(\tau\right)\mathrm{d}\tau^2=\eta_{\lambda\sigma}\left(\xi\right)\mathrm{d}\xi^{\lambda}\mathrm{d}\xi^{\sigma}\nonumber\\
\label{eq:eta}&\Rightarrow&
\eta\left(\tau\right)=\eta_{\lambda\sigma}\left(\xi\right)\dot{\xi}^{\lambda}\dot{\xi}^{\sigma}\equiv
\dot{x}_{\nu}\,\dot{x}^{\nu},
\end{eqnarray}
with the velocities $\frac{\mathrm{d}
\xi^{\mu}}{\mathrm{d}\tau}=\dot{\xi}^{\mu}$. Accordingly, the
world-line metric provides an arbitrary scale for velocities in
all coordinate systems. The Lagrangian in the new coordinates
becomes
\begin{eqnarray}\label{eq:Lxi}
L\left(\tau\right)=-m\sqrt{\eta\left(\tau\right)}=-m\sqrt{\eta_{\lambda\sigma}\left(\xi\right)\dot{\xi}^{\lambda}\dot{\xi}^{\sigma}}.
\end{eqnarray}
The momentum canonically conjugate to $\xi$ is defined by
\begin{eqnarray}\label{eq:varpi}
\pi_{\mu}=-\frac{\partial L}{\partial
\dot{\xi}^{\mu}}=\frac{m}{\sqrt{\eta\left(\tau\right)}}\eta_{\mu\lambda}\left(\xi\right)\dot{\xi}^{\lambda}=
\frac{m}{\sqrt{\eta\left(\tau\right)}}\frac{\partial x^{\sigma}
}{\partial\xi^{\mu}}\dot{x}_{\sigma}=\frac{\partial x^{\sigma}
}{\partial\xi^{\mu}}p_{\sigma},
\end{eqnarray} on
using~(\ref{eq:cm}). This is just the coordinate transform of the
momentum. They have to satisfy $\left\lbrace
\xi^{\mu},\pi^{\nu}\right\rbrace_{P}=-\eta^{\mu\nu}\left(\xi\right)$.
The canonical Hamiltonian as Legendre transform of the Lagrangian
becomes
\begin{eqnarray}\label{eq:Hc2}
H_c=\frac{\partial L}{\partial
\dot{\xi}^{\lambda}}\dot{\xi}^{\lambda}-L=-\pi_{\lambda}\dot{\xi}^{\lambda}-L=
-\frac{\sqrt{\eta\left(\tau\right)}}{m}\left(\eta^{\lambda\sigma}\left(\xi\right)\pi_{\lambda}
\pi_{\sigma}-m^2\right).
\end{eqnarray}
This vanishes on using~(\ref{eq:varpi}) and~(\ref{eq:Theta}),
\begin{eqnarray}\label{eq:pi^2}
\Theta(\pi)&=& \eta^{\lambda\sigma}\left(\xi\right)\pi_{\lambda}
\pi_{\sigma}-m^2=
\eta^{\lambda\sigma}\left(\xi\right)\frac{\partial
x^{\mu}}{\partial \xi^{\lambda}} \frac{\partial x^{\nu}}{\partial
\xi^{\sigma}}p_{\mu}p_{\nu}-m^2\nonumber\\&=&g^{\mu\nu}p_{\mu}p_{\nu}-m^2=0.
\end{eqnarray}
Thus, the canonical Hamiltonian vanishes in any coordinate system.
\\
A possible way to proceed is to make use of the Dirac-Bargmann
algorithm by introducing the primary constraint~(\ref{eq:p^2})
into the Hamiltonian by means of a Lagrangian multiplier
$\lambda$,
\begin{eqnarray}\label{eq:H}
H=H_c+\lambda\Theta\left(p\right),\quad \Theta\left(p\right):=p_{\lambda}\,p^{\lambda}-m^2.
\end{eqnarray}
The Hamiltonian equations of motion become, using~(\ref{eq:cm})
and $\left\lbrace x^{\mu},p^{\nu}\right\rbrace_{P}=-g^{\mu\nu}$,
\begin{eqnarray}\label{eq:Hem1}
\dot{x}^{\mu}&=&\left\lbrace x^{\mu},H\right\rbrace_{P}=-2\lambda p^{\mu},\\
\label{eq:Hem2} \dot{p}^{\mu}&=&\left\lbrace
p^{\mu},H\right\rbrace_{P}=0.
\end{eqnarray}
We see that (\ref{eq:Hem1}) contains the unknown Lagrange
multiplier $\lambda$ which makes the whole dynamics of the system
undetermined. Comparing (\ref{eq:Hem1}) with (\ref{eq:cm}) we
obtain
\begin{eqnarray}\label{eq:lambda}\lambda=
-\frac{\sqrt{\eta\left(\tau\right)}}{2m}.
\end{eqnarray}
To determine $\lambda$ or $\eta$, we have to fix a time. This is
achieved by imposing an auxiliary condition of the form
\begin{eqnarray}\label{eq:Xi}
\Xi\left(x;\tau\right)=0.
\end{eqnarray} Consistency
with~(\ref{eq:Hem1}) and~(\ref{eq:Hem2}) requires conservation in
time, leading to the stability condition
\begin{eqnarray}\label{eq:dotXi}
\dot{\Xi}=\frac{\partial \Xi}{\partial \tau}+ \left\lbrace
\Xi,H\right\rbrace_{P}=\frac{\partial \Xi}{\partial \tau}+\lambda
\left\lbrace \Xi,\Theta\right\rbrace_{P}=\frac{\partial
\Xi}{\partial \tau}-2\lambda p^{\mu}\frac{\partial \Xi}{\partial
x^{\mu}}\stackrel{!}{=}0.
\end{eqnarray}
Solving for $\lambda$ gives\footnote{ If~(\ref{eq:dotXi}) does not
determine $\lambda$, then we call it \emph{secondary constraint}
which has to be posed in additional to the primary
constraint~(\ref{eq:Theta}).}
\begin{eqnarray}\label{eq:lambda}
\lambda=- \frac{1}{\left\lbrace
\Xi,\Theta\right\rbrace_{P}}\frac{\partial \Xi}{\partial
\tau}=\frac{1}{2p^{\mu}\,\partial_{\mu} \Xi}\frac{\partial
\Xi}{\partial \tau}.
\end{eqnarray}
We see that $\Xi$ must depend explicitly on the time parameter
$\tau$ and at least one of the $x^{\mu}$ in order to get a finite
$\lambda$. This is equivalent with a non-vanishing Poisson bracket
$\left\lbrace \Xi,\Theta\right\rbrace_{P}$. Thus, (\ref{eq:Xi})
requires $\tau$ to be a function of $x^{\mu}$. Altogether, this
suggests $\Xi$ to have the form
\begin{eqnarray}\label{eq:Xi2}\Xi\left(x;\tau\right)=\tau-\xi^{0}\left(x\right).
\end{eqnarray}
We see from (\ref{eq:nv}), that
$p^{\mu}\,\partial_{\mu}\Xi=-p^{\mu}\,\partial_{\mu}\xi^{0}$ is
just the projection of $p^{\mu}$ onto the normal vector
$n^{\mu}\left(x\right)$ of the hypersurface $\Sigma_{\tau}$.
Inserting for $\lambda$ in (\ref{eq:Hem1}) gives
\begin{eqnarray}\label{eq:dotx}
\dot{x}^{\mu}=\frac{p^{\mu}}{p^{\lambda}\,n_{\lambda}\left(x\right)}.
\end{eqnarray}
Squaring gives the world-line metric as
\begin{eqnarray}\label{eq:eta2}\eta\left(\tau\right)=
\frac{m^2}{\left(p^{\lambda}\,n_{\lambda}\left(x\right)\right)^2}.
\end{eqnarray}
We know from~(\ref{eq:varpi}) how $\dot{x}^{\mu}$ and $p^{\mu}$
transform under coordinate transformations. Thus we can
immediately give the dynamics of the $\xi^{\mu}$ using~(\ref{eq:dotx})
\begin{eqnarray}\label{eq:dotxi}
\dot{\xi}^{\mu}=\frac{\partial \xi^{\mu}}{\partial
x^{\lambda}}\dot{x}^{\lambda}=\frac{\partial \xi^{\mu}}{\partial
x^{\lambda}}\frac{p^{\lambda}}{p^{\sigma}\,n_{\sigma}\left(x\right)}=\frac{\pi^{\mu}}{p^{\sigma}\,n_{\sigma}\left(x\right)}.
\end{eqnarray}
Then, the Hamiltonian $H_{\tau}$, i.e. the variable canonically
conjugate to $\tau$ generating $\tau$-evolution, is explicitly
given by
\begin{eqnarray}\label{eq:varpi0}H_{\tau}\equiv\pi^{0}=p^{\sigma}\,n_{\sigma}\left(x\right).
\end{eqnarray}
\section{Forms of Relativistic Dynamics}
\label{sec:frd} The Poincar\'e group $\mathcal{P}$ is the symmetry
group of any relativistic system. Consequently, the system
described above should be Poincar\'e invariant. Therefore, the
representations~(\ref{eq:P}) and~(\ref{eq:M}) should take into
account the constraint~(\ref{eq:Theta}), which guarantees
relativistic causality as it generates the dynamics. Proceeding as
before, choosing a time parameter $\tau$ leads to a particular
foliation of space-time into hypersurfaces $\Sigma$. A necessary
condition for causality is that the hypersurfaces should intersect
all possible world lines once and only once and therefore be
spacelike. For an arbitrary spacelike hypersurface
$\Sigma_{\tau}$ given by
$\tau=\xi^{0}\left(x\right)=\mathrm{const.}$, with $\tau$
arbitrary but fixed, we can analyze its transformation properties
under the action of the Poincar\'e generators~(\ref{eq:P})
and~(\ref{eq:M}). If a generator maps $\Sigma_{\tau}$ onto itself
for all $\tau$, i.e. if it leaves the hypersurfaces $\Sigma$
invariant, we call the generator \emph{kinematic}. Otherwise we
call it \emph{dynamic}. All kinematic generators span a subgroup
of $\mathcal{P}$, the so called \emph{stability group}
$\mathcal{P}_{\Sigma}$, whereas the dynamic generators are often
referred as \emph{Hamiltonians}. The latter map $\Sigma_{\tau}$
onto another hypersurface and thus involve evolution of the
system.
 When including interactions into the system (usually via an interaction term in the Lagrangian), only the dynamic generators will be affected whereas the kinematic generators stay interaction independent.\\
The higher the symmetry of the hypersurface, the larger will be
$\mathcal{P}_{\Sigma}$. There is a further requirement, namely
that any two points on $\Sigma$ can be connected by a
transformation generated by the stability
group~\cite{Heinzl:2000ht}. In his paper \emph{Forms of
Relativistic Dynamics}~\cite{Dirac:1949cp}, Dirac found three
different hypersurfaces with large stability groups of dimensions
$6,6$ and $7$ and called them \emph{instant}, \emph{point} and
\emph{front form}, respectively.\footnote{Two others were found
later, but they have smaller stability groups of dimension
4~\cite{Leutwyler:1977vy}. In the following only the first three
found by Dirac are discussed.}
\\
Transformations generated by an element of the stability group
must leave the hypersurface $\Sigma_{\tau}:
\tau=\xi^{0}\left(x\right)$ invariant. Consequently, we have for a
kinematic component $P^{\mu}$ of the four-vector $P$,
using~(\ref{eq:D}) for infinitesimal $a$, the condition
\begin{eqnarray}
\Omega\left(a,\mathbf{1}\right)\xi^{0}\left(x\right)&=&\xi^{0}\left(x\right)+ia_{\lambda}\,P^{\lambda}\xi^{0}\left(x\right)\stackrel{!}{=}\xi^{0}\left(x\right)\\
\label{eq:kP} \Rightarrow
P^{\mu}\xi^{0}\left(x\right)&=&i\partial^{\mu}\xi^{0}\left(x\right)=0.
\end{eqnarray}
Similarly, we have for a kinematic component $M^{\mu\nu}$ of the
tensor $\mathbf{M}$ and infinitesimal $\omega$
\begin{eqnarray}\label{eq:kM} M^{\mu\nu}\xi^{0}\left(x\right)=i\left(x^{\mu}\partial^{\nu}-x^{\nu}\partial^{\mu}\right)\xi^{0}\left(x\right)=0.
\end{eqnarray}
In terms of components of $x^{\mu}$ and $n^{\mu}\left(x\right)$ of
the vector normal on $\Sigma_{\tau}$~(\ref{eq:nv}) these equations
read
\begin{eqnarray}\label{eq:knv}
n^{\mu}\left(x\right)=0,\quad x^{\mu}n^{\nu}\left(x\right)-x^{\nu}n^{\mu}\left(x\right)=0.
\end{eqnarray}
If $\xi^{0}\left(x\right)$ has a non-trivial stability group,~(\ref{eq:knv}) is satisfied for at least one $\nu$ and/or $\mu$~\cite{Heinzl:2000ht}.
\subsection{Instant Form}
\label{sec:if} 
\begin{figure}
 \begin{center}
 \includegraphics[clip=7cm,width=8.9cm]{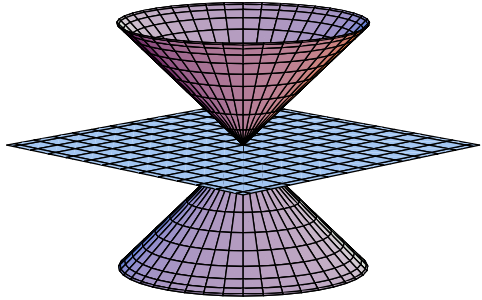}
 \caption{\label{fig:instantform2}A hypersurface in instant form
 is a hyperplane
 defining an instant in Minkowski time, here for $x^0=t=0$.}
 \end{center}
 \end{figure}

\begin{figure}
 \begin{center}
  \includegraphics[clip=7cm,width=8.9cm]{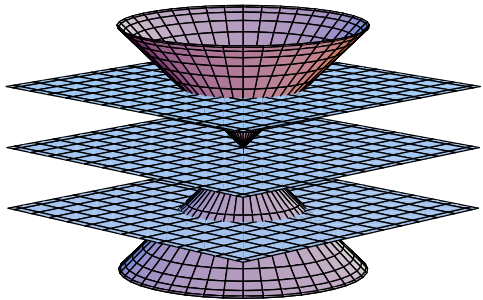}
 \caption{\label{fig:ifev3D}
 Time development in instant form generated by
 $P^0$.}
 \end{center}
 \end{figure}

\begin{figure}
 \begin{center}
\psfrag{St2}{$\Sigma_{t_2}$}
\psfrag{St1}{$\Sigma_{t_1}$}
\psfrag{S0}{$\Sigma_{t_0}$}
\psfrag{0}{$0$}
\psfrag{x0}{$x^0$}
\psfrag{vecx}{$\boldsymbol{x}$}
 \includegraphics[clip=100cm,width=8cm]{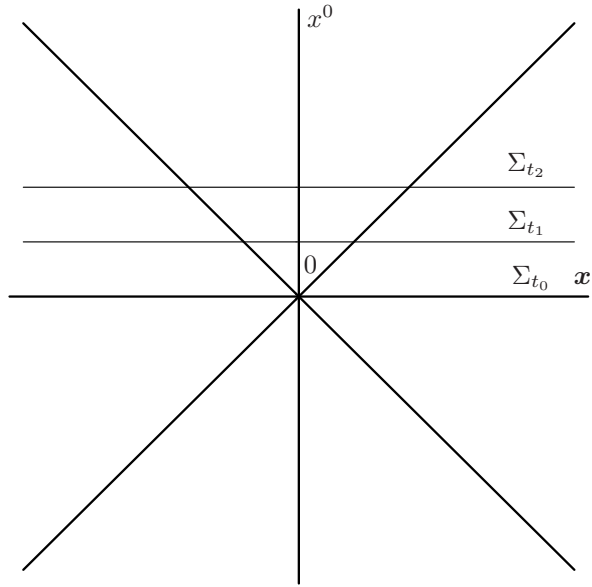}
 \caption{\label{fig:iftd}
 Time development in instant form generated by
 $P^0$ corresponds to evolution from one hyperplane $\Sigma_{t}:x^0=t$ to another $\Sigma_{t+\Delta t}:x^0=t+\Delta t$ (for $t,\Delta t=\mathrm{const.}$), here illustrated for
 $t_0=0<t_1<t_2$.}
 \end{center}
 \end{figure}

The most common choice for $\tau$ is the
Minkowskian time $\xi^{0}\left(x\right)=x^0=t$. The hypersurfaces
$\Sigma_{t}$ are planes isomorphic to $\mathbb{R}^3$ with the
normal vector $n=\left(1,\boldsymbol{0}\right)^T$ parallel to the
Minkowskian time coordinate, as shown in
Figure~\ref{fig:instantform2}.
From~(\ref{eq:kP}) and~(\ref{eq:kM}) we find
 \begin{eqnarray}\label{eq:kPi}
P^ix^0&=&0,\quad M^{ij}x^0=-M^{ji}x^0=0, \\
P^0x^0&\neq&0,\quad M^{i0}x^0=-M^{0i}x^0\neq0 \quad\forall\,
i,j=1,\ldots 3.
\end{eqnarray}
Thus the generators for space translations $P^i$ and space
rotations~$J^i=\epsilon_{ijk}M^{jk}$ are kinematic. The generator
for Minkowskian time evolution $H_{t}=P^0$ and the generators for
Lorentz boosts $B^i=M^{i0}$ become dynamic. Hence, time evolution
from $\Sigma_{t}$ to $\Sigma_{t+\Delta t}$ will be generated by
$P^0$~(cf.~Figures~\ref{fig:ifev3D} and~\ref{fig:iftd}).
Furthermore, $\Sigma_{t}$ will be invariant under space
translations and rotations,
 but not boost invariant, which is an expected result since boost mix space and time.
\newpage
\subsection{Front Form}
\begin{figure}
 \begin{center}
 \includegraphics[clip=7cm,width=8.9cm]{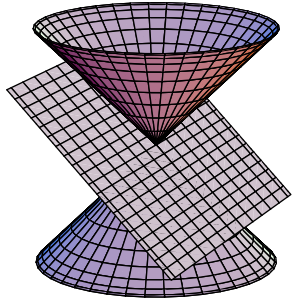}
 \caption{\label{fig:fform}
The hypersurface in front form is the null plane defining an
instant in $x^{-}$.
 }
 \end{center}
 \end{figure}

The choice $\tau=x^0+x^3=:x^{+}$ corresponds to a hypersurface
$\Sigma_{x^{+}}$ representing a hyperplane tangent to the light
cone, the, so called, \emph{null plane}~(cf.~Figure~\ref{fig:fform}). In front
form it is useful to introduce light-cone coordinates with a
metric tensor containing off-diagonal elements. Furthermore,
$n^{\mu}$ from~(\ref{eq:nv}) does not coincide with
$\dot{x}^{\mu}$ from~(\ref{eq:vel}), but $n_{\mu}\dot{x}^{\mu}=1$
still holds. At this point we won't go into details, but just to
mention the important feature of the front form having the largest
stability group with dimension 7.
\newpage
\subsection{Point Form}
\label{sec:pf} 
\begin{figure}
 \begin{center}
 \includegraphics[clip=7cm,width=8.9cm]{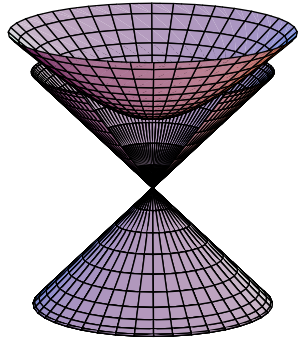}
 \caption{\label{fig:pform}
A hypersurface in point form is a hyperboloid defining an instant
in $\tau$.
 }
 \end{center}
 \end{figure}

\begin{figure}
 \begin{center}
  \includegraphics[clip=7cm,width=8.9cm]{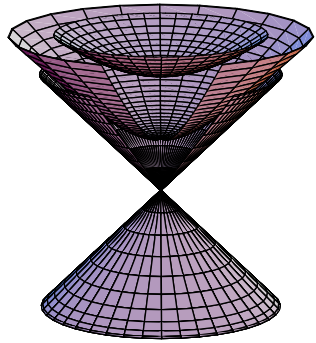}
 \caption{\label{fig:pfev3D}
 $\tau$-development generated by
 $D$.}
 \end{center}
 \end{figure}

\begin{figure}
 \begin{center}
\psfrag{St3}{$\Sigma_{\tau_3}$}
\psfrag{St2}{$\Sigma_{\tau_2}$}
\psfrag{St1}{$\Sigma_{\tau_1}$}
\psfrag{St0}{$\Sigma_{\tau_0}$}
\psfrag{0}{$0$}
\psfrag{x0}{$x^0$}
\psfrag{vecx}{$\boldsymbol{x}$}
 \includegraphics[clip=7cm,width=8cm]{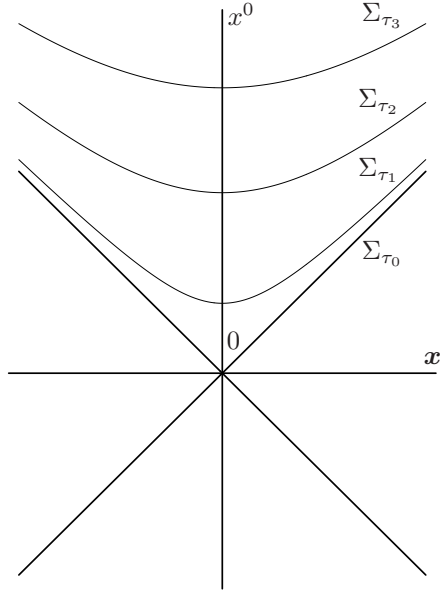}
 \caption{\label{fig:pftd}
$\tau$-development generated by the dilatation generator $D$
corresponds to evolution from one hyperboloid,
$\Sigma_{\tau}:x_{\lambda}\,x^{\lambda}=\tau^2$ to another
$\Sigma_{\tau+\Delta\tau}:x_{\lambda}\,x^{\lambda}=\left(\tau+\Delta\tau\right)^2$
(for $\tau,\Delta\tau=\mathrm{const.}$), here illustrated for
$\tau_0=0<\tau_1<\tau_2<\tau_3$. Note that $\Sigma_{\tau=0}$
represents the light cone.
 }
 \end{center}
 \end{figure}

The choice
$\xi^0\left(x\right)=\tau=\sqrt{x_{\lambda}\,x^{\lambda}}$ of time
corresponds to a curved equal-$\tau$ hypersurface
$\Sigma_{\tau}:x_{\lambda}\,x^{\lambda}=\tau^2$ describing a
hyperboloid in space-time~(Figure~\ref{fig:pform}).
The curvilinear coordinates parameterizing these hyperboloids are
introduced by the coordinate transformation
 \begin{eqnarray}\label{eq:xhc}
x^{\mu}\left(\alpha,\beta,\vartheta
,\varphi\right)=\mathrm{e}^{\alpha}
\left(%
\begin{array}{c}
  \mathrm{cosh}\beta\\
  \mathrm{sinh}\beta\,\mathrm{sin}\vartheta \,\mathrm{cos}\varphi\\
\mathrm{sinh}\beta\,\mathrm{sin}\vartheta \,\mathrm{sin}\varphi\\
\mathrm{sinh}\beta\,\mathrm{cos}\vartheta
\end{array}
\right)^{\mu},
\end{eqnarray} with $\mathrm{e}^{\alpha}=\tau$.
The metric is given by
\begin{eqnarray}\label{eq:etamunu}
\eta_{\mu\nu}\left(\tau,\beta,\vartheta ,\varphi\right)=
\left(%
\begin{array}{cccc}
1&0&0&0 \\
0&-\tau^2&0&0\\
0&0&-\tau^2\mathrm{sinh}^2\beta&0\\
0&0&0&-\tau^2\mathrm{sinh}^2\beta\,\mathrm{sin}^2\vartheta
\end{array}
\right)_{\mu\nu}.
\end{eqnarray}
This leads together with~(\ref{eq:eta}) to the world-line metric
\begin{eqnarray}\label{eq:etapf}
\eta\left(\tau\right)=1-\tau^2\left(\frac{\mathrm{d}\beta}{\mathrm{d}\tau}\right)^2-\tau^2\mathrm{sinh}^2\beta\left(\frac{\mathrm{d}\vartheta
}{\mathrm{d}\tau}\right)^2
-\tau^2\mathrm{sinh}^2\beta\,\mathrm{sin}^2\vartheta
\left(\frac{\mathrm{d}\varphi}{\mathrm{d}\tau}\right)^2.
\end{eqnarray}
From~(\ref{eq:nv}) we find for the normal vector on the
hyperboloid $\Sigma_{\tau}$,
$n^{\mu}\left(x\right)=\frac{x^{\mu}}{\tau}$. This is in that case
identical with the velocity
$\frac{\mathrm{d}x^{\mu}}{\mathrm{d}\tau}$~(\ref{eq:vel}) and can
be interpreted as a four-dimensional radial vector. The
Hamiltonian, i.e. the generator for $\tau$-evolution from
$\Sigma_{\tau}$ to $\Sigma_{\tau+\Delta\tau}$ is given
by~(\ref{eq:varpi0}) as
\begin{eqnarray}\label{eq:H}
H_{\tau}\equiv
D=n_{\lambda}\,P^{\lambda}=\frac{x_{\lambda}\,P^{\lambda}}{\tau},\quad
\tau\neq0.
\end{eqnarray}
$H_{\tau}$ is identified as the generator for dilatation
transformations denoted by $D$, which does not belong to the
Poincar\'e group but to the bigger \emph{conformal group}.
From~(\ref{eq:H}) we note the explicit $\tau$-dependence.
Furthermore, $D$ is only defined for $\tau>0$, thus
$\tau$-evolution is restricted to the forward light
cone~(cf.~Figures~\ref{fig:pfev3D} and~\ref{fig:pftd}). It becomes
rather difficult to describe $\tau$-evolution from the backward
light cone to the forward light cone, which is necessary when
formulating a scattering theory within
this approach (cf.~Chapter~\ref{cha:st}).\\
From~(\ref{eq:kP}) and~(\ref{eq:kM}) we find that
\begin{eqnarray}\label{eq:pmkpf}
P^{\mu}\sqrt{x_{\lambda}\,x^{\lambda}}\neq0,\quad
M^{\mu\nu}\sqrt{x_{\lambda}\,x^{\lambda}}=0.
\end{eqnarray}
This shows that in point form the generators for space-time
translations become dynamic, whereas the generators for Lorentz
transformations are kinematic. This manifest Lorentz covariance is
the typical feature of the point form.
\newpage
\section{General Evolution}
\label{sec:gtd} From~(\ref{eq:Pmu}) and~(\ref{eq:Mmunu}) we are
able to derive a general formula for a Hamiltonian of a system
with a given foliation $\tau=\xi^0\left(x\right)$ that generates
displacements of a hypersurface $\Sigma_{\tau}$. We define the
operator $G_{\zeta}$ by means of the energy-momentum
tensor~(\ref{eq:emt}) as
\begin{eqnarray}\label{eq:Gepsilon1}
G_{\zeta}=\int_{\Sigma}\mathrm{d}\Sigma_{\tau}^{\nu}\left(x\right)\zeta_{\mu}\left(x\right)\mathcal{T}^{\mu}_{\nu}\left(x\right),
\end{eqnarray}
where
\begin{eqnarray}\label{eq:dSigma}
\mathrm{d}\Sigma_{\tau}^{\mu}\left(x\right)&=&\mathrm{d}^4x\frac{\partial
\xi^0\left(x\right)}{\partial
x_{\mu}}\delta\left(\xi^0\left(x\right)-\tau\right)=\mathrm{d}^4x\,
n^{\mu}\left(x\right)\delta\left(\xi^0\left(x\right)-\tau\right)\nonumber\\&=&
n^{\mu}\left(x\right)\mathrm{d}\Sigma_{\tau}\left(x\right).
\end{eqnarray}
The quantity $G_{\zeta}$ generates the infinitesimal
transformation
\begin{eqnarray}
\label{eq:trafo} x^{\mu}\mapsto
x'^{\mu}=x^{\mu}+\zeta^{\mu}\left(x\right),\quad
\chi'\left(x\right)=
\chi\left(x\right)+\zeta^{\mu}\left(x\right)\partial_{\mu}\chi\left(x\right),
\end{eqnarray}
where $\zeta^{\mu}\left(x\right)$ is a function of $x$. If we
choose a particular space-time foliation, we find two classes of
operators from~(\ref{eq:Gepsilon1}), depending on the form of
$\zeta^{\mu}\left(x\right)$. The operators of the first class are
called \emph{kinematic}, if the hypersurface $\Sigma_{\tau}:~\tau=\xi^0\left(x\right)$ is invariant under
transformations~(\ref{eq:trafo}), i.e.
\begin{eqnarray}
\label{eq:trafo2}
\delta\xi^0\left(x\right):=\xi^0\left(x'\right)-\xi^0\left(x\right)=\zeta^{\mu}\left(x\right)\partial_{\mu}\xi^0\left(x\right)=\zeta^{\mu}\left(x\right)n_{\mu}\left(x\right)=0.
\end{eqnarray}
Otherwise, i.e. if
\begin{eqnarray}
\label{eq:trafo3}
\delta\xi^0\left(x\right)=\zeta^{\mu}\left(x\right)n_{\mu}\left(x\right)\neq0,
\end{eqnarray} then $G_{\zeta}$ belongs to the second class and is called \emph{dynamic}.
\\
When including interactions into the energy-momentum tensor via an
interaction Lagrangian
$\mathcal{L}_{\mathrm{int}}\left(x\right)$
\begin{eqnarray}
\label{eq:intenm}
\mathcal{T}^{\mu\nu}\left(x\right)\rightarrow\mathcal{T}^{\mu\nu}\left(x\right)-g^{\mu\nu}\mathcal{L}_{\mathrm{int}}\left(x\right)
,\end{eqnarray} we find that the kinematic operators are
independent of $\mathcal{L}_{\mathrm{int}}$. The interaction part
of $G_{\zeta}$ is
\begin{eqnarray}-\int_{\mathbb{R}^4}\mathrm{d}^4x\,\delta\left(\xi^0\left(x\right)-\tau\right)
n^{\mu}\left(x\right)\zeta_{\mu}\left(x\right)\mathcal{L}_{\mathrm{int}}\left(x\right)=0,
\end{eqnarray} if~(\ref{eq:trafo2}) holds.
We note the important result, that when including interactions
into the theory, the dynamic operators become interaction
dependent, whereas the kinematic operators stay interaction
free.\\
Furthermore, for particular choices of $\zeta^{\mu}\left(x\right)$
we recover the generators of the Poincar\'e group: $P^{\mu}$
corresponds to
$\zeta^{\mu\nu}\left(x\right)=g^{\mu\nu}$~(cf.~(\ref{eq:Pmu})) and
$M^{\mu\nu}$ to the choice
$\zeta^{\mu\nu\rho}\left(x\right)=x^{\mu}g^{\nu\rho}-x^{\nu}g^{\mu\rho}$~(cf.~(\ref{eq:Mmunu}))~\cite{Fubini:1972mf,Zelzer:2005}.\footnote{
For completeness we note that the dilatation generator
$D$~(\ref{eq:H}) corresponds to the choice
$\zeta^{\mu}\left(x\right)=x^{\mu}$, the choice
$\zeta^{\mu\nu}\left(x\right)=2x^{\mu}x^{\nu}-g^{\mu\nu}x_{\lambda}\,x^{\lambda}$
corresponds to the generator for special conformal transformations
$K^{\mu}$. Together with the Poincar\'e generators they obey
commutation relations, the so called conformal algebra of the 15
parameter conformal group. With $K^{\mu}$ and $P^{\mu}$ we find
another kinematic operator
$x_{\lambda}\,x^{\lambda}P^{\mu}-K^{\mu}$ leaving the hyperboloid
invariant~\cite{Fubini:1972mf}.}
\chapter{Covariant Canonical Quantization of Free Fields}
\label{cha:ccfq} The common procedure of canonical quantization
consists of posing canonical (anti)commutation relations on the
field operators at equal Minkowski times $x^0=y^0=t$. This
corresponds to quantization on the hyperplane $\Sigma_{t}$.
Adopting Dirac's nomenclature~\cite{Dirac:1949cp} we shall call it
therefore \emph{instant-form field quantization}. In
Chapter~\ref{cha:TP} we have found the freedom in choosing a
space-time foliation of Minkowski space as a characteristic
feature of relativistic parameterized systems. Generalizing these
ideas to a field theory leads to canonical (anti)commutation
relations imposed on an arbitrary spacelike hypersurface. This
corresponds to a particular choice of space-time foliation. In his
paper Schwinger~\cite{Schwinger:1948yk} proposes this way of
generalized canonical field quantization without making a
particular choice of time. In the following chapter we shall apply
this idea to the point form and quantize field theories on the
Lorentz-invariant hyperboloid $\Sigma_{\tau}$ by imposing
Lorentz-invariant canonical (anti)commutation relations. Therefore
we shall speak of \emph{point-form quantum field theory}. As we
have seen, the particular choice of space-time foliation should
not play a role for the dynamics of a relativistic theory. This is
expressed by the reparameterization invariance of the action
(cf.~Section \ref{sec:ri}). The Lie algebra~(\ref{eq:la})
demanding Poincar\'e invariance of the theory should hold for any
form of dynamics. In particular for a free theory, the Poincar\'e
generators should be essentially the same, which can be explicitly
shown using a common Fock basis. To see this equivalence, we will
make use of the, so called, \emph{Wigner basis} which consists of
simultaneous eigenstates of the three-momentum and spin.

\section{Complex Klein-Gordon Fields}
\label{sec:sf} We consider a free classical complex scalar field
theory in ($3+1$)-dimensional Minkowski space-time, with
$\phi\left(x\right),\phi^{\ast}\left(x\right)$ describing fields
with electric charge. Starting with a free Klein-Gordon Lagrangian
density \begin{eqnarray} \label{eq:kgld}
\mathcal{L}_{\mathrm{KG}}\left(x\right)=\partial_{\mu}\phi^{\ast}\left(x\right)\partial^{\mu}\phi\left(x\right)-m^2\phi^{\ast}\left(x\right)\phi\left(x\right),
\end{eqnarray} the classical action functional is defined by~(\ref{eq:action})
\begin{eqnarray} S_{\mathrm{KG}}\left[\phi,\phi^{\ast}\right]:=\int_{\mathbb{R}^4} \mathrm{d}^4 x\,\mathcal{L}_{\mathrm{KG}}\left(x\right)
.\end{eqnarray} The least action principle~(\ref{eq:elegen}) of varying the action
functional with respect to
$\phi\left(x\right),\phi^{\ast}\left(x\right)$ gives
\begin{eqnarray}
\label{eq:deltaaaction}
\frac{\delta S_{\mathrm{KG}}\left[\phi,\phi^{\ast}\right]}{\delta\phi\left(x\right)}\stackrel{!}{=}0,\quad \frac{\delta S_{\mathrm{KG}}\left[\phi,\phi^{\ast}\right]}{\delta\phi^{\ast}\left(x\right)}\stackrel{!}{=}0.
\end{eqnarray}
This is equivalent to the Euler-Lagrange equations
\begin{eqnarray}
\label{eq:kgel}
\frac{\partial\mathcal{L}_{\mathrm{KG}}\left(x\right)}{\partial\phi\left(x\right)}=\partial_{\mu}\frac{\partial\mathcal{L}_{\mathrm{KG}}\left(x\right)}{\partial\left(\partial_{\mu}\phi\left(x\right)\right)},\quad
\frac{\partial\mathcal{L}_{\mathrm{KG}}\left(x\right)}{\partial\phi^{\ast}\left(x\right)}=\partial_{\mu}\frac{\partial\mathcal{L}_{\mathrm{KG}}\left(x\right)}{\partial\left(\partial_{\mu}\phi^{\ast}\left(x\right)\right)}
 .
\end{eqnarray}
For the Lagrangian density~(\ref{eq:kgld}) these equations of motion are the Klein-Gordon equations
\begin{eqnarray}
\label{eq:KGE}
 &&\left(\square+m^2\right)\phi\left(x\right)=0,
\quad\left(\square+m^2\right)\phi^{\ast}\left(x\right)=0
.
\end{eqnarray}
\subsection{Invariant Scalar Product}
\label{sec:spkg} Let $\phi,\chi$ be arbitrary solutions of
the Klein-Gordon equation~(\ref{eq:KGE}). Their inner product can be defined by
\begin{eqnarray}
\label{eq:ip}\left(\phi,\chi\right)_{\Sigma}&:=&i\int_{\Sigma}\mathrm{d}\Sigma^{\mu}\left(x\right)\left[\phi^{\ast}
\left(x\right)\left(\partial_{\mu}\chi\left(x\right)\right)-\left(\partial_{\mu}\phi^{\ast}\left(x\right)\right)\chi\left(x\right)\right]\nonumber\\
&\equiv&i\int_{\Sigma}\mathrm{d}\Sigma^{\mu}\left(x\right)\phi^{\ast}
\left(x\right)\overleftrightarrow{\partial_{\mu}}\chi\left(x\right),
\end{eqnarray}
with $\Sigma$ denoting a spacelike hypersurface of Minkowski
space. It can be shown that the inner product~(\ref{eq:ip}) does
not depend on the particular choice of
$\Sigma$~\cite{Schweber:1961}.
\\
$\circ\quad$
 We make use of Gauss' theorem
\begin{eqnarray}
 \int_{\partial\mathcal{U}}\mathrm{d}\Sigma_{\mu}\zeta^{\mu}=\int_{\mathcal{U}}\mathrm{d}^4 x \,\partial_{\mu}\,\zeta^{\mu},
\end{eqnarray}
with $\mathcal{U}$ being a compact four-dimensional submanifold of
Minkowski space and $\zeta$ a vector field. Let
$\Sigma_1,\Sigma_2$ be two different, spacelike hypersurfaces and
let $\mathcal{U}$ be bounded by $\Sigma_1,\Sigma_2$ and by
suitable timelike hypersurfaces where $\phi=\chi=0$. Then we can
write
\begin{eqnarray}
\left(\phi,\chi\right)_{\Sigma_1}-\left(\phi,\chi\right)_{\Sigma_2}&=&i\int_{\partial\mathcal{U}}\mathrm{d}\Sigma^{\mu}\left[\phi^{\ast} \left(x\right)\left(\partial_{\mu}\chi\left(x\right)\right)-\left(\partial_{\mu}\phi^{\ast}\left(x\right)\right)\chi\left(x\right)\right]\nonumber\\&=&i\int_{\mathcal{U}}\mathrm{d}^4 x \,\partial^{\mu}\left[\phi^{\ast} \left(x\right)\left(\partial_{\mu}\chi\left(x\right)\right)-\left(\partial_{\mu}\phi^{\ast}\left(x\right)\right)\chi\left(x\right)\right]\nonumber\\&=&i\int_{\mathcal{U}}\mathrm{d}^4 x \left[\phi^{\ast} \left(x\right)\left(\square\,\chi\left(x\right)\right)-\left(\square\,\phi^{\ast}\left(x\right)\right)\chi\left(x\right)\right]\nonumber\\
\label{eq:s1-s2}&=&i\int_{\mathcal{U}}\mathrm{d}^4 x \left[m^2-m^2\right]\phi^{\ast} \left(x\right)\chi\left(x\right)=0,
\end{eqnarray}
where we have used in the last step that $\phi,\chi$ solve the
Klein-Gordon equation~(\ref{eq:KGE})~\cite{Schweber:1961}.$\quad\bullet$\\
We see that it is essential for the inner product to be
independent of the hypersurface, that $\phi$ and $\chi$ are
solutions of the Klein-Gordon equation. From~(\ref{eq:dSigma}) we
can write the hypersurface element as
\begin{eqnarray}\mathrm{d}\Sigma^{\mu}\left(x\right)=n^{\mu}\left(x\right)\mathrm{d}\Sigma\left(x\right).
\end{eqnarray}
For a hypersurface $\Sigma_{t}$ with fixed Minkowski time
$x^0=t=\mathrm{const.}$~(cf.~Section~\ref{sec:if}), we have
$\mathrm{d}x^0=0$. Therefore, we have for the hypersurface element
\begin{eqnarray}\label{eq:dSigmat}
\mathrm{d}\Sigma_t^{\mu}\left(x\right)&=&\left(
\begin{array}{c}\mathrm{d}x^1\mathrm{d}x^2\mathrm{d}x^3
  \\
 0\\
 0
 \\
  0\\
\end{array}
\right)^{\mu}=\left(
\begin{array}{c}1
  \\
 0\\
 0
 \\
  0\\
\end{array}
\right)^{\mu}\mathrm{d}^3x =g^{0\mu}\mathrm{d}^3x.
\end{eqnarray}
This inserted into~(\ref{eq:ip}) yields the well known scalar product
\begin{eqnarray}
\label{eq:ifsp}
\left(\phi,\chi\right)_{\Sigma_{t}}:=i\int_{\mathbb{R}^3}\mathrm{d}^3x\,\phi^{\ast}
\left(x\right)\overleftrightarrow{\partial_{0}}
\left.\chi\left(x\right)\right\vert_{x^0=t},
\end{eqnarray} which is independent of $t$. We shall call~(\ref{eq:ifsp}) \emph{instant-form scalar product}.\\
Every solution of the Klein-Gordon equation~(\ref{eq:KGE}) can be
expanded in terms of plane waves. This means that the functions
\begin{eqnarray}\label{eq:planew}
\phi_{\boldsymbol{p}}\left(x\right)=\frac{1}{\left(2\pi\right)^\frac32}\mathrm{e}^{-
ip_{\lambda}\,x^{\lambda}}\quad\text{and}\quad
\phi_{\boldsymbol{p}}^{\ast}\left(x\right)=\frac{1}{\left(2\pi\right)^\frac32}\mathrm{e}^{
ip_{\lambda}\,x^{\lambda}},
\end{eqnarray}
with $p_{\lambda}\,p^{\lambda}=m^2$ provide a complete set. Since
the constraint~(\ref{eq:p^2}) holds for~(\ref{eq:planew}), the
solutions of~(\ref{eq:KGE}) can be given a particle
interpretation. The modes $\phi_{\boldsymbol{p}}\left(x\right)$
 are said to be positive
energy and $\phi_{\boldsymbol{p}}^{\ast}\left(x\right)$ negative
energy solutions. The scalar product between these modes is
\begin{eqnarray}\label{eq:spbm1}
\left(\phi_{\boldsymbol{p}},\phi_{\boldsymbol{q}}\right)_{\Sigma_{t}}=2p^0\delta^3\left(\boldsymbol{p}-\boldsymbol{q}\right),\quad
\left(\phi_{\boldsymbol{p}}^{\ast},\phi_{\boldsymbol{q}}^{\ast}\right)_{\Sigma_{t}}=-2p^0\delta^3\left(\boldsymbol{p}-\boldsymbol{q}\right),\\
\label{eq:spbm2}
\left(\phi_{\boldsymbol{p}}^{\ast},\phi_{\boldsymbol{q}}\right)_{\Sigma_{t}}=\left(\phi_{\boldsymbol{p}},\phi_{\boldsymbol{q}}^{\ast}\right)_{\Sigma_{t}}=0
.
\end{eqnarray}
For the hypersurface $\Sigma_{\tau}$ of Section~\ref{sec:pf} with
fixed $x_{\lambda}\,x^{\lambda}=\tau^2=\mathrm{const.}$, we have for the
hypersurface element
\begin{eqnarray}\label{eq:dSigmatau}
\mathrm{d}\Sigma_{\tau}^{\mu}\left(x\right)=2\,\mathrm{d}^4
x\delta\left(x_{\lambda}\,x^{\lambda}-\tau^2\right)\theta\left(x^0\right)x^{\mu}.\end{eqnarray}
This is explicitly shown in Appendix~\ref{eq:app:hse}. The inner
product over the hyperboloid is given by
\begin{eqnarray}\label{eq:pfipsf} \left(\phi,\chi\right)_{\Sigma_{\tau}}&:=&i\int_{\mathbb{R}^4}2\,\mathrm{d}^4x\,\delta\left(x_{\lambda}\,x^{\lambda}-\tau^2\right)\theta\left(x^0\right)x^{\mu}\nonumber\\&&\times\phi^{\ast}
\left(x\right)\overleftrightarrow{\partial_{\mu}}\chi\left(x\right)
.
\end{eqnarray}
We have to show that the statement~(\ref{eq:s1-s2}) is true. This
means that this scalar product is independent of the
chosen hyperboloid characterized by $\tau$.
Again using plane waves~(\ref{eq:planew}) for $\phi,\chi$ the
scalar product reads
\begin{eqnarray}
\left(\phi_{\boldsymbol{p}},\phi_{\boldsymbol{q}}\right)_{\Sigma_{\tau}}&=&\frac{2}{\left(2\pi\right)^3}\int_{\mathbb{R}^4}\mathrm{d}^4x\,
\delta\left(x_{\lambda}\,x^{\lambda}-\tau^2\right)\theta\left(x^0\right)x^{\mu}
\left(p+q\right)_{\mu}\mathrm{e}^{ix^{\nu}\left(p-q\right)_{\nu}}\nonumber\\&=:&W\left(p+q,p-q\right)
.
\end{eqnarray}
This Lorentz-invariant distribution is calculated in
Appendix~(\ref{app:W}). Its value is
$W\left(p+q,p-q\right)=2p^0\delta^3\left(\boldsymbol{p}-\boldsymbol{q}\right)$.
Similarly, we have
\begin{eqnarray}
\left(\phi_{\boldsymbol{p}}^{\ast},\phi_{\boldsymbol{q}}^{\ast}\right)_{\Sigma_{\tau}}=-W\left(p+q,q-p\right)=-
2p^0\delta^3\left(\boldsymbol{p}-\boldsymbol{q}\right).
\end{eqnarray}
For the orthogonal plane waves we have
\begin{eqnarray}
\left(\phi_{\boldsymbol{p}},\phi_{\boldsymbol{q}}^{\ast}\right)_{\Sigma_\tau}&=&\frac{2}{\left(2\pi\right)^3}\int_{\mathbb{R}^4}\mathrm{d}^4x\,
\delta\left(x_{\lambda}\,x^{\lambda}-\tau^2\right)\theta\left(x^0\right)x^{\mu}
\left(p-q\right)_{\mu}\mathrm{e}^{ix^{\nu}\left(p+q\right)_{\nu}}\nonumber\\&=&W\left(p-q,p+q\right)=0
\end{eqnarray}
as calculated in the Appendix~(\ref{app:W}). Similarly we have
\begin{eqnarray}
\left(\phi^{\ast}_{\boldsymbol{p}},\phi_{\boldsymbol{q}}\right)_{\Sigma_\tau}=-
W\left(p-q,-p-q\right)=0.
\end{eqnarray}
Comparing these equations with~(\ref{eq:spbm1})
and~(\ref{eq:spbm2}) we see, that the inner product is indeed
independent of the chosen
spacelike hypersurface.\\
\subsection{Covariant Canonical Commutation Relations}
\label{sec:cccr} Now we want to perform the canonical quantization
of our fields. For this purpose, we replace the classical scalar
fields $\phi,\phi^{\ast}$ by field operators
$\hat{\phi},\hat{\phi}^{\dag}$. In order to impose quantization
conditions on these field operators,
Schwinger~\cite{Schwinger:1948yk} proposes covariant canonical
commutator relations on an arbitrary spacelike hypersurface
\begin{eqnarray}
\label{eq:cccrkg}&&\int_{\Sigma}\mathrm{d}\Sigma\left(x\right)\left[\hat{\phi}\left(y\right),\hat{\pi}\left(x\right)\right]_{x,y\in\Sigma}=\int_{\Sigma}\mathrm{d}\Sigma\left(x\right)\left[\hat{\phi}^{\dag}\left(y\right),\hat{\pi}^{\dag}\left(x\right)\right]_{x,y\in\Sigma}=
i,\nonumber\\
\\
&&
\left[\hat{\phi}\left(x\right),\hat{\phi}\left(y\right)\right]=\left[\hat{\phi}^{\dag}\left(x\right),\hat{\phi}^{\dag}\left(y\right)\right]=\left[\hat{\pi}\left(x\right),\hat{\pi}\left(y\right)\right]=\left[\hat{\pi}^{\dag}\left(x\right),\hat{\pi}^{\dag}\left(y\right)\right]\nonumber\\&&=\label{eq:cccrkg2}
\left[\hat{\phi}\left(x\right),\hat{\pi}^{\dag}\left(y\right)\right]=\left[\hat{\phi}^{\dag}\left(x\right),\hat{\pi}\left(y\right)\right]=0,\quad
x,y\in\Sigma.
\end{eqnarray} A generalization of these commutation relations to arbitrary $x$ and $y$ is given by
\begin{eqnarray}
\label{eq:pjfkg}
&&\left[\hat{\phi}\left(x\right),\hat{\phi}^{\dag}\left(y\right)\right]=i\Delta\left(x-y\right),\\
&&\left[\hat{\pi}\left(x\right),\hat{\pi}^{\dag}\left(y\right)\right]=i\,n_{\mu}\left(x\right)
n_{\nu}\left(y\right)\,\frac{\partial}{\partial
x_{\mu}}\,\frac{\partial}{\partial y_{\nu}}
\Delta\left(x-y\right), \label{eq:pjfkg2}\end{eqnarray} where
$\Delta\left(x-y\right)$ is the, so called, \emph{Pauli-Jordan
function}. The field operator $\hat{\pi}$ canonically conjugate to
$\hat{\phi}$ is given by
\begin{eqnarray}\label{eq:gccm}
\hat{\pi}\left(x\right)=n^{\mu}\left(x\right)\frac{\partial\hat{\mathcal{L}}_{\mathrm{KG}}\left(x\right)}{\partial\left(\partial^{\mu}\hat{\phi}\left(x\right)\right)}=
n^{\mu}\left(x\right)\partial_{\mu}\hat{\phi}^{\dag}\left(x\right)=\frac{\partial}{\partial{\xi^0}}\hat{\phi}^{\dag}\left(\xi\right)
.\nonumber\\
\end{eqnarray}
In the last step we have used the transformation properties of
$\partial_{\mu}$. Thus $\hat{\pi}$ is just the derivative of
$\hat{\phi}^{\dag}$ with respect to some timelike direction
$n^{\mu}\left(x\right)$ depending on the choice of space-time
foliation. The real distribution $\Delta\left(x-y\right)$
in~(\ref{eq:pjfkg}) and its second derivative with respect to a
chosen time parameter in~(\ref{eq:pjfkg2}) vanish for spacelike
$\left(x-y\right)$. They are given by\footnote{This explicit form
of $\Delta\left(x-y\right)$ will be clear after Fourier expanding
the fields and imposing canonical commutation relations in
momentum space~(cf.~Section~\ref{sec:ccrms}).}
\begin{eqnarray}\label{eq:Delta}
\Delta\left(x-y\right)&=&\frac{1}{i}\int_{\mathbb{R}^4}\frac{\mathrm{d}^4p}{\left(2\pi\right)^3}
\delta\left(p_{\lambda}\,p^{\lambda}-m^2\right)\mathrm{e}^{-ip_{\mu}\left(x-y\right)^{\mu}}\nonumber\\&&\times\left(\theta\left(p^0\right)
-\theta\left(-p^0\right )\right)
\end{eqnarray} and
\begin{eqnarray}
\label{eq:ddDelta}&&n_{\mu}\left(x\right)n_{\nu}\left(y\right)\,\frac{\partial}{\partial
x_{\mu}}\,\frac{\partial}{\partial
y_{\nu}}\Delta\left(x-y\right)\nonumber\\&&=
\frac{1}{i}\int_{\mathbb{R}^4}\frac{\mathrm{d}^4p}{\left(2\pi\right)^3}
 \delta\left(p_{\lambda}\,p^{\lambda}-m^2\right)\,n_{\mu}\left(x\right)\,p^{\mu}\, n_{\nu}\left(y\right)\,p^{\nu}\,\mathrm{e}^{-ip_{\rho}\left(x-y\right)^{\rho}}\left(\theta\left(p^0\right )-\theta\left(-p^0\right
 )\right).\nonumber\\
\end{eqnarray}
Since $p$ is timelike and~(\ref{eq:Delta}) and~(\ref{eq:ddDelta})
are Lorentz invariant, we can immediately conclude, that for
spacelike $\left(x-y\right)$ it follows that
\begin{eqnarray}\label{eq:Delta=0}
\Delta\left(x-y\right)=n_{\mu}\left(x\right)n_{\nu}\left(y\right)\,\frac{\partial}{\partial
x_{\mu}}\,\frac{\partial}{\partial
y_{\nu}}\Delta\left(x-y\right)=0.\end{eqnarray} This is a
consequence of locality and causality and explicitly proved in
Appendix~\ref{app:PJF}.
\\
Choosing in~(\ref{eq:cccrkg}) the instant-$t$ plane $\Sigma_{t}$
(\ref{eq:dSigmat}) we have
\begin{eqnarray}\int_{\mathbb{R}^3}\mathrm{d}^3 x\left[\hat{\phi}\left(y\right),\partial_{0}\hat{\phi}^{\dag}\left(x\right)\right]_{x^0=y^0=t}=i.
\end{eqnarray}
This relation is satisfied, if the commutator is equal
\begin{eqnarray}\label{eq:ifccr}
\left[\hat{\phi}\left(y\right),\partial_{0}\hat{\phi}^{\dag}\left(x\right)\right]_{x^0=y^0=t}=i\delta^3\left(\boldsymbol{x}-\boldsymbol{y}\right)
\end{eqnarray}
and we recover the equal-t canonical commutation relations.\\
If we choose in~(\ref{eq:cccrkg}) the hyperboloid
$\Sigma_{\tau}$~(\ref{eq:dSigmatau}), we have
\begin{eqnarray}\label{eq:pfccr}
 \int_{\mathbb{R}^4}2\mathrm{d}^4x\,\delta\left(x_{\lambda}\,x^{\lambda}-\tau^2\right)\theta\left(x^0\right)x^{\mu}\left[\hat{\phi}\left(y\right),\partial_{\mu}\hat{\phi}^{\dag}\left(x\right)\right]_{x^2=y^2=\tau^2}=i,
\end{eqnarray}
which yields
\begin{eqnarray}
\label{eq:liccr}
x^{\mu}\left[\hat{\phi}\left(y\right),\partial_{\mu}\hat{\phi}^{\dag}\left(x\right)\right]_{x^2=y^2=\tau^2}=ix^0\delta^3\left(\boldsymbol{x}-\boldsymbol{y}\right).
\end{eqnarray}
 These are the Lorentz-invariant canonical commutation relations when quantizing on a hyperboloid.
 The Lorentz invariance is explicitly seen by noting that the right hand side of~(\ref{eq:liccr})
 is the $W$ distribution~(cf.~Appendix~\ref{app:W}),
\begin{eqnarray}
x^0\delta^3\left(\boldsymbol{x}-\boldsymbol{y}\right)=\frac{1}{2}W\left(x+y,x-y\right),
\end{eqnarray}
which is Lorentz invariant by definition.\\
The commutation relations are covariant in the sense that no particular choice of a Minkowski time parameter has been made.\\
This result agrees with those of~\cite{Gromes:1974yu} as shown in
Appendix~\ref{app:cccr}. Furthermore we note that differentiation
of~(\ref{eq:pjfkg}) with respect to Minkowski time $x_0$
gives~\cite{Itzykson:2006}
\begin{eqnarray}\frac{\partial}{\partial
y^0}\Delta\left(x-y\right)\vert_{x^0=y^0=t}=\delta^3\left(\boldsymbol{x}-\boldsymbol{y}\right).\end{eqnarray}
This is nothing but the instant-form canonical commutation
relation~(\ref{eq:ifccr}).\\
The point-form analogue can be formulated as differentiation
of~(\ref{eq:pjfkg}) with respect to $\xi^0\left(x\right)=\tau$
using~$\tau\frac{\partial}{\partial{\tau}}=y^{\lambda}\frac{\partial}{\partial
y^{\lambda}}$, i.e.
\begin{eqnarray}
\label{eq:dxi0Delta}y^{\lambda}\frac{\partial}{\partial
y^{\lambda}}\Delta\left(x-y\right)\vert_{x^2=y^2=\tau^2}=x^0\delta^3\left(\boldsymbol{x}-\boldsymbol{y}\right).\end{eqnarray}
This is exactly the Lorentz-invariant commutation
relation~(\ref{eq:liccr}) we expected.\footnote{In the calculation
we have used the properties of $W^{\mu}\left(X,Y\right)$
(cf.~Appendix~\ref{app:Wmu}). After using
$W\left(Y,Y\right)=-W\left(Y,-Y\right)$
(cf.~(\ref{eq:WQQ})),~(\ref{eq:dxi0Delta}) follows immediately.}
\\
\subsection{Commutation Relations in Momentum Space}
\label{sec:ccrms} The general solutions $\phi$ and $\phi^{\dag}$
of the Klein-Gordon equations~(\ref{eq:KGE}) can be written as an
expansion in terms of a complete set of solutions. As shown
before, usual plane waves~(\ref{eq:planew}) are orthogonal with
respect to the invariant scalar product~(\ref{eq:ip}). Thus they
provide an appropriate basis. Expansion in terms of plane waves is
equivalent with a Fourier expansion. Therefore canonical
quantization is done by considering the Fourier coefficients as
field operators acting on a momentum Fock
space~(cf.~Section~\ref{sec:Focksp}). Then the field operators can
be written as
\begin{eqnarray}
\label{eq:kgfe}
\hat{\phi}\left(x\right)&=&\int_{\mathbb{R}^4}\mathrm{d}^4p\,\delta\left(p_{\lambda}\,p^{\lambda}-m^2\right)\theta\left(p^0\right)\left(\phi_{\boldsymbol{p}}\,\hat{a}\left(\boldsymbol{p}\right)+
\phi_{\boldsymbol{p}}^{\ast}\,\hat{b}^{\dag}\left(\boldsymbol{p}\right)
\right)\nonumber\\&=&
\frac{1}{\left(2\pi\right)^{\frac32}}\int_{\mathbb{R}^3}\frac{\mathrm{d}^3p}{2p_0}\left(\mathrm{e}^{-ip_{\mu}
x^{\mu}}\hat{a}\left(\boldsymbol{p}\right)+ \mathrm{e}^{ip_{\mu}
x^{\mu}}\hat {b}^{\dag}\left(\boldsymbol{p}\right)
\right),\\
\label{eq:kgfe2}
\hat{\phi}^{\dag}\left(x\right)&=&\int_{\mathbb{R}^4}\mathrm{d}^4p\,\delta\left(p_{\lambda}\,p^{\lambda}-m^2\right)\theta\left(p^0\right)\left(\phi_{\boldsymbol{p}}^{\ast}\,\hat{a}^{\dag}\left(\boldsymbol{p}\right)+
\phi_{\boldsymbol{p}}\,\hat{b}\left(\boldsymbol{p}\right)
\right)\nonumber\\&=&
\frac{1}{\left(2\pi\right)^{\frac32}}\int_{\mathbb{R}^3}\frac{\mathrm{d}^3p}{2p_0}\left(\mathrm{e}^{ip_{\mu}
x^{\mu}}\hat{a}^{\dag}\left(\boldsymbol{p}\right)+
\mathrm{e}^{-ip_{\mu} x^{\mu}}\hat {b}\left(\boldsymbol{p}\right)
\right).
\end{eqnarray}
The phase space measure for massive particles
\begin{eqnarray}
\mathrm{d}^4p\,\delta\left(p_{\lambda}\,p^{\lambda}-m^2\right)\theta\left(p^0\right)=\frac{\mathrm{d}^3p}{2p_0}\end{eqnarray}
is clearly Lorentz invariant with
$p^0=\sqrt{\boldsymbol{p}^2+m^2}>0$. The Fourier coefficients
$a,b$ being operators after canonical quantization are given
by\footnote{These relations are obvious since, e.g., for $\hat{a}$
we have
 \begin{eqnarray}\hat{a}\left(\boldsymbol{p}\right)&=&\left(\phi_{\boldsymbol{p}}, \hat{\phi}\right)_{\Sigma}=
\int_{\mathbb{R}^3}\frac{\mathrm{d}^3q}{2q^0}
\left[\left(\phi_{\boldsymbol{p}},
\phi_{\boldsymbol{q}}\right)_{\Sigma}\hat{a}\left(\boldsymbol{q}\right)+\underbrace{\left(\phi_{\boldsymbol{p}},
\phi^{\ast}_{\boldsymbol{q}}\right)_{\Sigma}}_{=0}\hat{b}^{\dag}\left(\boldsymbol{q}\right)\right]\nonumber\\&=&
\int_{\mathbb{R}^3}\frac{\mathrm{d}^3q}{2q^0}
2q^0\delta^3\left(\boldsymbol{p}-\boldsymbol{q}\right)\hat{a}\left(\boldsymbol{q}\right)
\nonumber,\end{eqnarray} where we have used the orthogonality
relations between plane waves~(\ref{eq:spbm1}),~(\ref{eq:spbm2}).}
 \begin{eqnarray}\label{eq:spfa}
&&\hat{a}\left(\boldsymbol{p}\right)=\left(\phi_{\boldsymbol{p}}, \hat{\phi}\right)_{\Sigma},\quad
\hat{a}^{\dag}\left(\boldsymbol{p}\right)=\left(\hat{\phi},\phi_{\boldsymbol{p}} \right)_{\Sigma},\\
\label{eq:spfb}
&&\hat{b}\left(\boldsymbol{p}\right)=-\left(\hat{\phi}, \phi_{\boldsymbol{p}}^{\ast}\right)_{\Sigma},\quad
\hat{b}^{\dag}\left(\boldsymbol{p}\right)=-\left(\phi_{\boldsymbol{p}}^{\ast}, \hat{\phi}\right)_{\Sigma}.
\end{eqnarray}
 These relations together with the canonical commutation relations~(\ref{eq:cccrkg}) and~(\ref{eq:cccrkg2}) imply
the harmonic-oscillator commutation relations
 \begin{eqnarray}\label{eq:ccrms}\left[\hat{a}\left(\boldsymbol{p}\right),\hat{a}^{\dag}\left(\boldsymbol{q}\right)\right]=
\left[\hat{b}\left(\boldsymbol{p}\right),\hat{b}^{\dag}\left(\boldsymbol{q}\right)\right]=
2p^0\delta^3\left(\boldsymbol{p}-\boldsymbol{q}\right),\\\label{eq:ccrms2}
\left[\hat{a}\left(\boldsymbol{p}\right),\hat{a}\left(\boldsymbol{q}\right)\right]=
\left[\hat{b}\left(\boldsymbol{p}\right),\hat{b}\left(\boldsymbol{q}\right)\right]=\left[\hat{a}\left(\boldsymbol{p}\right),\hat{b}\left(\boldsymbol{q}\right)\right]=\left[\hat{a}\left(\boldsymbol{p}\right),\hat{b}^{\dag}\left(\boldsymbol{q}\right)\right]=0.
\end{eqnarray}
In Appendix~(\ref{app:ccrms}) this is explicitly shown in point
form. These commutation relations in momentum space are the
Fourier transforms of~(\ref{eq:cccrkg}) and~(\ref{eq:cccrkg2}).
Since the operators
$\hat{a}\left(\boldsymbol{p}\right),\hat{b}\left(\boldsymbol{p}\right),\hat{a}^{\dag}\left(\boldsymbol{p}\right)$
and $\hat{b}^{\dag}\left(\boldsymbol{p}\right)$ satisfy the
commutation relations~(\ref{eq:ccrms}) and~(\ref{eq:ccrms2}), they
may be interpreted as annihilation or creation operators. By
acting on a Fock space constructed out of one-particle Hilbert
spaces (cf.~Section~\ref{sec:Focksp}), they annihilate or create
field quanta characterized by the continuous three-momentum vector
$\boldsymbol{p}$. The mass-shell constraint~(\ref{eq:p^2}) holds,
of course. These basis elements of the, so called, \emph{Wigner
basis} are eigenstates of the three-momentum operator
$\hat{\boldsymbol{P}}$.
\\
We shall note that the field expansions~(\ref{eq:kgfe})
and~(\ref{eq:kgfe2}) together with the commutation
relations~(\ref{eq:ccrms}) and~(\ref{eq:ccrms2}) imply the
explicit form of the Pauli-Jordan function
$\Delta\left(x-y\right)$ in~(\ref{eq:Delta}).
\subsection{Generators in Wigner Representation}
\label{sec:gwr} In order to show the equivalence between equal-$t$
and equal-$\tau$ field quantization, we represent the generators
of global gauge transformations and the Poincar\'e generators in
the Wigner basis. For free fields they are expected to be the same
in instant and point form~\cite{Zelzer:2005}.
\subsubsection{Global Gauge Transformations}
In Section~\ref{sec:Nt} we have seen that the invariance of the
Lagrangian density under a global $U\left(1\right)$ phase
transformation of the fields implies a conserved
current~(\ref{eq:cc}). Inserting the Lagrangian density for free
scalar fields~(\ref{eq:kgld}) yields after canonical quantization
the current operator
\begin{eqnarray}
\hat{\mathcal{J}}_{\mathrm{KG}}^{\mu}\left(x\right)=i:\hat{\phi}^{\dag}\left(x\right)\overleftrightarrow{\partial^{\mu}}\hat{\phi}\left(x\right):,\quad
\text{with}\quad\partial_{\mu}\,\hat{\mathcal{J}}_{\mathrm{KG}}^{\mu}\left(x\right)=0.
\end{eqnarray}
":...:" denotes the usual normal ordering, i.e. commuting all
creation operators to the left of the annihilation operators and
dropping the commutators in order to avoid infinite ground state
energies. Integration of the current operator as in~(\ref{eq:cch})
gives a conserved charge or symmetry operator
\begin{eqnarray}
\hat{Q}_{\mathrm{KG}}&=&\int_{\Sigma}\mathrm{d}\Sigma_{\mu}\left(x\right)\hat{\mathcal{J}}_{\mathrm{KG}}^{\mu}\left(x\right)\nonumber\\&=&
i\int_{\Sigma}\mathrm{d}\Sigma_{\mu}\left(x\right):\hat{\phi}^{\dag}\left(x\right)\overleftrightarrow{\partial^{\mu}}\hat{\phi}\left(x\right):.
\end{eqnarray}
 Inserting the field expansions~(\ref{eq:kgfe}) and~(\ref{eq:kgfe2}) and choosing the equal-t hyperplane
$\Sigma_{t}$ one ends up with the well known form for the charge
operator in Wigner representation
\begin{eqnarray}\label{eq:Qwrif}
\hat{Q}_{\mathrm{KG}}=\int_{\mathbb{R}^3}
\mathrm{d}^3x\,\hat{\mathcal{J}}_{\mathrm{KG}}^0\left(x\right)=
\int_{\mathbb{R}^3}
\frac{\mathrm{d}^3p}{2p_0}\left(\hat{a}^{\dag}\left(\boldsymbol{p}\right)\hat{a}\left(\boldsymbol{p}\right)-\hat{b}^{\dag}\left(\boldsymbol{p}\right)\hat{b}\left(\boldsymbol{p}\right)\right)
.
\end{eqnarray}
This result suggests to consider
$\hat{a}^{\dag}\left(\boldsymbol{p}\right)$ and
$\hat{a}\left(\boldsymbol{p}\right)$ as creation and annihilation
operators of particles with charge $+1$ and
$\hat{b}^{\dag}\left(\boldsymbol{p}\right)$ and
$\hat{b}\left(\boldsymbol{p}\right)$ as creation and annihilation
operators of antiparticles with charge $-1$, respectively.\\
If we choose the equal-$\tau$ hyperboloid $\Sigma_{\tau}$ as the
spacelike hypersurface, we have
\begin{eqnarray}\label{eq:pfQ2}
\hat{Q}_{\mathrm{KG}}&=&2i\int_{\mathbb{R}^4}\mathrm{d}^4x\,\delta\left(x_{\lambda}\,x^{\lambda}-\tau^2\right)\theta\left(x^0\right)x_{\mu}:\hat{\phi}^{\dag}\left(x\right)\overleftrightarrow{\partial^{\mu}}\hat{\phi}\left(x\right):
\nonumber\\
\label{eq:Qwrpf}&=&\int_{\mathbb{R}^3}
\frac{\mathrm{d}^3p}{2p_0}\left(\hat{a}^{\dag}\left(\boldsymbol{p}\right)\hat{a}\left(\boldsymbol{p}\right)-\hat{b}^{\dag}\left(\boldsymbol{p}\right)\hat{b}\left(\boldsymbol{p}\right)\right),\end{eqnarray}
as calculated in Appendix~\ref{app:cho}.
Comparing~(\ref{eq:Qwrif}) with~(\ref{eq:Qwrpf}) we see that the
charge operator integrated over the hyperboloid has the usual form
in Wigner representation.
\\
This result confirms~(\ref{eq:gfggt}) on using the canonical
commutation relations~(\ref{eq:ccrms}) and~(\ref{eq:ccrms2}), i.e.
\begin{eqnarray}
\hat{\phi}\left(x\right)=\left[\hat{\phi}\left(x\right),\hat{Q}_{\mathrm{KG}}\right].
\end{eqnarray}
\subsubsection{Translations}
We have seen in Section~\ref{sec:Nt} that a conserved current, the
energy-momentum tensor~(\ref{eq:emt}), follows from the invariance
of the action under displacements. Inserting for the Lagrangian
density~(\ref{eq:kgld}), the energy-momentum tensor becomes after
canonical quantization
\begin{eqnarray}
\label{eq:kgemt1}
\hat{\mathcal{T}}_{\mathrm{KG}}^{\mu\nu}\left(x\right)&=&\,:
\partial^{\mu}\hat{\phi}^{\dag}\left(x\right)\partial^{\nu}\hat{\phi}\left(x\right)+\partial^{\nu}\hat{\phi}^{\dag}\left(x\right)\partial^{\mu}\hat{\phi}\left(x\right)\nonumber\\&&-
g^{\mu\nu}\left(\partial_{\lambda}\hat{\phi}^{\dag}\left(x\right)\partial^{\lambda}\hat{\phi}\left(x\right)-m^2\hat{\phi}^{\dag}\left(x\right)\hat{\phi}\left(x\right)\right):,\nonumber\\
\\&&\text{with}\quad\partial_{\mu}\,\hat{\mathcal{T}}_{\mathrm{KG}}^{\mu\nu}\left(x\right)=0.
\end{eqnarray}
From equation~(\ref{eq:Gepsilon1}) we have obtained the
four-momentum operator~(\ref{eq:Pmu}) as integral over a spacelike hypersurface. Applying this to~(\ref{eq:kgemt1}) we have
\begin{eqnarray}\hat{P}_{\mathrm{KG}}^{\mu}=\int_{\Sigma}\mathrm{d}\Sigma_{\nu}\left(x\right)\hat{\mathcal{T}}_{\mathrm{KG}}^{\mu\nu}\left(x\right).
\end{eqnarray}
Inserting the field expansions~(\ref{eq:kgfe}),~(\ref{eq:kgfe2})
and taking the equal-$t$ hyperplane $\Sigma_{t}$, we obtain the
usual result for the translation generator in Wigner
representation
\begin{eqnarray}\hat{P}_{\mathrm{KG}}^{\mu}=\int_{\mathbb{R}^3}\mathrm{d}^3x\,
\hat{\mathcal{T}}_{\mathrm{KG}}^{\mu
0}\left(x\right)=\int_{\mathbb{R}^3}\frac{\mathrm{d}^3p}{2p_0}p^{\mu}\left(\hat
{a}^{\dag}\left(\boldsymbol{p}\right)\hat{a}\left(\boldsymbol{p}\right)+\hat{b}^{\dag}\left(\boldsymbol{p}\right)\hat{b}\left(\boldsymbol{p}\right)\right).\end{eqnarray} \\
Integration over the hyperboloid $\Sigma_{\tau}$ gives, after some
calculation (cf.~Appendix~\ref{app:Pmu}), the same result as in
instant form
\begin{eqnarray}\label{eq:pfP}
\hat{P}^{\mu}_{\mathrm{KG}}&=&\int_{\mathbb{R}^4}2\mathrm{d}^4x\,
\delta\left(x_{\lambda}\,x^{\lambda}-\tau^2\right)\theta\left(x^0\right)x_{\nu}\hat{\mathcal{T}}_{\mathrm{KG}}^{\mu\nu}\left(x\right)\nonumber\\&=&\int_{\mathbb{R}^3}\frac{\mathrm{d}^3p}{2p_0}p^{\mu}\left(\hat
{a}^{\dag}\left(\boldsymbol{p}\right)\hat{a}\left(\boldsymbol{p}\right)+\hat{b}^{\dag}\left(\boldsymbol{p}\right)\hat{b}\left(\boldsymbol{p}\right)\right)
.\end{eqnarray}
\\
We easily convince ourselves that $\hat{P}_{\mathrm{KG}}^{\mu}$
still transforms as a four-vector under Lorentz transformations:
\begin{eqnarray}\label{eq:tranfpmu}
\hat{U}\left(\Lambda\right)
\hat{P}^{\mu}_{\mathrm{KG}}\hat{U}\left(\Lambda\right)^{-1}&=&\int_{\mathbb{R}^3}\frac{\mathrm{d}^3p}{2p_0}p^{\mu}\left(\hat{a}^{\dag}\left(\boldsymbol{\Lambda
p}\right)\hat{a}\left(\boldsymbol{\Lambda
p}\right)+\hat{b}^{\dag}\left(\boldsymbol{\Lambda
p}\right)\hat{b}\left(\boldsymbol{\Lambda p}\right)\right)
\nonumber\\&=&\int_{\mathbb{R}^3}\frac{\mathrm{d}^3p}{2p_0}\left(\Lambda^{-1}
p\right)^{\mu}\left(\hat
{a}^{\dag}\left(\boldsymbol{p}\right)\hat{a}\left(\boldsymbol{p}\right)+\hat{b}^{\dag}\left(\boldsymbol{p}\right)\hat{b}\left(\boldsymbol{p}\right)\right)\nonumber\\&=&\left(\Lambda^{-1}\right)^{\mu}_{~\nu}\hat{P}^{\nu}_{\mathrm{KG}}
,\end{eqnarray} where $\boldsymbol{\Lambda p}$ means the spatial
component of $\Lambda p$. We have used Lorentz invariance of the
integration measure and the Lorentz-transformation properties of
single-particle states~\cite{Krassnigg:2001}
\begin{eqnarray}
 \hat{U}\left(\Lambda\right)\hat{a}^{\dag}\left(\boldsymbol{p}\right)
\hat{U}^{-1}\left(\Lambda\right)=\hat{a}^{\dag}\left(\boldsymbol{\Lambda
p}\right).\end{eqnarray}
\\
This Fock space representation of $\hat{P}_{\mathrm{KG}}^{\mu}$
together with the harmonic-oscillator commutation relations leads
to the conclusion, that the field quanta created by
$\hat{a}^{\dag}\left(\boldsymbol{p}\right)$ and
$\hat{b}^{\dag}\left(\boldsymbol{p}\right)$ are eigenstates of the
free four-momentum operator with eigenvalues $p^{\mu}$.
\\
Finally by using the canonical commutation
relations~(\ref{eq:ccrms}) and~(\ref{eq:ccrms2}) we
confirm~(\ref{eq:commP}), namely that \begin{eqnarray}
\partial^{\mu}\hat{\phi}\left(x\right)=i\left[\hat{
P}_{\mathrm{KG}}^{\mu},\hat{\phi}\left(x\right)\right].
\end{eqnarray}
\subsubsection{Lorentz Transformations}
In Section~\ref{sec:Nt} we have seen from the invariance of the
action under Lorentz transformations, that a conserved current
follows, the so called angular-momentum
density~(\ref{eq:Mmunusigma}). With the energy-momentum
tensor~(\ref{eq:kgemt1}) this gives an operator
\begin{eqnarray}
\hat{\mathcal{M}}_{\mathrm{KG}}^{\mu\nu\sigma}\left(x\right):=x^{\nu}\hat{\mathcal{T}}_{\mathrm{KG}}^{\mu\sigma}\left(x\right)-x^{\sigma}\hat{\mathcal{T}}_{\mathrm{KG}}^{\mu\nu}\left(x\right),\quad\text{with}\quad
\partial_{\mu}\,\hat{\mathcal{M}}_{\mathrm{KG}}^{\mu\nu\sigma}\left(x\right)=0.\end{eqnarray}
From~(\ref{eq:Gepsilon1}) we find the associated conserved
charges, the generators for Lorentz transformations as
\begin{eqnarray}\hat{M}_{\mathrm{KG}}^{\mu\nu}=\int_{\Sigma}\mathrm{d}\Sigma_{\lambda}\left(x\right)\hat{\mathcal{M}}_{\mathrm{KG}}^{\lambda\mu\nu}\left(x\right)
=\int_{\Sigma}\mathrm{d}\Sigma_{\lambda}\left(x\right)\left[x^{\mu}\hat{\mathcal{T}}_{\mathrm{KG}}^{\lambda\nu}\left(x\right)-x^{\nu}\hat{\mathcal{T}}_{\mathrm{KG}}^{\lambda\mu}\left(x\right)\right].
\end{eqnarray}
Inserting for the energy-momentum tensor~(\ref{eq:kgemt1}) and the
field expansions~(\ref{eq:kgfe}),~(\ref{eq:kgfe2}) and integrating
over the equal-t hyperplane $\Sigma_t$ gives the generators for
boosts and rotations in the Wigner
representation~\cite{Kugo:1997fs}
\begin{eqnarray}
\label{eq:gfbrkg}
\hat{M}_{\mathrm{KG}}^{\mu\nu}=
 \int_{\mathbb{R}^3}\frac{\mathrm{d}^3p}{2p_0}
\left(\hat{a}^{\dag}\left(\boldsymbol{p}\right)l^{\mu\nu}\left(\boldsymbol{p}\right)
\hat{a}\left(\boldsymbol{p}\right)+\hat{b}^{\dag}\left(\boldsymbol{p}\right)m^{\mu\nu}\left(\boldsymbol{p}\right)
\hat{b}\left(\boldsymbol{p}\right)
\right).
\end{eqnarray}
For $\mu,\nu=0,i$ we have the boost generators
$\hat{B}_{\mathrm{KG}}^i=\hat{M}_{\mathrm{KG}}^{0i}=-\hat{M}_{\mathrm{KG}}^{i0}$,
where $l^{0i}=i\frac{\partial}{\partial p_i}p^0$ and
$m^{0i}=ip^0\frac{\partial}{\partial p_i}$, with
$\frac{\partial}{\partial p_i}$ acting to the right. For
$\mu,\nu=j,k$ we have the generators for spatial rotations
$\hat{J}_{\mathrm{KG}}^i=\epsilon_{ijk}\hat{M}_{\mathrm{KG}}^{jk}$,
where
 $l^{ij}=m^{ij}=i\left(p^i\frac{\partial}{\partial p_j}-p^j\frac{\partial}{\partial p_i}\right)$.\\
The similar calculation by integrating over the hyperboloid
$\Sigma_{\tau}$ is more complicated but leads to the same result
as~(\ref{eq:gfbrkg}).
\\
Finally, we calculate~(\ref{eq:commM}) using the canonical
commutation relations~(\ref{eq:ccrms}) and~(\ref{eq:ccrms2}) as
\begin{eqnarray}
\left(x^{\mu}\partial^{\nu}-x^{\nu}\partial^{\mu}\right)\hat{\phi}\left(x\right)=
i\left[\hat{\phi}\left(x\right),\hat{M}_{\mathrm{KG}}^{\mu\nu}\right]
.\end{eqnarray}
\section{Dirac Fields}
Considering a free classical spin-$\frac12$ field theory in
($3+1$)-dimensional Minkowski space-time, we can proceed in an
analogous way as for scalar fields. We start with a Lagrangian
density for the four-component spinor fields
$\psi\left(x\right),\bar{\psi}\left(x\right)$,
\begin{eqnarray}\label{eq:Ld}
\mathcal{L}_{\mathrm{D}}\left(x\right)=\bar{\psi}\left(x\right)\left(i\gamma^{\lambda}\,\partial_{\lambda}-m\right)\psi\left(x\right),\quad\text{with}\quad\bar{\psi}
=\psi^{\dag}\gamma^0,
\end{eqnarray}where $\gamma^{\mu}$ are the $4\times4$ Dirac matrices\footnote{The four matrices $\gamma^{\mu}$ obey the following
relations:
\begin{eqnarray}
    \left\lbrace\gamma^{\mu},\gamma^{\nu}\right\rbrace=2g^{\mu\nu};\label{eq:gammamugammanu}\\
    \left(\gamma^0\right)^{\dag}=\gamma^0,\quad\left(\gamma^i\right)^{\dag}=-\gamma^i,\quad
    \gamma^0\left(\gamma^i\right)^{\dag}\gamma^0=\gamma^i;\\\label{eq:amugamma}
    a_{\mu}\,\gamma^{\mu}b_{\nu}\,\gamma^{\nu}=a_{\lambda}\,b^{\lambda}-i\sigma^{\mu\nu}a_{\mu}\,b_{\nu},\quad
    \text{with}\quad    \sigma^{\mu\nu}=\frac{i}{2}\left[\gamma^{\mu},\gamma^{\nu}\right].
\end{eqnarray}}
. The classical action functional is defined by~(\ref{eq:action})
as
\begin{eqnarray} S_{\mathrm{D}}\left[\psi,\bar{\psi}\right]:=\int_{\mathbb{R}^4} \mathrm{d}^4 x\,\mathcal{L}_{\mathrm{D}}\left(x\right)
.\end{eqnarray} The least action principle~(\ref{eq:elegen}) of
varying the action functional with respect to
$\psi\left(x\right),\bar{\psi}\left(x\right)$ gives
\begin{eqnarray} \frac{\delta S_{\mathrm{D}}\left[\psi,\bar{\psi}\right]}{\delta\psi\left(x\right)}\stackrel{!}{=}0,\quad \frac{\delta S_{\mathrm{D}}\left[\psi,\bar{\psi}\right]}{\delta\bar{\psi}\left(x\right)}\stackrel{!}{=}0
.\end{eqnarray} These equations are equivalent to the
Euler-Lagrange equations
\begin{eqnarray}
\frac{\partial\mathcal{L}_{\mathrm{D}}\left(x\right)}{\partial\psi\left(x\right)}=\partial_{\mu}\frac{\partial\mathcal{L}_{\mathrm{D}}\left(x\right)}{\partial\left(\partial_{\mu}\psi\left(x\right)\right)},\quad
\frac{\partial\mathcal{L}_{\mathrm{D}}\left(x\right)}{\partial\bar{\psi}\left(x\right)}=\partial_{\mu}\frac{\partial\mathcal{L}_{\mathrm{D}}\left(x\right)}{\partial\left(\partial_{\mu}\bar{\psi}\left(x\right)\right)} .
\end{eqnarray}
For the Lagrangian density~(\ref{eq:Ld}), these equations of
motion are the Dirac equations
\begin{eqnarray}\label{eq:De}
\left(i\gamma^{\lambda}\,\overrightarrow{\partial}_{\lambda}-m\right)\psi\left(x\right)&=&0\\
\label{eq:De2}
\bar{\psi}\left(x\right)\left(i\gamma^{\lambda}\,\overleftarrow{\partial}_{\lambda}+m\right)&=&0,\end{eqnarray}
with
$\bar{\psi}\overleftarrow{\partial}_{\lambda}\,\gamma^{\lambda}\equiv
\left(\partial_{\lambda}\,\bar{\psi}\right)\gamma^{\lambda}$.
\subsection{Invariant Scalar Product}
Let $\psi,\chi$ be arbitrary
solutions of the Dirac equation~(\ref{eq:De}). Then their inner product
on a spacelike hypersurface $\Sigma$ can be defined by
\begin{eqnarray}\label{eq:ipd}
\left(\psi,\chi\right)_{\Sigma}:=\int_{\Sigma}\mathrm{d}\Sigma^{\mu}\left(x\right)\bar{\psi}\left(x\right)\gamma_{\mu}\chi\left(x\right).
\end{eqnarray}
Similarly as in Section~\ref{sec:spkg},~(\ref{eq:ipd}) does not
depend on $\Sigma$~\cite{Schweber:1961}. Proceeding in the same
way as in Section~\ref{sec:spkg} this is proved as follows:
\begin{eqnarray}
\circ\quad\left(\psi,\chi\right)_{\Sigma_1}-\left(\phi,\chi\right)_{\Sigma_2}&=&\int_{\partial\mathcal{U}}\mathrm{d}\Sigma^{\mu}\left(x\right)\bar{\psi}\left(x\right)\gamma_{\mu}\chi\left(x\right)\nonumber\\&=&
\int_{\mathcal{U}}\mathrm{d}^4 x\,
\partial^{\mu}\left(\bar{\psi}\left(x\right)\gamma_{\mu}\chi\left(x\right)\right)\nonumber\\&=&
\int_{\mathcal{U}}\mathrm{d}^4 x
\left[\left(\partial^{\mu}\bar{\psi}\left(x\right)\right)\gamma_{\mu}\chi\left(x\right)+\bar{\psi}\left(x\right)\gamma_{\mu}\left(\partial^{\mu}\chi\left(x\right)\right)
\right] \nonumber\\
\label{eq:spfdf}
&=& i\int_{\mathcal{U}}\mathrm{d}^4 x
\left[m-m\right]\bar{\psi}
\left(x\right)\chi\left(x\right)=0,
\end{eqnarray}
where we have used in the last step that $\bar{\psi},\chi$ solve
the Dirac equations~(\ref{eq:De}) and~(\ref{eq:De2}),
respectively.$\quad\bullet$
\\The equal-$t$
hyperplane $\Sigma_t$ yields the usual instant-form scalar product
\begin{eqnarray}
\left(\psi,\chi\right)_{\Sigma_t}=\int_{\mathbb{R}^3}\mathrm{d}^3x\,
\bar{\psi}\left(x\right)\gamma_{0}\,\chi\left(x\right)=\int_{\mathbb{R}^3}\mathrm{d}^3x\,
\psi^{\dag}\left(x\right)\chi\left(x\right).
\end{eqnarray}
Every solution of the Dirac equations~(\ref{eq:De}) and~(\ref{eq:De2}) can be
written as an expansion of a set of orthogonal solutions. A complete set is given by the normalized four-spinors
\begin{eqnarray}\label{eq:dsp}
\psi_{\rho,\boldsymbol{p}}\left(x\right)=\phi_{\boldsymbol{p}}u_{\rho}\left(\boldsymbol{p}\right)\quad\text{and}\quad
\chi_{\rho,\boldsymbol{p}}\left(x\right)=\phi_{\boldsymbol{p}}^{\ast}v_{\rho}\left(\boldsymbol{p}\right),
\end{eqnarray}
 with $\phi_{\boldsymbol{p}}\left(x\right)$ given by~(\ref{eq:planew}) and $\rho=\pm\frac12$ being the spin-projection quantum number.
 The spinor modes
$\psi_{\rho,\boldsymbol{p}}\left(x\right)$
 are said to be positive
energy and $\chi_{\rho,\boldsymbol{p}}\left(x\right)$ negative
energy solutions.\\
In the following we want to show that these solutions are
orthogonal with respect to their scalar product~(\ref{eq:ipd}) and
that this scalar product is independent of the chosen
hypersurface.\\Since $\psi_{\rho,\boldsymbol{p}}\left(x\right)$
and $\chi_{\rho,\boldsymbol{p}}\left(x\right)$ solve the Dirac
equation\footnote{Clearly, each component of the solutions also
satisfies the Klein-Gordon equation~(\ref{eq:KGE}), expressing the
mass-shell constraint~(\ref{eq:p^2}).}, the four-spinors
$u_{\rho}\left(\boldsymbol{p}\right)$ and
$v_{\rho}\left(\boldsymbol{p}\right)$ have to satisfy the
momentum-space Dirac equations
\begin{eqnarray}\label{eq:mde1}
\left(\gamma^{\mu}p_{\mu}-m\right)u_{\rho}\left(\boldsymbol{p}\right)&=&0,\\\label{eq:mde2}
\left(\gamma^{\mu}p_{\mu}+m\right)v_{\rho}\left(\boldsymbol{p}\right)&=&0.\end{eqnarray}
The adjoint four-spinors
$\bar{u}_{\rho}\left(\boldsymbol{p}\right)$ and
$\bar{v}_{\rho}\left(\boldsymbol{p}\right)$ satisfy the adjoint
equations
\begin{eqnarray}\label{eq:amde1}
\bar{u}_{\rho}\left(\boldsymbol{p}\right)\left(\gamma^{\mu}p_{\mu}-m\right)&=&0,\\
\label{eq:amde2}
\bar{v}_{\rho}\left(\boldsymbol{p}\right)\left(\gamma^{\mu}p_{\mu}+m\right)&=&0.\end{eqnarray}
 They can
be written in terms of orthogonal two-component spinors
$\varsigma_{\rho},\boldsymbol{\varepsilon}\varsigma_{\rho}$ as
 \begin{eqnarray}
 u_{\rho}\left(\boldsymbol{p}\right)&=&\sqrt{p^0+m}\left(%
 \begin{array}{c}
   \varsigma_{\rho} \\
   \frac{\boldsymbol{\sigma}\cdot\boldsymbol{p}}{p^0+m}\varsigma_{\rho}
 \end{array}
 \right),%
 \\
  v_{\rho}\left(\boldsymbol{p}\right)&=&-\sqrt{p^0+m}\left(%
 \begin{array}{c}
   \frac{\boldsymbol{\sigma}\cdot\boldsymbol{p}}{p^0+m}\boldsymbol{\varepsilon}\varsigma_{\rho} \\
   \boldsymbol{\varepsilon}\varsigma_{\rho}
 \end{array}
 \right)%
 ,
 \end{eqnarray} with
\begin{eqnarray}
 \varsigma_{\rho}&=&\left(%
 \begin{array}{c}
   \frac12+\rho \\
   \frac12-\rho
 \end{array}
 \right)%
 ,\quad
\boldsymbol{\varepsilon}=\left(%
 \begin{array}{cc}
   0&1 \\
   -1&0
   \end{array}
 \right)%
  \end{eqnarray}
 and $\boldsymbol{\sigma}$ being the Pauli matrices\footnote{The Pauli spin matrices generate the 2-dimensional representation of the $SU\left(2\right)$ by the following Lie algebra:
\begin{eqnarray}
\left[\sigma_i,\sigma_j\right]=2i\epsilon_{ijk}\sigma_k.\end{eqnarray}
Furthermore, they have the following properties
\begin{eqnarray}
\mathrm{tr}\sigma_i=0;\quad\sigma_i=\sigma_i^{\dag};\quad\det \sigma_i=-1;\\
\left\lbrace\sigma_i,\sigma_j\right\rbrace=2\delta_{ij}\mathbf{1}_2,
 \end{eqnarray} from which we obtain the useful relation
\begin{eqnarray}\label{eq:sigma}
\left(\boldsymbol{\sigma}\cdot\boldsymbol{a}\right)
\left(\boldsymbol{\sigma}\cdot\boldsymbol{b}
\right)=\boldsymbol{a}\cdot\boldsymbol{b}+i\boldsymbol{\sigma}\cdot\left(\boldsymbol{a}\times
\boldsymbol{b}\right).\end{eqnarray}}. Since
$\varsigma_{\rho}^{\dag}\varsigma_{\sigma}=\delta_{\rho\sigma}$
with $\rho,~\sigma=\pm\frac12$ and
$\varsigma_{\rho}^{\dag}\boldsymbol{\varepsilon}\varsigma_{\rho}=0$,
we have, using~(\ref{eq:sigma})
\begin{eqnarray}&&\label{eq:ubaru}
 \bar{u}_{\rho}\left(\boldsymbol{p}\right)u_{\sigma}\left(\boldsymbol{p}\right)=-\bar{v}_{\rho}\left(\boldsymbol{p}\right)v_{\sigma}\left(\boldsymbol{p}\right)=2m\delta_{\rho\sigma}
,\\
&&\bar{u}_{\rho}\left(\boldsymbol{p}\right)v_{\sigma}\left(\boldsymbol{p}\right)=\bar{v}_{\rho}\left(\boldsymbol{p}\right)u_{\sigma}\left(\boldsymbol{p}\right)=0,\\
&&u_{\rho}^{\dag}\left(\boldsymbol{p}\right)v_{\sigma}\left(-\boldsymbol{p}\right)=v_{\rho}^{\dag}\left(\boldsymbol{p}\right)u_{\sigma}\left(-\boldsymbol{p}\right)=0.\nonumber\\
\end{eqnarray}
To proceed we use\footnote{
With~(\ref{eq:mde1}) and~(\ref{eq:amde1}) we can write
\begin{eqnarray}\bar{u}_{\rho}\left(\boldsymbol{p}\right)\gamma^{\mu}u_{\sigma}\left(\boldsymbol{p}\right)&=&
\frac{1}{2m}\bar{u}_{\rho}\left(\boldsymbol{p}\right)\left(p_{\nu}\,\gamma^{\nu}\,\gamma^{\mu}+\gamma^{\mu}\,\gamma^{\nu}\,p_{\nu}\right)u_{\sigma}\left(\boldsymbol{p}\right)
\nonumber\\&=&\frac{1}{m}\bar{u}_{\rho}\left(\boldsymbol{p}\right)p_{\nu}\,g^{\mu\nu}u_{\sigma}\left(\boldsymbol{p}\right)=\frac{p^{\mu}}{m}
\underbrace{\bar{u}_{\rho}\left(\boldsymbol{p}\right)u_{\sigma}\left(\boldsymbol{p}\right)}_{=2m\delta_{\rho\sigma},~(\ref{eq:ubaru})},\nonumber\end{eqnarray}
where we have used~(\ref{eq:gammamugammanu}).}
\begin{eqnarray}\label{eq:ugammau}
\bar{u}_{\rho}\left(\boldsymbol{p}\right)\gamma^{\mu}u_{\sigma}\left(\boldsymbol{p}\right)=\bar{v}_{\rho}\left(\boldsymbol{p}\right)\gamma^{\mu}v_{\sigma}\left(\boldsymbol{p}\right)=2p^{\mu}\delta_{\rho\sigma}.
\end{eqnarray} The instant-form scalar
product between the solutions
$\psi_{\rho,\boldsymbol{p}}\left(x\right)$ and
$\chi_{\rho,\boldsymbol{p}}\left(x\right)$ can now be calculated
as
\begin{eqnarray}\label{eq:ifsp1}
\left(\psi_{\rho,\boldsymbol{p}},\psi_{\sigma,\boldsymbol{q}}\right)_{\Sigma_t}=2p^0\delta_{\rho\sigma}\delta^3\left(\boldsymbol{p}-\boldsymbol{q}\right),\quad
\left(\chi_{\rho,\boldsymbol{p}},\chi_{\sigma,\boldsymbol{q}}\right)_{\Sigma_t}=2p^0\delta_{\rho\sigma}\delta^3\left(\boldsymbol{p}-\boldsymbol{q}\right),\\
\label{eq:ifsp2}
\left(\psi_{\rho,\boldsymbol{p}},\chi_{\sigma,\boldsymbol{q}}\right)_{\Sigma_t}=\left(\chi_{\rho,\boldsymbol{p}},\psi_{\sigma,\boldsymbol{q}}\right)_{\Sigma_t}=0.
\end{eqnarray}
The modes are orthogonal and the scalar product independent of
$t$. Thus, the scalar product between the modes should always give
the results of~(\ref{eq:ifsp1}) and~(\ref{eq:ifsp2}), independent
of the chosen spacelike hypersurface $\Sigma$. The scalar product
over the hyperboloid $\Sigma_{\tau}$ with fixed
$x_{\lambda}\,x^{\lambda}=\tau^2=\mathrm{const.}$ reads
\begin{eqnarray} \left(\psi,\chi\right)_{\Sigma_{\tau}}:=\int_{\mathbb{R}^4}2\mathrm{d}^4x\,\delta\left(x_{\lambda}\,x^{\lambda}-\tau^2\right)\theta\left(x^0\right)x^{\mu}
\bar{\psi}\left(x\right)\gamma_{\mu}\chi\left(x\right).
\end{eqnarray} Indeed we obtain the same result as above, namely
\begin{eqnarray}\label{eq:spbpfm}
\left(\psi_{\rho,\boldsymbol{p}},\psi_{\sigma,\boldsymbol{q}}\right)_{\Sigma_{\tau}}=2p^0\delta_{\rho\sigma}\delta^3\left(\boldsymbol{p}-\boldsymbol{q}\right),\quad
\left(\chi_{\rho,\boldsymbol{p}},\chi_{\sigma,\boldsymbol{q}}\right)_{\Sigma_{\tau}}=2p^0\delta_{\rho\sigma}\delta^3\left(\boldsymbol{p}-\boldsymbol{q}\right),\\\label{eq:spbmfm}
\left(\psi_{\rho,\boldsymbol{p}},\chi_{\sigma,\boldsymbol{q}}\right)_{\Sigma_{\tau}}=\left(\chi_{\rho,\boldsymbol{p}},\psi_{\sigma,\boldsymbol{q}}\right)_{\Sigma_{\tau}}=0,
\end{eqnarray} as shown in Appendix~\ref{app:pfspd}. Thus the inner product is independent of the chosen
spacelike hypersurface and~(\ref{eq:spfdf}) holds.\\
\subsection{Covariant Canonical Anticommutation Relations}
Canonical quantization is equivalent to considering the classical
 fields $\psi,\bar{\psi}$ as field operators
$\hat{\psi},\hat{\bar{\psi}}$. As in the scalar case
(cf.~Section~\ref{sec:cccr}), we can formulate covariant canonical
anticommutation relations over a spacelike hypersurface given
by\cite{Schwinger:1948yk}
\begin{eqnarray}
\label{eq:ccar}
&&\int_{\Sigma}\mathrm{d}\Sigma\left(x\right)\left\lbrace\hat{\psi}_{\alpha}\left(y\right),\hat{\varpi}_{\beta}\left(x\right)\right\rbrace_{x,y\in\Sigma}=\int_{\Sigma}\mathrm{d}\Sigma\left(x\right)\left\lbrace\hat{\bar{\psi}}_{\alpha}\left(y\right),\hat{\bar{\varpi}}_{\beta}\left(x\right)\right\rbrace_{x,y\in\Sigma}
=i\delta_{\alpha\beta},\nonumber\\
\\
&&\label{eq:ccar2}
\left\lbrace\hat{\psi}_{\alpha}\left(x\right),\hat{\psi}_{\beta}\left(y\right)\right\rbrace=\left\lbrace\hat{\varpi}_{\alpha}\left(x\right),\hat{\varpi}_{\beta}\left(y\right)\right\rbrace\nonumber\\&&=
\left\lbrace\hat{\psi}_{\alpha}\left(x\right),\hat{\bar{\varpi}}_{\beta}\left(y\right)\right\rbrace=\left\lbrace\hat{\bar{\psi}}_{\alpha}\left(x\right),\hat{\varpi}_{\beta}\left(y\right)\right\rbrace=0,\quad
x,y\in\Sigma.\end{eqnarray} For arbitrary $x$ and $y$ we may again
use the Pauli-Jordan function~(cf.~Section~\ref{sec:cccr})
\begin{eqnarray} \label{eq:Deltapsi1}
&&\left\lbrace\hat{\psi}_{\alpha}\left(x\right),\hat{\bar{\psi}}_{\beta}\left(y\right)\right\rbrace=i\left(i\gamma^{\mu}\,\frac{\partial}{\partial
x^{\mu}}+m\right)_{\alpha\beta}\Delta\left(x-y\right),\\&&
\left\lbrace\hat{\varpi}_{\alpha}\left(x\right),\hat{\bar{\varpi}}_{\beta}\left(y\right)\right\rbrace=i
\left[
n^{\lambda}\left(x\right)\gamma_{\lambda}\right]_{\beta}^{~\gamma}\left[n^{\nu}\left(y\right)\gamma_{\nu}\right]_{\alpha}^{~\delta}\left(i\gamma^{\mu}\,\frac{\partial}{\partial
x^{\mu}}+m\right)_{\gamma\delta}\Delta\left(x-y\right),\nonumber\\\label{eq:Deltapsi2}
\end{eqnarray}
with $\alpha,~\beta,~\gamma,~\delta=1,\ldots 4$ being the Dirac-spinor indices and $n^{\lambda}\left(x\right)$ the timelike
vector depending on the chosen space-time foliation. The field
operator $\hat{\varpi}$ canonically conjugate to $\hat{\psi}$ is
given by
\begin{eqnarray}\label{eq:cfD}
\hat{\varpi}\left(x\right)=n^{\mu}\left(x\right)\frac{\partial\hat{\mathcal{L}}_{\mathrm{D}}\left(x\right)}{\partial\left(\partial^{\mu}\hat{\psi}\left(x\right)\right)}=in^{\mu}\left(x\right)\hat{\bar{\psi}}\gamma_{\mu}\left(x\right)\\
\text{and}\quad \hat{\bar{\varpi}}=\gamma^{0}\hat{\varpi}^{\dag}.
\end{eqnarray}
Choosing the instant-t plane $\Sigma_t$ in~(\ref{eq:ccar}), we
have
\begin{eqnarray}\int_{\mathbb{R}^3}\mathrm{d}^3 x\left\lbrace \hat{\psi}_{\alpha}\left(y\right),i\left[\hat{\bar{\psi}}\left(x\right)\gamma_{0}\right]_{\beta}\right\rbrace  _{x^0=y^0=t}=i\delta_{\alpha\beta}.
\end{eqnarray}
Hence, we can immediately conclude that the equal-t canonical
anticommutation relations are
\begin{eqnarray}\left\lbrace \hat{\psi}_{\alpha}\left(y\right),\left[\hat{\bar{\psi}}\left(x\right)\gamma_{0}\right]_{\beta}\right\rbrace
_{x^0=y^0=t}=\left\lbrace
\hat{\psi}_{\alpha}\left(y\right),\hat{\psi^{\dag}}_{\beta}\left(x\right)\right\rbrace
_{x^0=y^0=t}=\delta_{\alpha\beta}\delta^3\left(\boldsymbol{x}-\boldsymbol{y}\right).\nonumber\\
\end{eqnarray}
If we chose the hyperboloid $\Sigma_{\tau}$ we have
\begin{eqnarray}
 2i\int_{\mathbb{R}^4}\mathrm{d}^4x\,\delta\left(x_{\lambda}\,x^{\lambda}-\tau^2\right)\theta\left(x^0\right)x^{\mu}\left\lbrace\hat{\psi}_{\alpha}\left(y\right),\left[ \hat{\bar{\psi}}\left(x\right)\gamma_{\mu}\right]_{\beta} \right\rbrace_{x^2=y^2=\tau^2}=i\delta_{\alpha\beta}\nonumber\\
\end{eqnarray}
and we can conclude
 that
\begin{eqnarray}x^{\mu}\left\lbrace\hat{\psi}_{\alpha}\left(y\right),\left[ \hat{\bar{\psi}}\left(x\right)\gamma_{\mu}\right]_{\beta} \right\rbrace_{x^2=y^2=\tau^2}=x^0\delta_{\alpha\beta}\delta^3\left(\boldsymbol{x}-\boldsymbol{y}\right).
\end{eqnarray}
Contracting with
$\left[x^{\nu}\,\gamma_{\nu}\right]_{\beta\gamma}$ and
using~(\ref{eq:amugamma}) gives
\begin{eqnarray}\label{eq:licacr}\left\lbrace\hat{\psi}_{\alpha}\left(y\right),\hat{\bar{\psi}}_{\gamma} \left(x\right)\right\rbrace_{x^2=y^2=\tau^2}=\left[x^{\mu}\,\gamma_{\mu}\right]_{\alpha\gamma}\frac{x^0}{x_{\lambda}\,x^{\lambda}}
\delta^3\left(\boldsymbol{x}-\boldsymbol{y}\right).
\end{eqnarray}
 These are the Lorentz-invariant canonical anticommutation relations when quantizing on a hyperboloid.
This result agrees with~\cite{Gromes:1974yu} as shown in
Appendix~\ref{app:ccacr}.\\
 For spacelike $\left(x-y\right)$ the Pauli-Jordan function
$\Delta\left(x-y\right)$ vanishes~(cf.~(\ref{eq:Delta=0})), thus,
when quantizing on a
spacelike hypersurface, only the derivative terms remain in~(\ref{eq:Deltapsi1}) and~(\ref{eq:Deltapsi2}).\\
 For equal Minkowski times
$x^0=y^0=t$ we recover~(\ref{eq:Deltapsi1}) as the usual
anticommutation relations~\cite{Itzykson:2006,Scheck:2001}
\begin{eqnarray}
\left\lbrace
\hat{\psi}_{\alpha}\left(x\right),\hat{\bar{\psi}}_{\beta}\left(y\right)\right\rbrace_{x^0=y^0=t}
&=&
-\gamma^{0}_{\alpha\beta}\,\partial_{0}\Delta\left(x-y\right)\left\vert_{x^0=y^0=t}\right.=\gamma^{0}_{\alpha\beta}\delta^3\left(\boldsymbol{x}-\boldsymbol{y}\right).\nonumber\\
\end{eqnarray}
When taking the hyperboloid
$x_{\lambda}\,x^{\lambda}=y_{\lambda}\,y^{\lambda}=\tau^2$, we obtain
for~(\ref{eq:Deltapsi1})
\begin{eqnarray}
&&\left\lbrace
\hat{\psi}_{\alpha}\left(x\right),\hat{\bar{\psi}}_{\beta}\left(y\right)\right\rbrace_{x^2=y^2=\tau^2}\nonumber\\&&=
-\left[ \gamma^{\mu}\,\partial_{\mu}\right]_{\alpha\beta}
\Delta\left(x-y\right)\left\vert_{x^2=y^2=\tau^2}\right.\nonumber\\&&=\left[
\gamma^{\mu}\right]_{\alpha\beta}
\int_{\mathbb{R}^4}\frac{\mathrm{d}^4p}{\left(2\pi\right)^3}\delta\left(p_{\lambda}\,p^{\lambda}-m^2\right)\theta\left(p^0\right)p_{\mu}
\left(\mathrm{e}^{-ip_{\nu}\left(x-y\right)^{\nu}}+\mathrm{e}^{ip_{\nu}\left(x-y\right)^{\nu}}\right)\left\vert_{x^2=y^2=\tau^2}\right.\nonumber\\\label{eq:deltadd2}&&=
\frac{\left[
\gamma^{\mu}\,X_{\mu}\right]_{\alpha\beta}}{X_{\lambda}\,X^{\lambda}}W\left(X,Y\right)\left\vert_{x^2=y^2=\tau^2}\right.=\left[
\gamma^{\mu}\,x_{\mu}\right]_{\alpha\beta}
\frac{x^0}{x_{\lambda}\,x^{\lambda}}
\delta^3\left(\boldsymbol{x}-\boldsymbol{y}\right),
\end{eqnarray}
where we have used the properties of the $W$
distribution~(cf.~Appendix~\ref{app:W}). For the other
anticommutator~(\ref{eq:Deltapsi2}) we obtain
\begin{eqnarray}
&&\left\lbrace\hat{\varpi}_{\alpha}\left(x\right),\hat{\bar{\varpi}}_{\beta}\left(y\right)\right\rbrace_{x^2=y^2=\tau^2}\nonumber\\&&=
\left[
n_{\rho}\left(x\right)\gamma^{\rho}\right]_{\beta}^{~\gamma}\left[n_{\nu}\left(y\right)\left(\gamma^{\nu}\right)\right]_{\alpha}^{~\delta}\left[
\gamma^{\mu}\,x_{\mu}\right]_{\gamma\delta}
\frac{x^0}{x_{\lambda}\,x^{\lambda}}
\delta^3\left(\boldsymbol{x}-\boldsymbol{y}\right)\left\vert_{x^2=y^2=\tau^2}\right.\nonumber\\\label{eq:deltadd1}
&&= \left[\gamma^{\nu}\,x_{\nu}\right]_{\alpha\beta}
\frac{x^0}{x_{\lambda}\,x^{\lambda}}
\delta^3\left(\boldsymbol{x}-\boldsymbol{y}\right),
\end{eqnarray}
where we have used the unit vector orthogonal on the hyperboloid
being
$n^{\mu}\left(x\right)=\frac{x^{\mu}}{\sqrt{x_{\lambda}\,x^{\lambda}}}$
~(cf.~Section~\ref{sec:pf}). We see that~(\ref{eq:deltadd2})
and~(\ref{eq:deltadd1}) are in agreement with~(\ref{eq:licacr}),
as was expected.
\\
\subsection{Anticommutation Relations in Momentum Space}\label{sec:acrms}
The general solutions $\psi$ and $\bar{\psi}$ of the Dirac
equations~(\ref{eq:De}) and~(\ref{eq:De2}) can be written as
expansions in terms of a complete set of modes.
$\psi_{\rho,\boldsymbol{p}}$ and $\chi_{\rho,\boldsymbol{p}}$
of~(\ref{eq:dsp}) provide an appropriate set, being orthogonal and
normalized with respect to the scalar product~(\ref{eq:ipd}).
After canonical quantization we have for the field operators
\begin{eqnarray}\hat{\psi}\left(x\right)&=&\sum_{\rho=\pm\frac12}\int_{\mathbb{R}^3}\frac{\mathrm{d}^3p}{2p^0}\left(\psi_{\rho,\boldsymbol{p}}\left(x\right)\hat{c}_{\rho}\left(\boldsymbol{p}\right)+\chi_{\rho,\boldsymbol{p}}\left(x\right)\hat{d}_{\rho}^{\dag}\left(\boldsymbol{p}\right)
\right)\nonumber\\&=&\label{eq:dfe1}
\frac{1}{\left(2\pi\right)^{\frac32}}\sum_{\rho=\pm\frac12}\int_{\mathbb{R}^3}\frac{\mathrm{d}^3p}{2p^0}\left(\mathrm{e}^{-ip_{\lambda}\,x^{\lambda}}u_{\rho}\left(\boldsymbol{p}\right)\hat{c}_{\rho}\left(\boldsymbol{p}\right)+\mathrm{e}^{ip_{\lambda}\,x^{\lambda}}v_{\rho}\left(\boldsymbol{p}\right)\hat{d}^{\dag}_{\rho}\left(\boldsymbol{p}\right)
\right),
\nonumber\\
\\
\hat{\bar{\psi}}\left(x\right)&=&\sum_{\rho=\pm\frac12}\int_{\mathbb{R}^3}\frac{\mathrm{d}^3p}{2p^0}\left(\bar{\psi}_{\rho,\boldsymbol{p}}\left(x\right)\hat{c}^{\dag}_{\rho}\left(\boldsymbol{p}\right)+\bar{\chi}_{\rho,\boldsymbol{p}}\left(x\right)\hat{d}_{\rho}\left(\boldsymbol{p}\right)
\right)\nonumber\\&=&\label{eq:dfe2}
\frac{1}{\left(2\pi\right)^{\frac32}}\sum_{\rho=\pm\frac12}\int_{\mathbb{R}^3}\frac{\mathrm{d}^3p}{2p^0}\left(\mathrm{e}^{ip_{\lambda}\,x^{\lambda}}\bar{u}_{\rho}\left(\boldsymbol{p}\right)\hat{c}^{\dag}_{\rho}\left(\boldsymbol{p}\right)+\mathrm{e}^{-ip_{\lambda}\,x^{\lambda}}\bar{v}_{\rho}\left(\boldsymbol{p}\right)\hat{d}_{\rho}\left(\boldsymbol{p}\right)
\right).\nonumber\\ \end{eqnarray} The operators $\hat{c},
\hat{c}^{\dag}, \hat{d}$ and $\hat{d}^{\dag}$ are given by the
invariant scalar product~(\ref{eq:ipd}) as\footnote{For $
\hat{c}_{\rho}$, e.g., we have
\begin{eqnarray}
\hat{c}_{\rho}\left(\boldsymbol{p}\right)&=&\left(\psi_{\rho,\boldsymbol{p}},
\hat{\psi}\right)_{\Sigma}= \sum_{\sigma=\pm\frac12}
\int_{\mathbb{R}^3}\frac{\mathrm{d}^3q}{2q^0}\left[\left(\psi_{\rho,\boldsymbol{p}},\psi_{\sigma,\boldsymbol{q}}
\right)_{\Sigma}\hat{c}_{\sigma}\left(\boldsymbol{q}\right)+\underbrace{\left(\psi_{\rho,\boldsymbol{p}},\chi_{\sigma,\boldsymbol{q}}
\right)_{\Sigma}}_{=0,~(\ref{eq:spbmfm})}\hat{d}^{\dag}_{\sigma}\left(\boldsymbol{q}\right)\right]\nonumber\\&&=
\sum_{\sigma=\pm\frac12}
\int_{\mathbb{R}^3}\frac{\mathrm{d}^3q}{2q^0}2q^0\delta_{\sigma\rho}\,\delta^3\left(\boldsymbol{p}-\boldsymbol{q}\right)\hat{c}_{\sigma}\left(\boldsymbol{q}\right),\nonumber
\end{eqnarray}
where we have used the orthogonality relations between plane wave
spinors~(\ref{eq:spbpfm}).}
 \begin{eqnarray}
&&\hat{c}_{\rho}\left(\boldsymbol{p}\right)=\left(\psi_{\rho,\boldsymbol{p}},
\hat{\psi}\right)_{\Sigma},\quad
\hat{c}_{\rho}^{\dag}\left(\boldsymbol{p}\right)=\left(\hat{\psi},\psi_{\rho,\boldsymbol{p}}\right)_{\Sigma},\\
&&
\hat{d}_{\rho}\left(\boldsymbol{p}\right)=\left(\hat{\psi},\chi_{\rho,\boldsymbol{p}}\right)_{\Sigma},\quad\hat{d}_{\rho}^{\dag}\left(\boldsymbol{p}\right)=\left(\chi_{\rho,\boldsymbol{p}},
\hat{\psi}\right)_{\Sigma}.
\end{eqnarray}
\\
These relations, together with the canonical anticommutation
relations~(\ref{eq:ccar}) and~(\ref{eq:ccar2}), imply the
harmonic-oscillator anticommutation relations in momentum space
\begin{eqnarray}\label{eq:acrms}
\left\lbrace
\hat{c}_{\rho}\left(\boldsymbol{p}\right),\hat{c}^{\dag}_{\sigma}\left(\boldsymbol{q}\right)\right\rbrace&=&
\left\lbrace \hat{d}_{\rho}\left(\boldsymbol{p}\right),\hat{d}^{\dag}_{\sigma}\left(\boldsymbol{q}\right)\right\rbrace=2p^0\delta_{\rho\sigma}\delta^3\left(\boldsymbol{p}-\boldsymbol{q}\right),\\
\left\lbrace
\hat{c}_{\rho}\left(\boldsymbol{p}\right),\hat{c}_{\sigma}\left(\boldsymbol{q}\right)\right\rbrace&=&\left\lbrace
\hat{c}^{\dag}_{\rho}\left(\boldsymbol{p}\right),\hat{c}^{\dag}_{\sigma}\left(\boldsymbol{q}\right)\right\rbrace
=\left\lbrace
\hat{d}_{\rho}\left(\boldsymbol{p}\right),\hat{d}_{\sigma}\left(\boldsymbol{q}\right)\right\rbrace=\left\lbrace
\hat{d}^{\dag}_{\rho}\left(\boldsymbol{p}\right),\hat{d}^{\dag}_{\sigma}\left(\boldsymbol{q}\right)\right\rbrace\nonumber\\&=&
\left\lbrace
\hat{c}_{\rho}\left(\boldsymbol{p}\right),\hat{d}^{\dag}_{\sigma}\left(\boldsymbol{q}\right)\right\rbrace=\left\lbrace
\hat{d}_{\rho}\left(\boldsymbol{p}\right),\hat{c}^{\dag}_{\sigma}\left(\boldsymbol{q}\right)\right\rbrace=0\label{eq:acrms2}
.
\end{eqnarray}
In Appendix~\ref{app:cacrms} this statement is explicitly shown in
point form.\\ Due to these anticommutation relations one may
interpret these operators as annihilation or creation operators on
a Fock space~(\ref{sec:Focksp}) annihilating or creating field
quanta characterized by the continuous three-momentum vector
$\boldsymbol{p}$ and the discrete spin-projection quantum number
$\rho$. Therefore the basis consists of simultaneous eigenstates
of
$\hat{\boldsymbol{P}}$ and a spin operator with eigenvalue $\rho$.\\
Finally, we shall note that the field expansions~(\ref{eq:dfe1})
and~(\ref{eq:dfe2}) together with the anticommutation relations in
momentum space~(\ref{eq:acrms}) and~(\ref{eq:acrms2}) imply the
anticommutation relation~(\ref{eq:Deltapsi1})
and~(\ref{eq:Deltapsi2}) and the explicit form of the Pauli-Jordan
function in~(\ref{eq:Delta}). The
anticommutator~(\ref{eq:Deltapsi1}), using the field
expansions~(\ref{eq:dfe1}) and~(\ref{eq:dfe2}) and the
anticommutation relations~(\ref{eq:acrms}) and~(\ref{eq:acrms2}),
is explicitly calculated as
\begin{eqnarray}
&&\left\lbrace
\hat{\psi}_{\alpha}\left(x\right),\hat{\bar{\psi}}_{\beta}\left(y\right)\right\rbrace\nonumber\\&&=
\int_{\mathbb{R}^3}\frac{\mathrm{d}^3p}{2p^0\left(2\pi\right)^3}\sum_{\rho=\pm\frac12}\nonumber\\&&\quad\times
\left(\mathrm{e}^{-ip_{\lambda}\left(x-y\right)^{\lambda}}\left[u_{\rho}\right]_{\alpha}\left(\boldsymbol{p}\right)\left[\bar{u}_{\rho}\right]_{\beta}\left(\boldsymbol{p}\right)
+\mathrm{e}^{ip_{\lambda}\left(x-y\right)^{\lambda}}\left[v_{\rho}\right]_{\alpha}\left(\boldsymbol{p}\right)\left[\bar{v}_{\rho}\right]_{\beta}\left(\boldsymbol{p}\right)
\right) \nonumber\\&&=
i\left(i\gamma^{\mu}\,\frac{\partial}{\partial x^{\mu}}+m\right)_{\alpha\beta}\Delta\left(x-y\right),
\end{eqnarray} where we have used the projectors~(\ref{eq:proj}) and~(\ref{eq:proj2}).
\subsection{Generators in Wigner Representation}
As in the scalar case~(cf.~Section~\ref{sec:gwr}) we want to show
the equivalence of instant- and point-form quantization by
representing the generators of our theory in the Wigner
basis~(cf.~Section~\ref{sec:acrms}). For free fields they can be
shown to have the same form.
\subsubsection{Global Gauge Transformations}
As we have seen in Section~\ref{sec:Nt}, the existence of a
conserved symmetry current~(\ref{eq:cc}) follows from the
invariance of the Lagrangian density
$\mathcal{L}_{\mathrm{D}}$~(\ref{eq:Ld})
 under the action of global $U\left(1\right)$-symmetry
group. After canonical quantization and normal ordering this
current reads
\begin{eqnarray}
\hat{\mathcal{J}}_{\mathrm{D}}^{\mu}\left(x\right)=\,:
\hat{\bar{\psi}}\left(x\right)\gamma^{\mu}\hat{\psi}\left(x\right):,\quad \text{with}\quad\partial_{\mu}\,\hat{\mathcal{J}}_{\mathrm{D}}^{\mu}\left(x\right)=0.
\end{eqnarray}
This current integrated over a spacelike hypersurface gives a conserved
charge or symmetry operator
\begin{eqnarray}
\hat{Q}_{\mathrm{D}}=\int_{\Sigma}\mathrm{d}\Sigma_{\mu}\left(x\right)\hat{\mathcal{J}}_{\mathrm{D}}^{\mu}\left(x\right)=
\int_{\Sigma}\mathrm{d}\Sigma_{\mu}\left(x\right):
\hat{\bar{\psi}}\left(x\right)\gamma^{\mu}\hat{\psi}\left(x\right):.
\end{eqnarray}
 Inserting the field expansions~(\ref{eq:dfe1}) and~(\ref{eq:dfe2}) and choosing the equal-t hyperplane
$\Sigma_{t}$ the resulting charge operator in Wigner
representation reads
\begin{eqnarray}
\hat{Q}_{\mathrm{D}}= \sum_{\rho=\pm\frac12}\int_{\mathbb{R}^3}
\frac{\mathrm{d}^3p}{2p_0}\left(\hat{c}_{\rho}^{\dag}\left(\boldsymbol{p}\right)\hat{c}_{\rho}\left(\boldsymbol{p}\right)-\hat{d}_{\rho}^{\dag}\left(\boldsymbol{p}\right)\hat{d}_{\rho}\left(\boldsymbol{p}\right)\right).
\end{eqnarray}
As in the scalar case this suggests to consider
$\hat{c}_{\rho}^{\dag}\left(\boldsymbol{p}\right)$ and
$\hat{c}_{\rho}\left(\boldsymbol{p}\right)$ as creation and
annihilation operators of particles with charge $+1$ and
$\hat{d}_{\rho}^{\dag}\left(\boldsymbol{p}\right)$ and
$\hat{d}_{\rho}\left(\boldsymbol{p}\right)$ as creation and
annihilation
operators of antiparticles with charge $-1$, respectively.\\
 If we choose the
equal-$\tau$ hyperboloid $\Sigma_{\tau}$ as the spacelike
hypersurface we obtain the same result as above, namely
\begin{eqnarray}\label{eq:pfQ}
\hat{Q}_{\mathrm{D}}&=&\int_{\mathbb{R}^4}2\mathrm{d}^4x\,\delta\left(x_{\lambda}\,x^{\lambda}-\tau^2\right)\theta\left(x^0\right)x^{\mu}:
\hat{\bar{\psi}}\left(x\right)\gamma^{\mu}\hat{\psi}\left(x\right):
\nonumber\\&=&\sum_{\rho=\pm\frac12}\int_{\mathbb{R}^3}
\frac{\mathrm{d}^3p}{2p_0}\left(\hat{c}_{\rho}^{\dag}\left(\boldsymbol{p}\right)\hat{c}_{\rho}\left(\boldsymbol{p}\right)-\hat{d}_{\rho}^{\dag}\left(\boldsymbol{p}\right)\hat{d}_{\rho}\left(\boldsymbol{p}\right)\right).\end{eqnarray}
This is calculated in Appendix~\ref{app:ggt}.
\\
Finally we confirm~(\ref{eq:gfggt}) by using the canonical
anticommutation relations~(\ref{eq:acrms}) and~(\ref{eq:acrms2}),
\begin{eqnarray}
\hat{\psi}\left(x\right)=\left[\hat{\psi}\left(x\right),\hat{Q}_{\mathrm{D}}\right].
\end{eqnarray}
\subsubsection{Translations}
In Section~\ref{sec:Nt} we have seen, that the energy-momentum
tensor~(\ref{eq:emt}) follows from the invariance of the action
under displacements. Inserting~(\ref{eq:Ld}) for the Lagrangian
density $\mathcal{L}_{\mathrm{D}}$, the energy-momentum tensor
operator for Dirac fields becomes after canonical quantization
\begin{eqnarray}\label{eq:demt1}
\hat{\mathcal{T}}_{\mathrm{D}}^{\mu\nu}\left(x\right)=
\label{eq:demt1}\frac{i}{2}:\left[\hat{\bar{\psi}}\left(x\right)\gamma^{\mu}\left(\partial^{\nu}\hat{\psi}\left(x\right)\right)-\left(\partial^{\nu}\hat{\bar{\psi}}\left(x\right)\right)\gamma^{\mu}\hat{\psi}\left(x\right)\right]:,\\
\text{with}\quad\partial_{\mu}\,\hat{\mathcal{T}}_{\mathrm{D}}^{\mu\nu}\left(x\right)=0.\end{eqnarray}
From equation~(\ref{eq:Gepsilon1}) we obtain
the four-momentum operator
\begin{eqnarray}\hat{P}_{\mathrm{D}}^{\mu}=\int_{\Sigma}\mathrm{d}\Sigma_{\nu}\left(x\right)\hat{\mathcal{T}}_{\mathrm{D}}^{\mu\nu}\left(x\right).
\end{eqnarray}
Inserting the field expansions~(\ref{eq:dfe1}),~(\ref{eq:dfe2}) and taking the
equal-$t$ hyperplane $\Sigma_{t}$ we obtain the usual result for
the translation generator in Wigner representation,
\begin{eqnarray}
\hat{P}_{\mathrm{D}}^{\mu}=\sum_{\rho=\pm\frac12}\int_{\mathbb{R}^3}\frac{\mathrm{d}^3p}{2p_0}p^{\mu}\left(\hat
{c}_{\rho}^{\dag}\left(\boldsymbol{p}\right)\hat{c}_{\rho}\left(\boldsymbol{p}\right)+\hat{d}_{\rho}^{\dag}\left(\boldsymbol{p}\right)\hat{d}_{\rho}\left(\boldsymbol{p}\right)\right).\end{eqnarray} \\
As shown in Appendix~(\ref{app:PmuD}), integration over the
hyperboloid $\Sigma_{\tau}$ gives the same result,
\begin{eqnarray}
\hat{P}_{\mathrm{D}}^{\mu}&=&\int_{\mathbb{R}^4}2\mathrm{d}^4x\,
\delta\left(x_{\lambda}\,x^{\lambda}-\tau^2\right)\theta\left(x^0\right)x_{\nu}\hat{\mathcal{T}}_{\mathrm{D}}^{\mu\nu}\left(x\right)\nonumber\\\label{eq:PmuD}&=&\sum_{\rho=\pm\frac12}\int_{\mathbb{R}^3}\frac{\mathrm{d}^3p}{2p_0}p^{\mu}\left(\hat
{c}_{\rho}^{\dag}\left(\boldsymbol{p}\right)\hat{c}_{\rho}\left(\boldsymbol{p}\right)+\hat{d}_{\rho}^{\dag}\left(\boldsymbol{p}\right)\hat{d}_{\rho}\left(\boldsymbol{p}\right)\right).\end{eqnarray}
$\hat{P}_{\mathrm{D}}^{\mu}$, represented in this basis, still
transforms as a four-vector under Lorentz transformations,
\begin{eqnarray}\label{eq:tranfpmu}
\hat{U}\left(\Lambda\right)
\hat{P}^{\mu}_{\mathrm{D}}\hat{U}\left(\Lambda\right)^{-1}&=&\sum_{\rho,\sigma,\lambda=\pm\frac12}
\int_{\mathbb{R}^3}\frac{\mathrm{d}^3p}{2p_0}p^{\mu}\left(\hat
{c}_{\sigma}^{\dag}\left(\boldsymbol{\Lambda
p}\right)\hat{c}_{\lambda}\left(\boldsymbol{\Lambda
p}\right)D^{\frac12}_{\sigma\rho}\left(R\left(p,\Lambda\right)\right)D^{\frac12\ast}_{\lambda\rho}\left(R\left(
p,\Lambda\right)\right)\right.\nonumber\\&&+\left.\hat
{d}_{\sigma}^{\dag}\left(\boldsymbol{\Lambda
p}\right)\hat{d}_{\lambda}\left(\boldsymbol{\Lambda
p}\right)D^{\frac12}_{\sigma\rho}\left(R\left(p,\Lambda\right)\right)D^{\frac12\ast}_{\lambda\rho}\left(R\left(
p,\Lambda\right)\right)\right)\nonumber\\&=&
\sum_{\rho,\sigma,\lambda=\pm\frac12}\int_{\mathbb{R}^3}\frac{\mathrm{d}^3p}{2p_0}\left(\Lambda^{-1}
p\right)^{\mu} \nonumber\\&&\times\left(\hat
{c}_{\sigma}^{\dag}\left(\boldsymbol{p}\right)\hat{c}_{\lambda}\left(\boldsymbol{
p}\right)D^{\frac12}_{\sigma\rho}\left(R\left(\Lambda^{-1}p,\Lambda\right)\right)D^{\frac12}_{\rho\lambda}\left(R^{-1}\left(\Lambda^{-1}p,\Lambda\right)\right)\right.\nonumber\\&&\quad+\left.\hat
{d}_{\sigma}^{\dag}\left(\boldsymbol{p}\right)\hat{d}_{\lambda}\left(\boldsymbol{
p}\right)D^{\frac12}_{\sigma\rho}\left(R\left(\Lambda^{-1}p,\Lambda\right)\right)D^{\frac12}_{\rho\lambda}\left(R^{-1}\left(\Lambda^{-1}p,\Lambda\right)\right)\right)\nonumber\\&=&
\left(\Lambda^{-1}\right)^{\mu}_{~\nu}\hat{P}^{\nu}_{\mathrm{D}}.
\end{eqnarray} Here, we have used Lorentz invariance of the integration measure and the Lorentz-transformation
properties of single-particle
states~\cite{Krassnigg:2001}\begin{eqnarray}
 \hat{U}\left(\Lambda\right)\hat{c}_{\rho}^{\dag}\left(\boldsymbol{p}\right)
\hat{U}\left(\Lambda\right)^{-1}=\sum_{\sigma\pm\frac12}\hat{c}_{\sigma}^{\dag}\left(\boldsymbol{\Lambda
p}\right)D^{\frac12}_{\sigma\rho}\left(R\left(
p,\Lambda\right)\right).\end{eqnarray} The
$D^{\frac12}_{\sigma\rho}$ are the matrix elements of the
\emph{Wigner D-functions}\footnote{Note that
$D^{\frac12\ast}_{\lambda\rho}\left(R\left(
p,\Lambda\right)\right)=D^{\frac12}_{\rho\lambda}\left(R^{-1}\left(
p,\Lambda\right)\right)$.}. $R$ denotes a Wigner rotation given by
\begin{eqnarray}
R\left(p,\Lambda\right)=\Lambda\left(\Lambda\left(\omega\right)
v\right)^{-1}\Lambda\left(\omega\right)\Lambda\left(v\right),
\end{eqnarray} with $\Lambda\left(\omega\right)$ being a general Lorentz transformation and
$\Lambda\left(v=\frac{p}{m}\right)$ a Lorentz boost.\\This
Fock-space representation of $\hat{P}_{\mathrm{D}}^{\mu}$ together
with the harmonic-oscillator anticommutation
relations~(\ref{eq:acrms}),~(\ref{eq:acrms2}) leads to the
conclusion, that the field quanta created by
$\hat{c}_{\rho}^{\dag}\left(\boldsymbol{p}\right)$ and
$\hat{d}_{\rho}^{\dag}\left(\boldsymbol{p}\right)$ are eigenstates
of the free four-momentum operator with eigenvalues $p^{\mu}$.
\\
Finally, on using the canonical commutation
relations~(\ref{eq:acrms}) and~(\ref{eq:acrms2}), we confirm that
\begin{eqnarray}
\partial^{\mu}\hat{\psi}\left(x\right)=i\left[\hat{
P}_{\mathrm{D}}^{\mu},\hat{\psi}\left(x\right)\right].
\end{eqnarray}
\subsubsection{Lorentz Transformations} From Neother's theorem in
Section~\ref{sec:Nt}, we have found a conserved current under the
assumption of the invariance of the action under Lorentz
transformations. This current is the, so called, angular-momentum
density~(\ref{eq:Mmunusigma}). After canonical quantization it is
given by
\begin{eqnarray}
\hat{\mathcal{M}}_{\mathrm{D}}^{\mu\nu\sigma}\left(x\right):=x^{\nu}\hat{\mathcal{T}}_{\mathrm{D}}^{\mu\sigma}\left(x\right)-x^{\sigma}\hat{\mathcal{T}}_{\mathrm{D}}^{\mu\nu}\left(x\right),\quad\text{with}\quad
\partial_{\mu}\,\hat{\mathcal{M}}_{\mathrm{D}}^{\mu\nu\sigma}\left(x\right)=0.\end{eqnarray}
Then, the associated conserved charge operator reads
\begin{eqnarray}\hat{M}_{\mathrm{D}}^{\mu\nu}=\int_{\Sigma}\mathrm{d}\Sigma_{\lambda}\left(x\right)\hat{\mathcal{M}}_{\mathrm{D}}^{\lambda\mu\nu}\left(x\right)
=\int_{\Sigma}\mathrm{d}\Sigma_{\lambda}\left(x\right)\left[x^{\mu}\hat{\mathcal{T}}_{\mathrm{D}}^{\lambda\nu}\left(x\right)-x^{\nu}\hat{\mathcal{T}}_{\mathrm{D}}^{\lambda\mu}\left(x\right)\right].
\end{eqnarray}
Inserting for the energy-momentum tensor~(\ref{eq:demt1}), the field expansions~(\ref{eq:dfe1}),~(\ref{eq:dfe2}) and integrating over the
equal-t hyperplane $\Sigma_t$ gives~\cite{Kugo:1997fs}
\begin{eqnarray}
\hat{M}_{\mathrm{D}}^{\mu\nu}&=&\sum_{\rho,\sigma=\pm\frac12}
 \int_{\mathbb{R}^3}\frac{\mathrm{d}^3p}{2p_0}
\left\lbrace\hat{c}_{\rho}^{\dag}\left(\boldsymbol{p}\right)u^{\dag}_{\rho}\left(\boldsymbol{p}\right)\left[l^{\mu\nu}\left(\boldsymbol{p}\right)+\frac12\sigma^{\mu\nu}\right]u_{\sigma}
\left(\boldsymbol{p}\right)\hat{c}_{\sigma}\left(\boldsymbol{p}\right)\right.\nonumber\\&&+\left.\hat{d}_{\rho}^{\dag}\left(\boldsymbol{p}\right)v^{T}_{\rho}\left(\boldsymbol{p}\right)\left[m^{\mu\nu}\left(\boldsymbol{p}\right)-\frac12\left(\sigma^T\right)^{\mu\nu}\right]v^{\ast}_{\sigma}
\left(\boldsymbol{p}\right)\hat{d}_{\sigma}\left(\boldsymbol{p}\right)\right\rbrace\label{eq:gfbr},
\end{eqnarray}
with $l^{\mu\nu}$ and $m^{\mu\nu}$ given in Section~\ref{sec:gwr}.
The similar calculation by integrating over the hyperboloid
$\Sigma_{\tau}$ is more complicated but leads to the same result
as~(\ref{eq:gfbr}).
\\
Finally we find, using the canonical anticommutation
relations~(\ref{eq:acrms}) and~(\ref{eq:acrms2}), that
\begin{eqnarray}
\left[\left(x^{\mu}\partial^{\nu}-x^{\nu}\partial^{\mu}\right)\delta^{\alpha}_{~\beta}-\frac{i}{2}\left[\sigma^{\mu\nu}\right]^{\alpha}_{~\beta}
\right]\hat{\psi}^{\beta}\left(x\right)=i\left[\hat{\psi}^{\alpha}\left(x\right),\hat{M}_{\mathrm{D}}^{\mu\nu}\right].
\end{eqnarray}
\chapter{Covariant Scattering Theory for Interacting Fields}
\label{cha:st} As we have already mentioned, a necessary condition
for the formulation of a scattering theory is a time development
that covers the whole Minkowski space. We assume that the
interaction is local and decreases fast enough at infinity. This
ensures that we can define asymptotic states and a S operator that
maps between these states of non-interacting
particles~(cf.~Section~\ref{sec:sop}). Therefore the
Hamiltonian~(\ref{eq:H}) that generates $\tau$-development from
one hyperboloid to another does not seem to be very useful, since it only covers the forward light cone.\\
Looking for an evolution that covers the whole Minkowski space we
make the choice as in~\cite{Zelzer:2005}. That is, we keep $\tau$
fixed and shift the hyperboloid along a timelike path. The
generators for this evolution are the components of the
four-momentum operator. It should, however, be noted that this
kind of evolution is clearly not perpendicular to the quantization
surface. But as we will see, this fact does not play a significant
role for the formulation of a scattering theory.
\section{Poincar\'e Generators}
When including interactions into a free theory, the kinematic
generators stay interaction free, whereas the dynamic generators
will contain interaction terms~(cf.~Section~\ref{sec:gtd}).\\ In
order to include interactions, we add an interaction term to the
free Lagrangian density,
\begin{eqnarray}
\hat{\mathcal {L}}\left(x\right)=\hat {\mathcal
{L}}_{\mathrm{free}}\left(x\right)+\hat{\mathcal
{L}}_{\mathrm{int}}\left(x\right).
\end{eqnarray}
In~(\ref{eq:emt}) we saw that, as long as $\mathcal
{L}_{\mathrm{int}}$ does not contain derivatives of the fields, we
can write the interaction part of the energy-momentum tensor as
\begin{eqnarray}
\hat{\mathcal{T}}_{\mathrm{int}}^{\mu\nu}\left(x\right)=-g^{\mu\nu}:\hat{\mathcal
{L}}_{\mathrm{int}}\left(x\right):.\end{eqnarray} Therefore we
have from~(\ref{eq:Pmu}) for the interacting part of the
four-momentum operator
\begin{eqnarray}\hat{P}^{\mu}_{\mathrm{int}}= \int_{\Sigma}\mathrm{d}\Sigma_{\nu}\left(x\right)\hat{\mathcal{T}}_{\mathrm{int}}^{\nu\mu}\left(x\right)=
-\int_{\Sigma}\mathrm{d}\Sigma_{\nu}\left(x\right)g^{\nu\mu}:\hat{\mathcal
{L}}_{\mathrm{int}}\left(x\right):.\end{eqnarray} Choosing the
equal-t hyperplane $\Sigma_t$ we see immediately, that
$\hat{\mathcal {L}}_{\mathrm{int}}$ does not enter the
three-components of the four-momentum operator, i.e.
\begin{eqnarray}\hat{P}^{\mu}_{\mathrm{int}}=\int_{\mathbb{R}^3}\mathrm{d}^3x\, g_{~\nu}^0g^{\nu\mu}:\hat{\mathcal {L}}_{\mathrm{int}}\left(x\right):=
\int_{\mathbb{R}^3}\mathrm{d}^3x \,g^{0\mu}:\hat{\mathcal
{L}}_{\mathrm{int}}\left(x\right):= \left(
\begin{array}{c}
  \hat{P}^{0}_{\mathrm{int}}\\
\boldsymbol{0}
\end{array}\right)^{\mu}.\end{eqnarray} On the other hand, if we take the interacting
part of the Lorentz generator $\hat{M}^{\mu\nu}$ and integrate
over $\Sigma_t$ we have
\begin{eqnarray}\hat{M}^{\mu\nu}_{\mathrm{int}}&=&\int_{\mathbb{R}^3}\mathrm{d}^3x\, g_{~\sigma}^0\left(x^{\mu}
g^{\sigma\nu}-x^{\nu} g^{\sigma\mu}\right):\hat{\mathcal
{L}}_{\mathrm{int}}\left(x\right):\nonumber\\&=&
\int_{\mathbb{R}^3}\mathrm{d}^3x\, \left(x^{\mu} g^{0\nu}-x^{\nu}
g^{0\mu}\right):\hat{\mathcal
{L}}_{\mathrm{int}}\left(x\right):.\end{eqnarray} This expression
does not vanish, if either $\mu=k\,\wedge\,\nu=0$ or
$\nu=k\,\wedge\,\mu=0$ for $k=1,\ldots3$, i.e. for boost
generators $\hat{B}^k=\hat{M}^{0k}=-\hat{M}^{k0}$. Hence, we see
explicitly that the boost generators together with $\hat{P}^0$
become interaction dependent in instant form, which is exactly the
statement in Section~\ref{sec:gtd}.
\\
\\
In point form we have the equal-$\tau$ hyperboloid $\Sigma_{\tau}$
giving
\begin{eqnarray}
\hat{P}^{\mu}_{\mathrm{int}}&=&
\int_{\mathbb{R}^4}2\,\mathrm{d}^4x\,\delta\left(x_{\lambda}\,x^{\lambda}-\tau^2\right)\theta\left(x^0\right)x_{\nu}\hat{\mathcal{T}}_{\mathrm{int}}^{\nu\mu}\left(x\right)\nonumber\\&=&
-
\int_{\mathbb{R}^4}2\,\mathrm{d}^4x\,\delta\left(x_{\lambda}\,x^{\lambda}-\tau^2\right)\theta\left(x^0\right)x^{\mu}:\hat{\mathcal
{L}}_{\mathrm{int}}\left(x\right):.\label{eq:Pmuint}\end{eqnarray}
We see explicitly that all components of the four-momentum
operator become interaction dependent. On the other hand, the
antisymmetric tensor $\hat{M}^{\mu\nu}$ stays interaction free,
i.e. the interaction dependent part of this tensor vanishes
\begin{eqnarray}\hat{M}^{\mu\nu}_{\mathrm{int}}&=&2 \int_{\mathbb{R}^4}\mathrm{d}^4x\,\delta\left(x_{\lambda}\,x^{\lambda}-\tau^2\right)\theta\left(x^0\right)x_{\sigma}\left(x^{\mu}
\hat{\mathcal{T}}_{\mathrm{int}}^{\sigma\nu}\left(x\right)-x^{\nu}
\hat{\mathcal{T}}_{\mathrm{int}}^{\sigma\mu}\left(x\right)\right)\nonumber\\&=&-2
\int_{\mathbb{R}^4}\mathrm{d}^4x\,\delta\left(x_{\lambda}\,x^{\lambda}-\tau^2\right)\theta\left(x^0\right)x_{\sigma}\left(x^{\mu}
g^{\sigma\nu}-x^{\nu} g^{\sigma\mu}\right):\hat{\mathcal
{L}}_{\mathrm{int}}\left(x\right):\nonumber\\&=&0.\end{eqnarray}
Again we observe that the statement in Section~\ref{sec:gtd}
holds.\\ Finally, it should be noted that quantization on the
hyperboloid provides a representation of the Poincar\'e
algebra~(\ref{eq:la}) expressed by the, so called,
\emph{point-form equations}~\cite{Krassnigg:2001,Krassnigg:2003}
\begin{eqnarray}\label{eq:pfeq}
\left[\hat{P}^{\mu},\hat{P}^{\nu}\right]&=&0,\\
\hat{U}\left(\Lambda\right)
\hat{P}^{\mu}\hat{U}\left(\Lambda\right)^{-1}&=&\left(\Lambda^{-1}\right)^{\mu}_{~\nu}\hat{P}^{\nu},
\label{eq:pfeq2}\end{eqnarray} where $\hat{P}^{\mu}$ is the total
four-momentum operator (including all interactions).\footnote{If
we have
$\hat{P}^{\mu}=\hat{P}^{\mu}_{\mathrm{free}}+\hat{P}^{\mu}_{\mathrm{int}}$,
then
$\left[\hat{P}_{\mathrm{int}}^{\mu},\hat{P}_{\mathrm{int}}^{\nu}\right]=0$
follows from microscopic causality~(cf.~Section~\ref{sec:mc}).
$\left[\hat{P}_{\mathrm{int}}^{\mu},\hat{P}_{\mathrm{free}}^{\nu}\right]+\left[\hat{P}_{\mathrm{free}}^{\mu},\hat{P}_{\mathrm{int}}^{\nu}\right]=0$
and~(\ref{eq:pfeq2}) follow from the transformation properties of
$\hat{\mathcal {L}}_{\mathrm{int}}\left(x\right)$ under
translations and Lorentz transformations,
respectively\cite{Krassnigg:2001}.}
\section{Covariant Interaction Picture}
As we have seen, the effect of quantization on the hyperboloid
$x_{\lambda}\,x^{\lambda}=\tau^2$ is that interactions described
by the $\hat{\mathcal {L}}_{\mathrm{int}}$ enter all components of
the four-momentum operator. Thus, we can write the total
four-momentum operator as the sum of a free and an interacting
part
\begin{eqnarray}\hat{P}^{\mu}=\hat{P}^{\mu}_{\mathrm{free}}+\hat{P}^{\mu}_{\mathrm{int}}.\end{eqnarray}
Since all 4 components of the translation generator are
interaction dependent, we can adapt a covariant interaction
picture. It is covariant in the sense that it describes evolution
into arbitrary timelike space-time directions. In a covariant
interaction picture both, operators and states, are $x$-dependent.
The operators have an evolution generated by the free
four-momentum operator $\hat{P}_{\mathrm{free}}$, whereas the
states have an $x$-dependence generated by the interaction
four-momentum $\hat{P}_{\mathrm{int}}$ \cite{Aitchison:2003}. Let
$\hat{O}$ be an operator and $\vert\Phi\rangle$ be a state
specified on the quantization surface
$x_{\lambda}\,x^{\lambda}=\tau^2$. Then we have
 \begin{eqnarray}
\label{eq:evolution of operator}
\hat{O}\left(x\right)=\mathrm{e}^{i\hat{P}_{\mathrm{free}}^{\lambda}\,x_{\lambda}}\,
\hat{O}\,\mathrm{e}^{-i\hat{P}_{\mathrm{free}}^{\nu}\,x_{\nu}},\quad\text{with
}\hat{O}\left(x=0\right)= \hat{O}
\end{eqnarray} and
\begin{eqnarray}\label{eq:Phix}
\left\vert\Phi\right.\left(x\right)\rangle=\mathrm{e}^{i\hat{P}_{\mathrm{free}}^{\mu}\,x_{\mu}}\mathrm{e}^{-i\hat{P}^{\nu}\,
x_{\nu}}\left\vert\Phi\right.\rangle,\quad\text{with
}\left\vert\Phi\right.\left(x=0\right)\rangle=\left\vert\Phi\right.\rangle.
 \end{eqnarray}
Then the equations of motions describing evolution of the system
into the $x$-direction are given by
 \begin{eqnarray}
i\partial^{\mu}\hat{O}\left(x\right)=\left[\hat{O}\left(x\right),\hat{P}_{\mathrm{free}}^{\mu}\right]
\end{eqnarray}
and
\begin{eqnarray}\label{eq:DmuPhi}
i\partial^{\mu}\left\vert\Phi\right.\left(x\right)\rangle=\hat{P}_{\mathrm{int}}^{\mu}\left(x\right)\left\vert\Phi\right.\left(x\right)\rangle,\end{eqnarray}
 with
 \begin{eqnarray}\label{eq:evolutionPint}
\hat{P}_{\mathrm{int}}^{\mu}\left(x\right)=\mathrm{e}^{i\hat{P}_{\mathrm{free}}^{\lambda}\,x_{\lambda}}\hat{P}_{\mathrm{int}}^{\mu}\mathrm{e}^{-i\hat{P}_{\mathrm{free}}^{\nu}\,x_{\nu}}.\end{eqnarray}
Evolution of the state from $y$ to $x$ is described by an
evolution operator $\hat{U}\left(x,y\right)$, such that

\begin{eqnarray}\label{eq:UPhi}
\hat{U}\left(x,y\right)\left\vert\Phi\right.\left(y\right)\rangle=\left\vert\Phi\right.\left(x\right)\rangle,
\end{eqnarray}
with the boundary condition $\hat{U}\left(x,x\right)=\hat{1}$.\\
Then the asymptotic states
$\left\vert\Phi_{\mathrm{in}}\right.\rangle$ and
$\left\vert\Phi_{\mathrm{out}}\right.\rangle$ are given by
\begin{eqnarray}\label{eq:asymtotst}
\left\vert\Phi_{\mathrm{in}}\right.\rangle=\lim_{x^2\rightarrow\infty}\hat{U}\left(x,y\right)\left\vert\Phi\right.\left(y\right)\rangle,\quad
x^0<0
\end{eqnarray} and
\begin{eqnarray}\label{eq:asymtotst2}
\left\vert\Phi_{\mathrm{out}}\right.\rangle=\lim_{x^2\rightarrow\infty}\hat{U}\left(x,y\right)\left\vert\Phi\right.\left(y\right)\rangle,\quad
x^0>0.
\end{eqnarray}The limits are taken in such a way,
that $x$ is timelike, lying in the forward or backward light cone
for $x^0>0$ or $x^0<0$, respectively. At $x^2\rightarrow\infty$,
we assume $\hat{\mathcal {L}}_{\mathrm{int}}\left(x\right)$ and
therefore $\hat{P}_{\mathrm{int}}^{\mu}\left(x\right)$ to be
negligible. Then we see from~(\ref{eq:DmuPhi}) that
$\left\vert\Phi_{\mathrm{in}}\right.\rangle$ and
$\left\vert\Phi_{\mathrm{out}}\right.\rangle$ are constant and
eigenstates of $\hat{P}_{\mathrm{free}}^{\mu}$. Thus
$\left\vert\Phi_{\mathrm{in}}\right.\rangle$ and
$\left\vert\Phi_{\mathrm{out}}\right.\rangle$ describe non-interacting particles with (physical masses and) definite momenta.\\
 Inserting~(\ref{eq:UPhi})
into the equation of motion~(\ref{eq:DmuPhi}) leads to the
differential equation for $\hat{U}\left(x,y\right)$,
\begin{eqnarray}\label{eq:DmuU}
i\partial^{\mu}\hat{U}\left(x,y\right)=\hat{P}_{\mathrm{int}}^{\mu}\left(x\right)\hat{U}\left(x,y\right).\end{eqnarray}
This equation can be integrated along an arbitrary smooth path
$\mathcal{C}\left(x,y\right)$ joining $x$ and $y$.
$\mathcal{C}\left(x,y\right)$ can be parameterized in the
following way:
\begin{eqnarray}
w_{\mu}\left(s\right)=y_{\mu}+s\left(x-y\right)_{\mu},\quad 0\leq
s\leq 1,\end{eqnarray} with
\begin{eqnarray}\mathrm{d}w_{\mu}=\left(x-y\right)_{\mu}\mathrm{d}s\quad
\text{and}\quad \frac{\partial}{\partial
w_{\mu}}=\frac{\left(x-y\right)^{\mu}}{\left(x-y\right)^2}\frac{\partial}{\partial
s}. \end{eqnarray} Integrating the equation of
motion~(\ref{eq:DmuU}) using this parameterization gives the
integral equation
\begin{eqnarray}\label{eq:Ieq}
i\int_{\mathcal{C}\left(x,y\right)}\mathrm{d}w_{\mu}\frac{\partial}{\partial
w_{\mu}}\,\hat{U}\left(w,y\right)&=& i\int_{0}^{1}\mathrm{d}s
\frac{\partial}{\partial
s}\,\hat{U}\left(y+s\left(x-y\right),y\right)=\left.i\hat{U}\left(y+s\left(x-y\right),y\right)\right\vert_{0}^1\nonumber\\&=&
i\hat{U}\left(x,y\right)-i\hat{U}\left(y,y\right)=
\int_{\mathcal{C}\left(x,y\right)}\mathrm{d}w_{\mu}\hat{P}_{\mathrm{int}}^{\mu}\left(w\right)\hat{U}\left(w,y\right).\nonumber\\\end{eqnarray}
Then the solution of this integral equation with the boundary
condition can be written as
\begin{eqnarray}
\hat{U}\left(x,y\right)&=&\hat{1}-i\int_{\mathcal{C}\left(x,y\right)}\mathrm{d}w_{\mu}\,
\hat{P}_{\mathrm{int}}^{\mu}\left(w\right)\hat{U}\left(w,y\right)\nonumber\\&=&\hat{1}-i\int_{\mathcal{C}\left(x,y\right)}
\mathrm{d}w_{1\mu}\hat{P}_{\mathrm{int}}^{\mu}\left(w_1\right)\nonumber\\&&+\left(i\right)^2\int_{\mathcal{C}\left(x,y\right)}
\mathrm{d}w_{1\mu}\,\int_{\mathcal{C}\left(w_1,y\right)}
\mathrm{d}w_{2\nu}\,\hat{P}_{\mathrm{int}}^{\mu}\left(w_1\right)\hat{P}_{\mathrm{int}}^{\nu}\left(w_2\right)+\ldots.
\nonumber\\\end{eqnarray} Thus the formal solution of the integral
equation~(\ref{eq:Ieq}) can be written as path-ordered exponential
\begin{eqnarray}\hat{U}\left(x,y\right)=\mathcal{P}\,\mathrm{exp}\left(-i\int_{\mathcal{C}\left(x,y\right)}
\mathrm{d}w_{\mu}\hat{P}_{\mathrm{int}}^{\mu}\left(w\right)\right),\end{eqnarray}where
$\mathcal{P}$ denotes the path ordering.
\section{Lorentz-Invariant Scattering
Operator} 
\begin{figure}
 \begin{center}
  \includegraphics[clip=7cm,width=8.9cm]{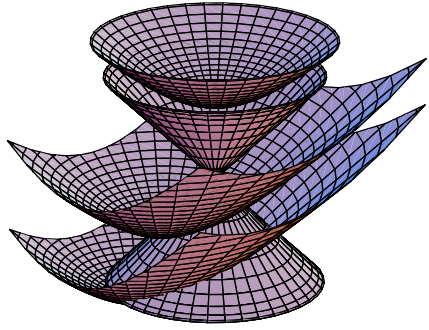}
 \caption{\label{fig:pfsc3D}
 $s$-development into arbitrary timelike directions.}
 \end{center}
 \end{figure}

\begin{figure}
 \begin{center}
\psfrag{Ss3}{$\Sigma_{\tau}\left(s_3\right)$}
\psfrag{Ss2}{$\Sigma_{\tau}\left(s_2\right)$}
\psfrag{Ss1}{$\Sigma_{\tau}\left(s_1\right)$}
\psfrag{ws}{$w\left(s\right)$}
\psfrag{0}{$0$}
\psfrag{x0}{$x^0$}
\psfrag{vecx}{$\boldsymbol{x}$}
 \includegraphics[clip=7cm,width=8cm]{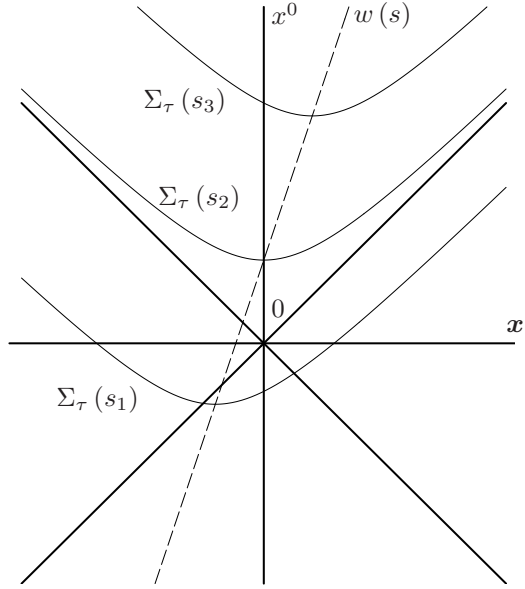}
 \caption{\label{fig:pfst}The scattering process described by $w\left(s\right)$
 of~(\ref{eq:w(s)}) corresponds to shifting a hyperboloid with
 fixed but arbitrary $\tau$ into an arbitrary timelike space-time
 direction. This is illustrated for $s_1<s_2<s_3$. Note that the development in $s$ is, unlike $\tau$-development, not perpendicular
 to the hyperboloid.
 }
 \end{center}
 \end{figure}

Covariant scattering may be described as evolution from
$\left\vert\Phi_{\mathrm{in}}\right.\rangle$ to
$\left\vert\Phi_{\mathrm{out}}\right.\rangle$. That is, one starts
with non-interacting particles at $x^2\rightarrow\infty,\,x^0<0$
described by $\left\vert\Phi_{\mathrm{in}}\right.\rangle$, then
these particles approach each other, scatter and finally one ends
up with non-interacting particles at
$x^2\rightarrow\infty,\,x^0>0$ described by
$\left\vert\Phi_{\mathrm{out}}\right.\rangle$. From the definition
of the scattering operator~(\ref{eq:scattopdef}) together
with~(\ref{eq:asymtotst}) and~(\ref{eq:asymtotst2}) we find
\begin{eqnarray}
\label{eq:S-operator3}
\left\vert\Phi_{\mathrm{out}}\right.\rangle&=&\lim_{x^2\rightarrow\infty}\left\vert\Phi\right.\left(x\right)\rangle=
\lim_{x^2\rightarrow\infty}\lim_{y^2\rightarrow\infty}\hat{U}\left(x,y\right)\left\vert\Phi\right.\left(y\right)\rangle\nonumber\\&=&
\lim_{x^2\rightarrow\infty}\lim_{y^2\rightarrow\infty}\hat{U}\left(x,y\right)\left\vert\Phi_{\mathrm{in}}\right.\rangle\quad\text{such
that}\quad x^0>0,\,y^0<0.\end{eqnarray} Consequently, the
scattering operator can be written as
\begin{eqnarray}\label{eq:S-operator}
\hat{S}=\lim_{x^2\rightarrow\infty}\lim_{y^2\rightarrow\infty}
\hat{U}\left(x,y\right).\end{eqnarray}
 As we have mentioned
before, the path $\mathcal{C}\left(x,y\right)$ of the scattering
process can be chosen arbitrarily. For simplicity, we take a
straight line joining $x$ and $y$. Then, the path may be
parameterized as
\begin{eqnarray}\label{eq:w(s)}w_{\mu}\left(s\right)=a_{\mu}+s\,k_{\mu}. \end{eqnarray}
$a$ is a constant arbitrary four-vector in Minkowski space and
 $k$ denotes a timelike
four-vector normalized to unity describing the direction of the
scattering process,
\begin{eqnarray}k:=\lim_{x^2\rightarrow\infty}\lim_{y^2\rightarrow\infty}\frac{x-y}{\sqrt{\left(x-y\right)^2}},
\quad \text{with}\quad k_{\lambda}\,k^{\lambda}=1.\end{eqnarray}
This is illustrated in Figures~\ref{fig:pfsc3D}
and~\ref{fig:pfst}.
With this parameterization the S operator~(\ref{eq:S-operator})
becomes a simple $s$-ordered exponential
\begin{eqnarray}
\label{eq:t-orderedexponential}\hat{S}=\mathcal{S}\,\mathrm{exp}\left(-i\int_{-\infty}^{\infty}\mathrm{d}s\,
k_{\mu}\hat{P}_{\mathrm{int}}^{\mu}\left(w\left(s\right)\right)\right).\end{eqnarray}
The $\mathcal{S}$ in front of the exponential denotes the
$s$-ordering.\\
 Expanding the exponential in powers of the
interaction we obtain
\begin{eqnarray}\hat{S}&=&\hat{1}-i\int_{-\infty}^{\infty}\mathrm{d}s\, k_{\mu}
\hat{P}_{\mathrm{int}}^{\mu}\left(w\left(s\right)\right)\nonumber\\&&+\frac{\left(i\right)^2}{2}\int_{-\infty}^{\infty}\mathrm{d}s_1\,
k_{\mu}\int_{-\infty}^{\infty}\mathrm{d}s_2\,
k_{\nu}\,\mathcal{S}\left[
\hat{P}_{\mathrm{int}}^{\mu}\left(w\left(s_1\right)\right)
\hat{P}_{\mathrm{int}}^{\nu}\left(w\left(s_2\right)\right)\right]+\ldots.\nonumber\\\end{eqnarray}
It follows from~(\ref{eq:Pmuint}) and~(\ref{eq:evolutionPint})
that the evolution of $\hat{P}_{\mathrm{int}}^{\mu}$ in the
interaction picture is
\begin{eqnarray}
\hat{P}_{\mathrm{int}}^{\mu}\left(w\right)&=&-2
\int_{\mathbb{R}^4}\mathrm{d}^4x\,\delta\left(x_{\lambda}\,x^{\lambda}-\tau^2\right)\theta\left(x^0\right)x^{\mu}\mathrm{e}^{i\hat{P}_{\mathrm{free}}^
{\mu}\, w_{\mu}}:\hat{\mathcal
{L}}_{\mathrm{int}}\left(x\right):\mathrm{e}^{-i\hat{P}_{\mathrm{free}}^
{\nu}\, w_{\nu}}\nonumber\\&=& -2
\int_{\mathbb{R}^4}\mathrm{d}^4x\,\delta\left(x_{\lambda}\,x^{\lambda}-\tau^2\right)\theta\left(x^0\right)x^{\mu}:\hat{\mathcal
{L}}_{\mathrm{int}}\left(x+w\right):
\label{eq:Pmuint2}.\end{eqnarray} Inserting~(\ref{eq:Pmuint2})
into~(\ref{eq:t-orderedexponential}) yields
\begin{eqnarray}\hat{S}&=&\mathcal{S}\,\mathrm{exp}\left(2i\int_{-\infty}^{\infty}\mathrm{d}s\, k_{\mu} \int_{\mathbb{R}^4}\mathrm{d}^4x\,\delta\left(x_{\lambda}\,x^{\lambda}-\tau^2\right)\theta\left(x^0\right)x^{\mu}:\hat{\mathcal {L}}_{\mathrm{int}}\left(x+s\,k+a\right):\right)\nonumber\\&=&
\hat{1}+2i\int_{-\infty}^{\infty}\mathrm{d}s\, k_{\mu}
\int_{\mathbb{R}^4}\mathrm{d}^4x\,\delta\left(x_{\lambda}\,x^{\lambda}-\tau^2\right)\theta\left(x^0\right)x^{\mu}:\hat{\mathcal
{L}}_{\mathrm{int}}\left(x+s\,k+a\right):\nonumber\\&&+\frac{\left(2i\right)^2}{2}\int_{-\infty}^{\infty}\mathrm{d}s_1\,
k_{\mu}\int_{-\infty}^{\infty}\mathrm{d}s_2\,
k_{\nu}\,\int_{\mathbb{R}^4}\mathrm{d}^4x_1\,\delta\left(x_{1\lambda}\,x_1^{\lambda}-\tau^2\right)\theta\left(x_1^0\right)x_1^{\mu}\nonumber\\&&\quad\times\int_{\mathbb{R}^4}\mathrm{d}^4x_2\,\delta\left(x_{2\rho}\,x_2^{\rho}-\tau^2\right)\theta\left(x_2^0\right)x_2^{\nu}\nonumber\\&&\quad\times\,\mathcal{S}\left[:\hat{\mathcal
{L}}_{\mathrm{int}}\left(x_1+s_1\,k+a\right)::\hat{\mathcal
{L}}_{\mathrm{int}}\left(x_2+\,s_2\,k+a\right):\right]+\ldots.\nonumber\\\label{eq:Des1}
\end{eqnarray}
This expansion of the S operator in orders of the Lagrangian
density can be shown to be equivalent to the usual instant-form
expansion
\begin{eqnarray}\hat{S}&=&\mathcal{S}\,\mathrm{exp}\left(i\int_{-\infty}^{\infty}\mathrm{d}s\, \int_{\mathbb{R}^3}\mathrm{d}^3x:\hat{\mathcal {L}}_{\mathrm{int}}\left(s,\boldsymbol{x}\right):\right)\nonumber\\&=&
\hat{1}+i\int_{-\infty}^{\infty}\mathrm{d}s\,
\int_{\mathbb{R}^3}\mathrm{d}^3x:\hat{\mathcal
{L}}_{\mathrm{int}}\left(s,\boldsymbol{x}\right):\nonumber\\&&+\frac{\left(i\right)^2}{2}\int_{-\infty}^{\infty}\mathrm{d}s_1\,
\int_{-\infty}^{\infty}\mathrm{d}s_2\,
\,\int_{\mathbb{R}^3}\mathrm{d}^3x_1\int_{\mathbb{R}^3}\mathrm{d}^3x_2\,\mathcal{S}\left[:\hat{\mathcal
{L}}_{\mathrm{int}}\left(s_1,\boldsymbol{x}_1\right)::\hat{\mathcal
{L}}_{\mathrm{int}}\left(s_2,\boldsymbol{x}_2\right):\right]+\ldots.\nonumber\\\label{eq:Des2}
\end{eqnarray}
The latter corresponds to scattering theory in the
$(s=x_0)$-direction, i.e. $k=\left(1,0,0,0\right)^{T}$. This
equivalence is explicitly shown in Appendix~\ref{app:st}.
Therefore, this manifest covariant formulation of scattering
theory and the resulting series expansion of the S
operator~(\ref{eq:Des1}) leads to the usual perturbative results.
Hence, the consequences like overall four-momentum conservation at
the vertex is guaranteed, although three-momentum conservation at
the vertex does, in general, not hold in point-form quantum field
theory.
\chapter{Summary and Outlook}
Canonical field quantization is usually formulated at equal times.
In addition, also quantization on the light front has been
investigated extensively. These quantization procedures can be
found in common text books about quantum field theory. Only a few
papers exist about quantization on the space-time hyperboloid
$x_{\lambda}\,x^{\lambda}=\tau^2$, although this, so called,
point-form quantum field theory has some attractive features. In
point form the dynamic generators of the Poincar\'e group,
generating evolution of the system away from the quantization
surface, can be combined to a four-vector $P^{\mu}$. On the other
hand, the generators for Lorentz transformations $B^{i}$ and
$J^{i}$, $i=1,\ldots 3$, are purely kinematic and can be combined
to a second-order tensor $M^{\mu\nu}$. This makes it possible to
formulate canonical field quantization in a manifestly Lorentz
covariant way, without making reference to a particular time
parameter. \\In the earlier papers about point-form quantum field
theory evolution in $\tau$, generated by the dilatation operator,
has been studied and a Fock basis related to the generators of the
Lorentz group, the Lorentz basis, has been used. However,
$\tau$-evolution together with the
Lorentz basis lead to a number of conceptual difficulties.\\
\\In this diploma thesis we have developed a formalism for
quantization on the forward hyperboloid which makes use of the
usual momentum-state basis. Our main objective was then to study
evolution of the system generated by $P^{\mu}$.\\ For free massive
spin-$0$ and free massive spin-$\frac12$ quantum fields we have
shown that the Fock-space representation of the Poincar\'e
generators in the momentum basis is identical with their
Fock-space representation when quantizing at equal times.
Furthermore, (anti)commutation relations on the hyperboloid have
been found which are Lorentz invariant. These field
(anti)commutators are in agreement with the general
Schwinger-Tomonaga quantization conditions, which apply to
arbitrary (spacelike) quantization surfaces. All necessary
integrations over the hyperboloid have been performed in Cartesian
coordinates by means of an appropriately defined
distribution\footnote{The idea how to calculate this distribution
goes back to F. Coester.}. \\For interacting theories a
generalized interaction picture has been suggested which makes no
preference of a particular space-time direction. Within this
covariant interaction picture it is possible to define a
Lorentz-invariant scattering operator and to formulate a covariant
scattering theory. The expansion of the generalized scattering
operator in powers of the interaction was shown to be equivalent
to the usual
time-ordered perturbation theory.\\
\\
As a next step the consequences of these results and their
applications to point-form quantum mechanical models with a finite
number of degrees of freedom should be further investigated. In
this context one can think of deriving effective interactions and
(conserved) current operators for application in relativistic
few-body systems. Another field of application of point-form
quantum field theory are gauge theories. Due to the manifest
Lorentz covariance, gauge transformations and gauge invariance
can be naturally incorporated into the theory. Therefore, by
viewing quantum chromodynamics as a point-form quantum field
theory may lead to new insights into the nature of gauge fixing
and other properties of non-Abelian gauge theories.
  \thispagestyle{empty}\pagebreak
\thispagestyle{empty}
\begin{appendix}
\chapter{Hypersurface Element}
\label{eq:app:hse}
We have to show that
\begin{eqnarray}
\mathrm{d}\Sigma^{\mu}_{\tau}\left(x\right)=n^{\mu}\left(x\right)\mathrm{d}\Sigma_{\tau}\left(x\right)=2\,\mathrm{d}^4
x\,\delta\left(x_{\lambda}\,x^{\lambda}-\tau^2\right)\theta\left(x^0\right)x^{\mu},\end{eqnarray}
where $n^{\mu}$ is the timelike unit vector orthogonal on the
spacelike hyperboloid $\Sigma_{\tau}:
x_{\lambda}\,x^{\lambda}=\tau^2$. We will use hyperbolic
coordinates
$\left(\tau,\xi\equiv\mathrm{cosh}\beta,\vartheta,\varphi\right)$
defined by
the coordinate transformation~(\ref{eq:xhc}).\\
$\circ\quad$
For the volume element we have
\begin{eqnarray}
\mathrm{d}^4x\,=\left\vert\frac{\partial\left(x^0,x^1,x^2,x^3\right)}{\partial\left(\tau,\xi,\vartheta,\varphi\right)}\right\vert\,\mathrm{d}\tau\,\mathrm{d}\xi\,\mathrm{d}\vartheta\,\mathrm{d}\varphi=
\tau^3\sqrt{\xi^2-1}\,\mathrm{sin}\vartheta\,\mathrm{d}\tau\,\mathrm{d}\xi\,\mathrm{d}\vartheta\,\mathrm{d}\varphi.\end{eqnarray}
Since $x_{\lambda}\,x^{\lambda}=\tau^2$, we can write
\begin{eqnarray}
\delta\left(x_{\lambda}\,x^{\lambda}-\tau_0^2\right)=\delta\left(\tau^2-\tau_0^2\right)=\frac{\delta\left(\tau-\tau_0\right)+\delta\left(\tau+\tau_0\right)}{2\tau}.
\end{eqnarray} Since
$\xi=\mathrm{cosh}\beta>0,\,\forall\beta\in\mathbb{R}$, we can
write
\begin{eqnarray}\theta\left(x^0\right)=\theta\left(\tau\xi\right)=\theta\left(\tau\right).\end{eqnarray}
Therefore, we can write
\begin{eqnarray}\delta\left(\tau_0-\tau\right)=2\tau\,\delta\left(x_{\lambda}\,x^{\lambda}-\tau_0^2\right)\theta\left(x^0\right)
\end{eqnarray} and
 we have for the hypersurface element
\begin{eqnarray} \mathrm{d}\Sigma_{\tau}&=&2\,\mathrm{d}^4
x\,\delta\left(x_{\lambda}\,x^{\lambda}-\tau^2\right)\theta\left(x^0\right)\sqrt{x_{\lambda}\,x^{\lambda}}=\delta\left(\tau_0-\tau\right)\tau_0^3\sqrt{\xi^2-1}\,\mathrm{sin}\vartheta\,\mathrm{d}\tau_0\,\mathrm{d}\xi\,\mathrm{d}\vartheta\,\mathrm{d}\varphi\nonumber\\\label{eq:hse}
&=&
\tau^3\sqrt{\xi^2-1}\,\mathrm{sin}\vartheta\,\mathrm{d}\xi\,\mathrm{d}\vartheta\,\mathrm{d}\varphi.
\end{eqnarray} The oriented hypersurface element can be
written in Cartesian coordinates as
\begin{eqnarray}
 \mathrm{d}\Sigma^{\mu}\left(x\right)=\left(
\begin{array}{c}\mathrm{d}x^1\,\mathrm{d}x^2\,\mathrm{d}x^3
  \\
 \mathrm{d}x^0\,\mathrm{d}x^2\,\mathrm{d}x^3\\
 \mathrm{d}x^0\,\mathrm{d}x^1\,\mathrm{d}x^3
 \\
  \mathrm{d}x^0\,\mathrm{d}x^1\,\mathrm{d}x^2\\
\end{array}
\right)^{\mu}.
\end{eqnarray}
A short calculation yields
\begin{eqnarray}
\,\mathrm{d}x^1\,\mathrm{d}x^2\,\mathrm{d}x^3=\left\vert\frac{\partial\left(x^1,x^2,x^3\right)}{\partial\left(\xi,\vartheta,\varphi\right)}
\right\vert\,\mathrm{d}\xi\,\mathrm{d}\vartheta\,\mathrm{d}\varphi=\tau^3\xi\sqrt{\xi^2-1}
\,\mathrm{sin}\vartheta\,\mathrm{d}\xi\,\mathrm{d}\vartheta\,\mathrm{d}\varphi;\nonumber\\
\,\mathrm{d}x^0\,\mathrm{d}x^2\,\mathrm{d}x^3=\left\vert\frac{\partial\left(x^0,x^2,x^3\right)}{\partial\left(\xi,\vartheta,\varphi\right)}
\right\vert\,\mathrm{d}\xi\,\mathrm{d}\vartheta\,\mathrm{d}\varphi=\tau^3\left(\xi^2-1\right)
\,\mathrm{sin}^2\vartheta \,\mathrm{cos}\varphi\,\mathrm{d}\xi\,\mathrm{d}\vartheta\,\mathrm{d}\varphi;\nonumber\\
\,\mathrm{d}x^0\,\mathrm{d}x^1\,\mathrm{d}x^3=\left\vert\frac{\partial\left(x^0,x^1,x^3\right)}{\partial\left(\xi,\vartheta,\varphi\right)}
\right\vert\,\mathrm{d}\xi\,\mathrm{d}\vartheta\,\mathrm{d}\varphi=\tau^3\left(\xi^2-1\right)
\,\mathrm{sin}^2\vartheta \,\mathrm{sin}\varphi\,\mathrm{d}\xi\,\mathrm{d}\vartheta\,\mathrm{d}\varphi;\nonumber\\
\,\mathrm{d}x^0\,\mathrm{d}x^1\,\mathrm{d}x^2=\left\vert\frac{\partial\left(x^0,x^1,x^2\right)}{\partial\left(\xi,\vartheta,\varphi\right)}
\right\vert\,\mathrm{d}\xi\,\mathrm{d}\vartheta\,\mathrm{d}\varphi=\tau^3\left(\xi^2-1\right)
\,\mathrm{sin}\vartheta
\,\mathrm{cos}\vartheta\,\mathrm{d}\xi\,\mathrm{d}\vartheta\,\mathrm{d}\varphi\nonumber.\end{eqnarray}
Thus, we finally obtain
\begin{eqnarray}
 \,\mathrm{d}\Sigma_{\tau}^{\mu}&=&\left(
\begin{array}{c}\tau\xi\\
 \tau \sqrt{\xi^2-1}
\,\mathrm{sin}\vartheta \,\mathrm{cos}\varphi\\ \tau \sqrt{\xi^2-1}
\,\mathrm{sin}\vartheta \,\mathrm{sin}\varphi
 \\
 \tau\sqrt{\xi^2-1}
\,\mathrm{cos}\vartheta
\end{array}
\right)^{\mu}
\tau^2\sqrt{\xi^2-1}\,\mathrm{sin}\vartheta\,\mathrm{d}\xi\,\mathrm{d}\vartheta
\,\mathrm{d}\varphi\nonumber\\&=&2\,\mathrm{d}^4
x\,\delta\left(x_{\lambda}\,x^{\lambda}-\tau^2\right)\theta\left(x^0\right)x^{\mu}=n^{\mu}\,\mathrm{d}\Sigma_{\tau},
\end{eqnarray} where
\begin{eqnarray}\frac{x^{\mu}}{\sqrt{x_{\lambda}\,x^{\lambda}}}=n^{\mu}\left(x\right),
\end{eqnarray} which is in agreement with~(\ref{eq:nv}) and~(\ref{eq:dSigma}).$\quad\bullet$
\chapter {$W$ and $W^{\mu}$ Distribution}
\label{app:WWmu} Integrations over the space-time hyperboloid
$x_{\lambda }\,x^{\lambda}=\tau^2$ are easily performed in
Cartesian coordinates on using the distributions $W$ and
$W^{\mu}$.
\section {$W$ Distribution}
\label{app:W} For $W$ we have to show that
\begin{eqnarray}\label{eq:Wdistr}
W\left(p+q,p-q\right)=2p^0\delta^3\left(\boldsymbol{p}-\boldsymbol{q}\right)
\end{eqnarray}and
\begin{eqnarray}
W\left(p-q,p+q\right)=0,
\end{eqnarray}
for $p_{\lambda}\,p^{\lambda}=q_{\lambda}\,q^{\lambda}=m^2$.\\
$\circ\quad$ The $W$ distribution is defined as
\begin{eqnarray}\label{eq:irepW}
&&W\left(p+q,p-q\right)=W\left(P,Q\right)\nonumber\\&&:=\frac{2}{\left(2\pi\right)^3}\int_{\mathbb{R}^4}\mathrm{d}^4x\,
\delta\left(x_{\lambda}\,x^{\lambda}-\tau^2\right)\theta\left(x^0\right)x^{\mu}
P_{\mu}\,\mathrm{e}^{ix^{\nu}Q_{\nu}},
\end{eqnarray} where we have introduced new variables
\begin{eqnarray}P:=p+q\quad\text{and}\quad Q:=p-q.\end{eqnarray} We see that the constraint
\begin{eqnarray}\label{eq:PQ=0}P_{\lambda}\,Q^{\lambda}=0\end{eqnarray} is equivalent to the mass-shell constraint~(\ref{eq:p^2}),
\begin{eqnarray}p_{\lambda}\,p^{\lambda}=q_{\lambda}\,q^{\lambda}=m^2.\end{eqnarray}
Thus, the timelike four-vector $P$ is orthogonal to the
spacelike $Q$. Since $P$ is timelike, it can be written as a
boost transform of a vector $\tilde{P}$ that has a time component
only, namely
\begin{eqnarray}
P=\Lambda\left(v\right)\tilde{P}=\Lambda\left(v\right)\left(%
\begin{array}{c}
  \tilde{P}^0 \\
  \boldsymbol{0} \\
\end{array}
\right),%
\quad\text{with}\quad P_{\lambda}\,P^{\lambda}=\tilde{P}_0
\,\tilde{P}^0=M^2. \label{eq:Lambda}
\end{eqnarray}
$\Lambda\left(v\right)$ is a rotationless canonical boost with a
four-velocity $v=\frac{P}{M}$ and is explicitly given
by~\cite{Klink:1998qf}
\begin{eqnarray}
\Lambda\left(v\right)=\left(%
\begin{array}{cc}
  v^0&\boldsymbol{v}^T\\
  \boldsymbol{v}&\boldsymbol{1}+\frac{v^0-1}{\boldsymbol{v}^2}\boldsymbol
{v} \boldsymbol{v}^T\\
\end{array}
\right)%
=
\left(%
\begin{array}{cc}
  \frac{P^0}{M}&\frac{\boldsymbol{P}^T}{M}\\
  \frac{\boldsymbol{P}}{M}&\boldsymbol{1}+\frac{P^0/M-1}{\boldsymbol{P}^2}\boldsymbol
{P} \boldsymbol{P}^T\\
\end{array}
\right).
\end{eqnarray}
Since $Q$ is orthogonal to $P$, it can be written as a boost
transform of a vector $\tilde{Q}$, that has only spatial components,
i.e.
\begin{eqnarray}\label{eq:Q}
Q=\Lambda\left(v\right)\tilde{Q}=\Lambda\left(v\right)\left(%
\begin{array}{c}
  0 \\
  \tilde{\boldsymbol{Q}} \\
\end{array}
\right)%
,\quad Q_{\lambda}\,Q^{\lambda}=-\tilde{\boldsymbol{Q}}^2 ,
 \end{eqnarray} such that
\begin{eqnarray}
\tilde{P}_{\lambda}\,\tilde{Q}^{\lambda}=0\end{eqnarray} holds.
Inverting~(\ref{eq:Q}) gives
\begin{eqnarray}\label{eq:Qtilde}
\tilde{Q}=\Lambda^{-1}\left(P/M\right)Q=
\left(
\begin{array}{cc}
  \frac{P^0}{M}&-\frac{\boldsymbol{P}^T}{M}\\
  -\frac{\boldsymbol{P}}{M}&\boldsymbol{1}+\frac{P^0/M-1}{\boldsymbol{P}^2}\boldsymbol
{P} \boldsymbol{P}^T\\
\end{array}\right)
\left(\begin{array}{c}Q^0\\
\boldsymbol{Q}\\\end{array}\right) .
 \end{eqnarray}
From $P_{\lambda}\,Q^{\lambda}=0$, we can express $Q^0$ as
$Q_0=\frac{\boldsymbol{P}\cdot\boldsymbol{Q}}{P^0} $. Using this
relation together with~(\ref{eq:Qtilde}) we calculate
\begin{eqnarray}
\tilde{\boldsymbol{Q}}&=&-\frac{\boldsymbol{P}}{M}Q^0+\left(\boldsymbol{1}+\frac{P^0/M-1}{\boldsymbol{P}^2}\boldsymbol{P}\boldsymbol{P}^T\right)\boldsymbol{Q}\nonumber\\
&=&-\frac{\boldsymbol{P}}{M}\frac{\boldsymbol{P}\cdot\boldsymbol{Q}}{P^0}+\left(\boldsymbol{1}+\frac{P^0/M-1}{\boldsymbol{P}^2}\boldsymbol{P}\boldsymbol{P}^T\right)\boldsymbol{Q}\nonumber\\
&=&\left[\boldsymbol{1}-\left(\frac{\boldsymbol{1}}{MP^0}-\frac{P^0-M}{M\boldsymbol{P}^2}\right)\boldsymbol{P}\boldsymbol{P}^T\right]\boldsymbol{Q}\nonumber\\
&=&\left(\boldsymbol{1}-\frac{\boldsymbol{P}\boldsymbol{P}^T}{P^0\left(P^0+M\right)}\right)
\boldsymbol{Q}=\mathbf{N}\boldsymbol{Q}. \end{eqnarray}
$\mathbf{N}$ denotes a $3\times3$ matrix with determinant
\begin{eqnarray}\det \boldsymbol{\bold{N}}=\det\left(
\boldsymbol{1}-\frac{\boldsymbol{P}\boldsymbol{P}^T}{P^0\left(P^0+M\right)}\right)=1-\frac{\boldsymbol{P}^2}{P^0\left(P^0+M\right)}=\frac{M}{P^0}
.\end{eqnarray} Since $W\left(P,Q\right)$ is Lorentz
invariant by definition, we have
\begin{eqnarray}
W\left(P,Q\right)&=&W\left(\tilde{P},\tilde{Q}\right)=\frac{2}{\left(2\pi\right)^3}\int_{\mathbb{R}^4}\mathrm{d}^4\tilde{x}
\,\delta\left(\tilde{x}_{\lambda}\,\tilde{x}^{\lambda}-\tau^2\right)\theta\left(\tilde{x}^0\right)\tilde{x}^{0}
\tilde{P}_{0}\,\mathrm{e}^{-i\tilde{\boldsymbol{x}}\cdot
\tilde{\boldsymbol{Q}}}\nonumber\\&=&\frac{1}{\left(2\pi\right)^3}\int_{\mathbb{R}^4}\mathrm{d}^4\tilde{x}\,\frac{\delta\left(\tilde{x}^{0}-\sqrt{\tilde{\boldsymbol{x}}^2+\tau^2}\right)}{\tilde{x}^{0}}\,\tilde{x}^{0}
M\,\mathrm{e}^{-i\tilde{\boldsymbol{x}}\cdot
\tilde{\boldsymbol{Q}}}\nonumber\\&=&\frac{1}{\left(2\pi\right)^3}\int_{\mathbb{R}^3}\mathrm{d}^3\tilde{x}M
\mathrm{e}^{-i\tilde{\boldsymbol{x}}\cdot \tilde{\boldsymbol{Q}}}=
M\,\delta^{3}\left(\tilde{\boldsymbol{Q}}\right) .\end{eqnarray}
In the original frame this has finally the form
\begin{eqnarray}
W\left(P,Q\right)=M\delta^{3}\left(\tilde{\boldsymbol{Q}}\right)=M\delta^{3}\left(\mathbf{N}\boldsymbol{Q}\right)=\frac{M}{\det \boldsymbol{\mathbf{N}}}\delta^{3}\left(\boldsymbol{Q}\right)=P^0\delta^{3}\left(\boldsymbol{Q}\right)
.\quad\bullet\end{eqnarray}  \\
If $P$ and $Q$ are interchanged $W\left(Q,P\right)$ becomes zero:
\\ $\circ\quad$Similarly as above we have
\begin{eqnarray}
W\left(Q,P\right)&=&W\left(\tilde{Q},\tilde{P}\right)=-\frac{2}{\left(2\pi\right)^3}\int_{\mathbb{R}^4}\mathrm{d}^4\tilde{x}\,
\delta\left(\tilde{x}_{\lambda}\,\tilde{x}^{\lambda}-\tau^2\right)\theta\left(\tilde{x}^0\right)\tilde{\boldsymbol{x}}\cdot
\tilde{\boldsymbol{Q}}\,\mathrm{e}^{i\tilde{x}^{0}
\tilde{P}_{0}}\nonumber\\&=&-\frac{1}{\left(2\pi\right)^3}\int_{\mathbb{R}^3}\mathrm{d}^3\tilde{x}\,\frac{\tilde{\boldsymbol{x}}\cdot
\tilde{\boldsymbol{Q}}}{\sqrt{\tilde{\boldsymbol{x}}^2+\tau^2}}\,
\mathrm{e}^{i\sqrt{\tilde{\boldsymbol{x}}^2+\tau^2} M}=0,
\end{eqnarray} since the integrand is odd in
$\tilde{\boldsymbol{x}}.$ $\quad\bullet$
\section {$W^{\mu}$ Distribution}
\label{app:Wmu}
 We define the
Lorentz vector $W^{\mu}$ as
\begin{eqnarray}
\label{eq:Wmu3}
W^{\mu}\left(P,Q\right):=\frac{P^{\mu}}{P_{\lambda}\,P^{\lambda}}W\left(P,Q\right)+\frac{Q^{\mu}}{Q_{\lambda}\,Q^{\lambda}}W\left(Q,Q\right)
.\end{eqnarray} We want to show that $W^{\mu}\left(P,Q\right)$ is identical with
\begin{eqnarray}\label{eq:irepWmu}
W_{\tau}^{\mu}\left(Q\right)=
\frac{2}{\left(2\pi\right)^3}\int_{\mathbb{R}^4}\mathrm{d}^4
x\,\delta\left(x_{\lambda}\,x^{\lambda}-\tau^2\right)\theta\left(x^0\right)x^{\mu}
\mathrm{e}^{ix_{\sigma}\, Q^{\sigma}}\end{eqnarray} This can be
proved as follows:
\\
$\circ\quad$ We have 8 variables, $P^{\mu}$ and $Q^{\mu}$ and the
constraint $P_{\lambda}\,Q^{\lambda}=0$, therefore 7 independent
variables. If $P^{\mu}$ is timelike, this constraint is equivalent to
$Q^{\mu}$ spacelike. For timelike $P^{\mu}$
we have $P_{\lambda}\,P^{\lambda}=M^2$.
 In the following calculation we take $Q^{\mu}$ and $\boldsymbol{P}$ as independent and $M=\sqrt{\left(P^0\right)^2-\boldsymbol{P}^2}$.
Differentiation of the Lorentz-invariant $W$
distribution~(\ref{eq:irepW}) with respect to the timelike
variable $P^{\mu}$ gives
\begin{eqnarray}
\frac{\partial}{\partial
P^ {\mu}}W\left(P,Q\right)&=&\frac{\partial}{\partial
P^{\mu}}\frac{2}{\left(2\pi\right)^3}\int_{\mathbb{R}^4}\mathrm{d}^4
x\,\delta\left(x_{\lambda}\,x^{\lambda}-\tau^2\right)\theta\left(x^0\right)x_{\sigma}\,
P^{\sigma} \,\mathrm{e}^{ix_{\nu}\,Q^{\nu}}\nonumber\\&=& \frac{\partial}{\partial
P^{\mu}}\frac{2}{\left(2\pi\right)^3}\int_{\mathbb{R}^4}\mathrm{d}^4
x\,\delta\left(x_{\lambda}\,x^{\lambda}-\tau^2\right)\theta\left(x^0\right)\left[\Lambda\left(v=P/M\right)\tilde{x}\right]_{\sigma}\, P^{\sigma}\,\mathrm{e}^{ix_{\nu}\,Q^{\nu}}\nonumber\\&=&\frac{2}{\left(2\pi\right)^3}\int_{\mathbb{R}^4}\mathrm{d}^4
x\,\delta\left(x_{\lambda}\,x^{\lambda}-\tau^2\right)\theta\left(x^0\right)x_{\mu}\,
\mathrm{e}^{ix_{\nu}\, Q^{\nu}}\nonumber\\&&+
\frac{2}{\left(2\pi\right)^3}\int_{\mathbb{R}^4}\mathrm{d}^4
x\,\delta\left(x_{\lambda}\,x^{\lambda}-\tau^2\right)\theta\left(x^0\right)\left[\frac{\partial}{\partial
P^{\mu}}\Lambda\left(\boldsymbol{v}\right)\tilde{x}\right]_{\sigma}\,P^{\sigma}\, \mathrm{e}^{ix_{\nu}\,Q^{\nu}},\nonumber\\
\label{eq:dp0W}
\quad\qquad
\end{eqnarray}
with
\begin{eqnarray}
\tilde{x}_{\mu}=\left[\Lambda\left(-\boldsymbol{v}\right)x\right]_{\mu}=
\left(%
\begin{array}{cc}
  \frac{P^0}{\sqrt{P_{\lambda}\,
P^{\lambda}}}&-\frac{\boldsymbol{P}^T}{\sqrt{P_{\lambda}\,
P^{\lambda}}}\\
  -\frac{\boldsymbol{P}}{\sqrt{P_{\lambda}\,
P^{\lambda}}}&\mathbf{1}+\frac{P^0/\sqrt{P_{\lambda}\,
P^{\lambda}}-1}{\boldsymbol{P}^2}\boldsymbol
{P} \boldsymbol{P}^T\\
\end{array}
\right)_{\mu}^{~\nu}\,x_{\nu}
.\end{eqnarray}
On using $ \frac{\partial}{\partial
P^{\mu}}\frac{1}{M}=-\frac{P_{\mu}}{M^3}$ a short calculation yields
\begin{eqnarray}\left[\frac{\partial}{\partial
P^{\mu}}\Lambda\left(\boldsymbol{v}\right)\tilde{x}\right]_{\sigma}\,P^{\sigma}&=&\left(%
\begin{array}{c}
  -\frac{\boldsymbol{P}\cdot\boldsymbol{\tilde{x}}}{\sqrt{P_{\lambda}\,
P^{\lambda}}}\\
  \boldsymbol{\tilde{x}}+\frac{P^0/\sqrt{P_{\lambda}\,
P^{\lambda}}-1}{\boldsymbol{P}^2}\boldsymbol
{P} \boldsymbol{P}\cdot\boldsymbol{\tilde{x}}\\
\end{array}
\right)_{\mu}%
\nonumber\\&=&\Lambda\left(-\boldsymbol{v}\right)_{\mu}^{~\nu}\left(%
\begin{array}{c}
  0\\
  \boldsymbol{\tilde{x}}\\
\end{array}
\right)_{\nu}.
\end{eqnarray}
The second integral in~(\ref{eq:dp0W}) then becomes
\begin{eqnarray}&&\Lambda\left(-\boldsymbol{v}\right)_{\mu}^{~\nu}
\frac{2}{\left(2\pi\right)^3}\int_{\mathbb{R}^4}\mathrm{d}^4
\tilde{x}\,\delta\left(\tilde{x}_{\lambda}\,\tilde{x}^{\lambda}-\tau^2\right)\theta\left(\tilde{x}^0\right)\left(%
\begin{array}{c}
  0\\
  \boldsymbol{\tilde{x}}\\
\end{array}
\right)_{\nu}\mathrm{e}^{-i\boldsymbol{\tilde{x}}\cdot
\tilde{\boldsymbol{Q}}}.
\end{eqnarray}
To evaluate this integral we perform the following spatial rotation:\begin{eqnarray}\Lambda\left(\boldsymbol{\varphi}\right)_{\mu}^{~\nu}
\left(
\begin{array}{c}
  0\\
  \tilde{\boldsymbol{Q}}\\
\end{array}
\right)_{\nu}=\left(
\begin{array}{c}
  0
  \\0\\0\\
\tilde{Q}'\\
\end{array}
\right)_{\mu},\quad
\Lambda\left(\boldsymbol{\varphi}\right)_{\mu}^{~\nu} \left(
\begin{array}{c}
  0\\
  \boldsymbol{\tilde{x}}\\
\end{array}
\right)_{\nu}=\left(
\begin{array}{c}
  0\\
  \boldsymbol{\tilde{x}}'\\
\end{array}
\right)_{\mu}.
\end{eqnarray}
Then the integral becomes
\begin{eqnarray}&&\Lambda\left(-\boldsymbol{v}\right)_{\mu}^{~\nu}\Lambda\left(-\boldsymbol{\varphi}\right)_{\nu}^{~\sigma}
\frac{2}{\left(2\pi\right)^3}\int_{\mathbb{R}^4}\mathrm{d}^4
\tilde{x}'\,\delta\left(\tilde{x}'_{\lambda}\,\tilde{x}'^{\lambda}-\tau^2\right)\theta\left(\tilde{x}'^0\right)
\left(%
\begin{array}{c}
  0\\
  \boldsymbol{\tilde{x}}'\\
\end{array}
\right)_{\sigma}\mathrm{e}^{-i\tilde{x}'_3
\tilde{Q}'}\nonumber\\&&=
\Lambda\left(-\boldsymbol{v}\right)_{\mu}^{~\nu}\Lambda\left(-\boldsymbol{\varphi}\right)_{\nu}^{~\sigma}\left(\begin{array}{c}
  0\\
0\\0\\
  1\\
\end{array}\right)_{\sigma}
\frac{2}{\left(2\pi\right)^3}\int_{\mathbb{R}^4}\mathrm{d}^4
\tilde{x}'\,\delta\left(\tilde{x}'_{\lambda}\,\tilde{x}'^{\lambda}-\tau^2\right)\theta\left(\tilde{x}'^0\right)
\tilde{x}_3'\,\mathrm{e}^{-i\tilde{x}'_3 \tilde{Q}'}\nonumber\\&&=
\Lambda\left(-\boldsymbol{v}\right)_{\mu}^{~\nu}\frac{1}{\sqrt{\tilde{\boldsymbol{Q}}^2}}\left(\begin{array}{c}
  0\\
\tilde{\boldsymbol{Q}}
\end{array}
\right)_{\nu}
\frac{2}{\left(2\pi\right)^3}\int_{\mathbb{R}^4}\mathrm{d}^4
\tilde{x}\,\delta\left(\tilde{x}_{\lambda}\,\tilde{x}^{\lambda}-\tau^2\right)\theta\left(\tilde{x}^0\right)\frac{\tilde{\boldsymbol{Q}}\cdot
\boldsymbol{\tilde{x}}}{\sqrt{\tilde{\boldsymbol{Q}}^2}}
\mathrm{e}^{-i\tilde{\boldsymbol{Q}}\cdot
\boldsymbol{\tilde{x}}}\nonumber\\&&=
\frac{Q_{\mu}}{\sqrt{Q_{\sigma}\,
Q^{\sigma}}}\frac{2}{\left(2\pi\right)^3}\int_{\mathbb{R}^4}\mathrm{d}^4
x\,\delta\left(x_{\lambda}\,x^{\lambda}-\tau^2\right)\theta\left(x^0\right)\frac{-x_{\rho}\,Q^{\rho}}{\sqrt{Q_{\sigma}\,
Q^{\sigma}}} \,\mathrm{e}^{ ix_{\nu}\, Q^{\nu}}\nonumber\\&&=
-\frac{Q_{\mu}}{Q_{\sigma}\,Q^{\sigma}}W\left(Q,Q\right),
\end{eqnarray}
where we have used in the second step that
\begin{eqnarray}\int_{\mathbb{R}^3}\frac{\mathrm{d}^3
x}{\sqrt{\boldsymbol{x}^2+\tau^2}}\,\boldsymbol{x}=\boldsymbol{0}.
\end{eqnarray}
\\
On the other hand, differentiation of~(\ref{eq:Wdistr}) gives
 \begin{eqnarray}\frac{\partial}{\partial
P^{\mu}}W\left(P,Q\right)=\frac{\partial}{\partial
P^{\mu}}P^0\,\delta^3\left(\boldsymbol{Q}\right)=\frac{\partial}{\partial
P^{\mu}}M\,\delta^3\left(\tilde{\boldsymbol{Q}}\right).
\end{eqnarray}
We have 8 variables, $M,\tilde{\boldsymbol{Q}}$ and $v^{\mu}$ with
the constraint~(\ref{eq:PQ=0})
 \begin{eqnarray}P_{\lambda}\,Q^{\lambda} =\Lambda\left(v_0,-\boldsymbol{v}\right)^{~\sigma}_{\lambda}\left(\begin{array}{c}
  M\\
\boldsymbol{0}
\end{array}\right)_{\sigma}\,\Lambda\left(v_0,-\boldsymbol{v}\right)_{~\nu}^{\lambda}\left(\begin{array}{c}
  0\\
\tilde{\boldsymbol{Q}}
\end{array}\right)^{\nu}=0.
\end{eqnarray}
If we take $\boldsymbol{v},M,\tilde{\boldsymbol{Q}}$ as
independent and $v_0=\sqrt{1+\boldsymbol{v}^2}$, then we have
\begin{eqnarray}
\frac{\partial}{\partial
P^{0}}M\delta^3\left(\boldsymbol{q}\right)=\left(\frac{\partial M}{\partial
P^{0}}\frac{\partial}{\partial
M}+\frac{\partial \tilde{Q}^i}{\partial
P^{0}}\frac{\partial}{\partial
\tilde{Q}^i}+\frac{\partial v^i}{\partial
P^{0}}\frac{\partial}{\partial
v^i}\right)M\delta^3\left(\tilde{\boldsymbol{Q}}\right),\quad i=1,\ldots3.\nonumber\\
\end{eqnarray}
Since $P^0=M v_0= M \sqrt{1+\boldsymbol{v}^2}$ we have
$\frac{\partial q^i}{\partial P^{0}}=0$ and therefore the second
term vanishes. Since the $\boldsymbol{v},M,\tilde{\boldsymbol{Q}}$
are independent we have $\frac{\partial} {\partial
v^i}M\delta^3\left(\tilde{\boldsymbol{Q}}\right)=0$ and the third
term vanishes also. Only the first term survives giving
\begin{eqnarray}\label{eq:wmu1}\frac{
\partial M}{\partial
P^{0}}\,\delta^3\left(\tilde{\boldsymbol{Q}}\right)=\frac{
\partial \sqrt{\left(P^0\right)^2-\boldsymbol{P}^2}}{\partial
P^{0}}\,\delta^3\left(\tilde{\boldsymbol{Q}}\right)=\frac{P^0}{M}\,\delta^3\left(\tilde{\boldsymbol{Q}}\right)=\frac{\left(P^0\right)^2}{M^2}\,\delta^3\left(\boldsymbol{Q}\right).
\end{eqnarray}
For the spatial components we have
\begin{eqnarray}
\frac{\partial}{\partial
P^{i}}M\delta^3\left(\tilde{\boldsymbol{Q}}\right)=\left(\frac{\partial M}{\partial
P^{i}}\frac{\partial}{\partial
M}+\frac{\partial \tilde{Q}^j}{\partial
P^{i}}\frac{\partial}{\partial
\tilde{Q}^j}+\frac{\partial v^j}{\partial
P^{i}}\frac{\partial}{\partial
v^j}\right)M\delta^3\left(\tilde{\boldsymbol{Q}}\right).
\end{eqnarray}
Since $\boldsymbol{P}=M\boldsymbol{v}$ again only the first term
survives giving
\begin{eqnarray}\frac{
\partial M}{\partial
P^{i}}\,\delta^3\left(\tilde{\boldsymbol{Q}}\right)=\frac{
\partial \sqrt{\left(P^0\right)^2-\boldsymbol{P}^2}}{\partial
P^{i}}\,\delta^3\left(\tilde{\boldsymbol{Q}}\right)=-\frac{P^i}{M}\,\delta^3\left(\tilde{\boldsymbol{Q}}\right)=\frac{P_i}{M^2}P^0\,\delta^3\left(\boldsymbol{Q}\right).\nonumber\\
\label{eq:wmu2}
\end{eqnarray}
Thus, we can write~(\ref{eq:wmu1}) and~(\ref{eq:wmu2}) as
components of a four-vector
\begin{eqnarray}&&\frac{\partial}{\partial
P^{\mu}}P_0\,\delta^3\left(\boldsymbol{Q}\right)=\frac{P_{\mu}}{M^2}P_0\,\delta^3\left(\boldsymbol{Q}\right)=\frac{P_{\mu}}{P_{\lambda}\,P^{\lambda}}W\left(P,Q\right)
.\end{eqnarray} Therefore, we have shown that
$W^{\mu}\left(P,Q\right)\equiv W_{\tau}^{\mu}\left(Q\right)$ and
the proof is completed.~~~$\bullet$\\
This can be seen also as follows: \\
$\circ\quad$ We introduce 2 additional spacelike four-vectors
$R$ and $S$, such that they form together with $P$ and $Q$ an
orthogonal basis of Minkowski space. Representing $x$ in terms of
this basis, we can write~(\ref{eq:irepWmu}) as
\begin{eqnarray}&&
\frac{2}{\left(2\pi\right)^3}\int_{\mathbb{R}^4}\mathrm{d}^4
x\,\delta\left(x_{\lambda}\,x^{\lambda}-\tau^2\right)\theta\left(x^0\right)x^{\mu}
\mathrm{e}^{ix_{\nu}\,
Q^{\nu}}\nonumber\\&&=\frac{P^{\mu}}{P_{\lambda}\,P^{\lambda}}W\left(P,Q\right)+\frac{Q^{\mu}}{Q_{\lambda}\,Q^{\lambda}}W\left(Q,Q\right)+\frac{R^{\mu}}{R_{\lambda}\,R^{\lambda}}W\left(R,Q\right)
+\frac{S^{\mu}}{S_{\lambda}\,S^{\lambda}}W\left(S,Q\right).\nonumber\\\label{eq:expx}\end{eqnarray}
For the calculation of $W\left(R,Q\right)$ and $W\left(S,Q\right)$
we can, using Lorentz invariance, perform a spatial rotation of
our coordinate system for the integration variables, so that the
new spatial coordinate axes for $x'^1,$ $x'^2$ and $x'^3$ coincide
with $Q$, $R$ and $S$, respectively. For this choice of
coordinates the integrands of $W\left(R,Q\right)$ and
$W\left(S,Q\right)$ are odd in $x'^2$ and $x'^3$, respectively.
Thus we can conclude that $W\left(R,Q\right)=W\left(S,Q\right)=0$,
which proves~(\ref{eq:Wmu3}).$\quad\bullet$\\
\\
If we interchange P and Q in $W^{\mu}\left(P,Q\right)$, we
immediately obtain
\begin{eqnarray}
W^{\mu}\left(Q,P\right)
&=&\frac{P^{\mu}}{P_{\lambda}\,P^{\lambda}}W\left(P,P\right)+\frac{Q^{\mu}}{Q_{\lambda}\,Q^{\lambda}}W\left(Q,P\right)
\nonumber\\
&=&\frac{P^{\mu}}{P_{\lambda}\,P^{\lambda}}W\left(P,P\right),
\end{eqnarray}
with the corresponding integral representation
\begin{eqnarray}
W_{\tau}^{\mu}\left(P\right) =
\frac{2}{\left(2\pi\right)^3}\int_{\mathbb{R}^4}\mathrm{d}^4
x\,\delta\left(x_{\lambda}\,x^{\lambda}-\tau^2\right)\theta\left(x^0\right)x^{\mu}
\mathrm{e}^{ix_{\nu}\, P^{\nu}}.\end{eqnarray} $\circ\quad$ As
in~(\ref{eq:expx}), we expand $x$ in terms of $P,\,Q,\,R$ and $S$.
Then we perform a boost $\Lambda\left(v\right)$~(\ref{eq:Lambda})
in order to find
$W\left(Q,P\right)=W\left(R,P\right)=W\left(S,P\right)=0.\quad\bullet$\\
In addition, we note that
\begin{eqnarray}\label{eq:WQQ}
W\left(P,-Q\right)=W\left(P,Q\right),\quad W\left(Q,-Q\right)=-W\left(Q,Q\right).
\end{eqnarray}
$\circ\quad$ The first relation follows immediately
from~(\ref{eq:Wdistr}). The second can be shown as follows:
\begin{eqnarray}
W\left(Q,-Q\right)&=&W\left(\tilde{Q},-\tilde{Q}\right)=
-\frac{1}{\left(2\pi\right)^3}\int_{\mathbb{R}^3}\frac{\mathrm{d}^3\tilde{x}}{\tilde{x}^0}
\tilde{\boldsymbol{Q}}\cdot \boldsymbol{\tilde{x}}\, \mathrm{e}^{i\tilde{\boldsymbol{Q}}\cdot \boldsymbol{\tilde{x}}}\nonumber\\&=&\frac{1}{\left(2\pi\right)^3}\int_{\mathbb{R}^3}\frac{\mathrm{d}^3\tilde{x}}{\tilde{x}^0}
\tilde{\boldsymbol{Q}}\cdot \boldsymbol{\tilde{x}}\, \mathrm{e}^{-i\tilde{\boldsymbol{Q}}\cdot \boldsymbol{\tilde{x}}}=-W\left(\tilde{Q},\tilde{Q}\right)\nonumber\\&=&-W\left(Q,Q\right).
\quad\bullet
\end{eqnarray}
\chapter{Complex Klein-Gordon Fields}
\label{app:cs0f}
\section{Pauli-Jordan Function}
\label{app:PJF} For spacelike $\left(x-y\right)$,
$\Delta\left(x-y\right)$ and its second derivative with respect to some timelike directions
vanish.
\\
$\circ\quad$ In a similar way as before we can write the
spacelike vector $Y:=x-y$ as a boost transform of a vector which
has spatial components only, i.e.
 \begin{eqnarray}
Y=\Lambda\left(v\right)\tilde{Y}=\Lambda\left(v\right)\left(%
\begin{array}{c}
  0 \\
  \tilde{\boldsymbol{Y}} \\
\end{array}
\right)%
\quad\text{with}\quad
Y_{\lambda}\,Y^{\lambda}=-\tilde{\boldsymbol{Y}}^2\\
\quad\text{and}\quad v_{\lambda}\,Y^{\lambda}=0.
 \end{eqnarray}
Inverting this equation and making use of Lorentz invariance
yields
\begin{eqnarray}\Delta\left(Y\right)&=&\frac{1}{i}\int_{\mathbb{R}^4}\frac{\mathrm{d}^4\tilde{p}}{\left(2\pi\right)^3}
\,\delta\left(\tilde{p}_{\lambda}\,\tilde{p}^{\lambda}-m^2\right)\theta\left(\tilde{p}^0\right)\left(\mathrm{e}^{i\tilde{\boldsymbol{p}}\cdot\tilde{\boldsymbol{Y}}}-\mathrm{e}^{-i\tilde{\boldsymbol{p}}\cdot\tilde{\boldsymbol{Y}}}\right)\nonumber\\
&=&\frac{1}{\left(2\pi\right)^3}\int_{\mathbb{R}^3}\mathrm{d}^3\tilde{p}\,\frac{\mathrm{sin}\left(\tilde{\boldsymbol{p}}\cdot\tilde{\boldsymbol{Y}}\right)}{\sqrt{\tilde{\boldsymbol{p}}^2+m^2}}=0
,\quad\forall\, Y_{\lambda}\,Y^{\lambda}<0,\end{eqnarray}
since the integrand is odd in $\tilde{\boldsymbol{p}}$.\\
For the second derivative of $\Delta$ we have in addition the
scalar products $n_{\lambda}\left(x\right)\,p^{\lambda}\,
n_{\sigma}\left(y\right)\,p^{\sigma}$ in the integrand.
If we choose $x,y$ to be on the hyperboloid
$x^2=y^2=\tau^2=\left(\xi^{0}\left(x\right)\right)^2=\left(\xi^{0}\left(y\right)\right)^2$,
we have then from~(\ref{eq:nv})
$n^{\mu}\left(x\right)=\frac{x^{\mu}}{\tau}$ and $
n^{\mu}\left(y\right)=\frac{y^{\mu}}{\tau}$. Then a timelike
$X$ is given by $X=x+y$ and with
$v^{\mu}=\frac{X^{\mu}}{\sqrt{X_{\lambda}\,X^{\lambda}}}$,
$X_{\lambda}Y^{\lambda}=0$ holds. In these new variables $X,Y$ we have
$n\left(x\right)=\frac{X+Y}{2\tau}$ and
$n\left(y\right)=\frac{X-Y}{2\tau}$ and thus
\begin{eqnarray}
n_{\lambda}\left(x\right)\,p^{\lambda}\,
n_{\sigma}\left(y\right)\,p^{\sigma}=\frac{1}{4\tau^2}\left[\left(X_{\lambda}\,p^{\lambda}\right)^2-\left(Y_{\lambda}\,p^{\lambda}\right)^2\right].
\end{eqnarray} In the boosted frame this has the form
\begin{eqnarray}
\left(X_{\lambda}\,p^{\lambda}\right)^2-\left(Y_{\lambda}\,p^{\lambda}\right)^2=\left(\tilde{X}_{0}\,\tilde{p}^{0}\right)^2-\left(
\tilde{\boldsymbol{Y}}\cdot\tilde{\boldsymbol{p}}\right)^2
.\end{eqnarray} Proceeding in a similar way as above we obtain
 \begin{eqnarray}&&n^{\lambda}\left(x\right)n^{\sigma}\left(y\right)\frac{\partial}{\partial
x^{\lambda}}\frac{\partial}{\partial
y^{\sigma}}\Delta\left(x-y\right)\nonumber\\&&=\frac{1}{4\tau^2\left(2\pi\right)^3}\int_{\mathbb{R}^3}\mathrm{d}^3\tilde{p}\,\frac{\left[\left(\tilde{X}_{0}\,\sqrt{\tilde{\boldsymbol{p}}^2+m^2}\right)^2-\left(
\tilde{\boldsymbol{Y}}\cdot\tilde{\boldsymbol{p}}\right)^2\right]
\mathrm{sin}\left(\tilde{\boldsymbol{p}}\cdot\tilde{\boldsymbol{Y}}\right)}{\sqrt{\tilde{\boldsymbol{p}}^2+m^2}}=0,\nonumber\\&&\forall\,
Y_{\lambda}Y^{\lambda}<0,\end{eqnarray} since the integrand is odd
in $\tilde{\boldsymbol{p}}$.
$\quad\bullet$\\
 This result can be generalized to arbitrary spacelike
hypersurfaces.
\section{Covariant Canonical Commutation Relations}
\label{app:cccr} We make use of the hyperbolic coordinates
$\left(\tau,\xi\equiv\mathrm{cosh}\beta,\vartheta,\varphi\right)$
defined by
the coordinate transformation~(\ref{eq:xhc}) with the hypersurface element~(\ref{eq:hse}).\\
$\circ\quad$ The Lagrangian operator for free complex scalar
fields in hyperbolic coordinates reads
\begin{eqnarray}
\hat{L}_{\mathrm{KG}}&=&
\int_{\Sigma_{\tau}}\tau^3\sqrt{\xi^2-1}\,\mathrm{d}\xi\,
\mathrm{d}\mathrm{cos}\vartheta  \,\mathrm{d}\varphi\left\lbrace
\frac{\partial\hat{\phi}^{\dag}}{\partial\tau}
\frac{\partial\hat{\phi}}{\partial\tau}-\frac{\xi^2-1}{\tau^2}\frac{\partial\hat{\phi}^{\dag}}{\partial\xi}
\frac{\partial\hat{\phi}}{\partial\xi}-\frac{1}{\tau^2\left(\xi^2-1\right)}\frac{\partial\hat{\phi}^{\dag}}{\partial\vartheta
}
\frac{\partial\hat{\phi}}{\partial\vartheta }\right.\nonumber\\
&&-\left.\frac{1}{\tau^2\left(\xi^2-1\right)\mathrm{sin}^2\vartheta
}\frac{\partial\hat{\phi}^{\dag}}{\partial\varphi}
\frac{\partial\hat{\phi}}{\partial\varphi}-
m^2\hat{\phi}^{\dag}\hat{\phi} \right\rbrace =
\int_{\Sigma_{\tau}} \mathrm{d}\Sigma_{\tau}
\hat{\mathcal{L}}_{\mathrm{KG}}\left(\tau,\xi,\vartheta
,\varphi\right),
\end{eqnarray}where $\hat{\mathcal{L}}_{\mathrm{KG}}\left(\tau,\xi,\vartheta ,\varphi\right)$ denotes the Lagrangian density in hyperbolic coordinates.
The equal-$\tau$ canonical commutators for the field operators
$\hat{\phi},\hat{\phi}^{\dag}$ are given by~\cite{Gromes:1974yu}
\begin{eqnarray}
&&\left[\hat{\phi}\left(y\right),
\frac{\partial}{\partial\tau}\hat{\phi}^{\dag}\left(x\right)\right]_{\tau\left(x\right)=\tau\left(y\right)=\tau}\nonumber\\&&=i\frac{1}{\tau^3\sqrt{\left(\xi\left(x\right)\right)^2-1}}\,\delta\left(\xi\left(x\right)-\xi\left(y\right)\right)
\delta\left(\mathrm{cos}\,\vartheta
\left(x\right)-\mathrm{cos}\,\vartheta \left(y\right)\right)
\delta\left(\varphi\left(x\right)-\varphi\left(y\right)\right).
\quad\qquad
\end{eqnarray}
This is in agreement with~(\ref{eq:gccm}), if we transform
\begin{eqnarray}\label{eq:derivt2}
\frac{\partial}{\partial x^{\mu}}=\frac{\partial\xi^{0}}{\partial
x^{\mu}}\frac{\partial}{\partial
\xi^{0}}=n_{\mu}\left(x\right)\frac{\partial}{\partial
\tau},\end{eqnarray} with $\tau\equiv\xi^0$ and
$n_{\lambda}\left(x\right)\,n^{\lambda}\left(x\right)=1.$\footnote{The notation for hyperbolic coordinates $\xi^{\mu}=(\tau,\xi,\vartheta ,\varphi)^{\mu}$ should be clear from the context.}\\
From~(\ref{eq:dSigma}) and
\begin{eqnarray}\label{eq:derivt1}
\frac{\partial}{\partial
\tau}=n^{\mu}\left(x\right)\frac{\partial}{\partial
x^{\mu}}=\frac{x^{\mu}}{\sqrt{x_{\lambda}\,x^{\lambda}}}\frac{\partial}{\partial
x^{\mu}}\end{eqnarray} we see that the covariant commutation
relation~(\ref{eq:cccrkg}) is fulfilled,
\begin{eqnarray}&&
\int_{\Sigma}\mathrm{d}\Sigma^{\mu}\left(x\right)\left[\hat{\phi}\left(y\right),\frac{\partial}{\partial x^{\mu}}\hat{\phi}^{\dag}\left(x\right)\right]_{x,y\in\Sigma}\nonumber\\&&=
\int_{\Sigma_{\tau}}\tau^3\sqrt{\left(\xi\left(x\right)\right)^2-1}\,\mathrm{d}\xi\left(x\right) \,\mathrm{d}\mathrm{cos}\,\vartheta \left(x\right) \,\mathrm{d}\varphi\left(x\right)\left[\hat{\phi}\left(y\right),\frac{\partial}{\partial\tau}\hat{\phi}^{\dag}\left(x\right)\right]_{x^2=y^2=\tau^2}=i.\quad\bullet\nonumber\\
\end{eqnarray}
From \begin{eqnarray}
x^{\mu}\left[\hat{\phi}\left(y\right),\frac{\partial}{\partial
x^{\mu}}\hat{\phi}^{\dag}\left(x\right)\right]_{x^2=y^2=\tau^2}=\tau\left[\hat{\phi}\left(y\right),
\frac{\partial}{\partial\tau}\hat{\phi}^{\dag}\left(x\right)\right]_{\tau\left(x\right)=\tau\left(y\right)=\tau}
\end{eqnarray} we obtain the useful relation
 \begin{eqnarray}\label{eq:userel}&&x^0\delta^3\left(\boldsymbol{x}-\boldsymbol{y}\right)\nonumber\\&&=\frac{1}{\tau^2\sqrt{\left(\xi\left(x\right)\right)^2-1}}\,\delta\left(\xi\left(x\right)-\xi\left(y\right)\right) \delta\left(\mathrm{cos}\,\vartheta \left(x\right)-\mathrm{cos}\,\vartheta \left(y\right)\right)
 \delta\left(\varphi\left(x\right)-\varphi\left(y\right)\right).\nonumber\\
\end{eqnarray}
\section{Commutation Relations in Momentum Space}
\label{app:ccrms}
We want to show that the commutation relations in momentum space,~(\ref{eq:ccrms}) and~(\ref{eq:ccrms2}), follow from~(\ref{eq:cccrkg2}) and~(\ref{eq:liccr}). We will do this by quantizing on the hyperboloid using~(\ref{eq:liccr}).\\
$\circ\quad$ For the first commutator~(\ref{eq:ccrms}) we have,
using~(\ref{eq:spfa}) and~(\ref{eq:pfipsf}),
\begin{eqnarray}
&&\left[\hat{a}\left(\boldsymbol{p}\right),\hat{a}^{\dag}\left(\boldsymbol{q}\right)\right]\nonumber\\&&=\frac{4}{\left(2\pi\right)^3}
\int_{\mathbb{R}^4}\mathrm{d}^4x\,\delta\left(x_{\lambda}\,x^{\lambda}-\tau^2\right)\theta\left(x^0\right)x^{\mu}\mathrm{e}^{ip_{\rho}\,x^{\rho}}\int_{\mathbb{R}^4}\mathrm{d}^4y\,\delta\left(y_{\lambda}\,y^{\lambda}-\tau^2\right)\theta\left(y^0\right)y^{\nu}\mathrm{e}^{-iq_{\sigma}\,y^{\sigma}}
\nonumber\\&&\quad\times\left\lbrace
\underbrace{\left[\frac{\partial}{\partial
x^{\mu}}\hat{\phi}\left(x\right),\frac{\partial}{\partial
y^{\nu}}\hat{\phi}^{\dag}\left(y\right)\right]_{x^2=y^2}}_{=0,~(\ref{eq:cccrkg2})}-ip_{\mu}\left[\hat{\phi}\left(x\right),\frac{\partial}{\partial
y^{\nu}}\hat{\phi}^{\dag}\left(y\right)\right]_{x^2=y^2}\right.\nonumber\\
&&\qquad+\left.iq_{\nu}\left[\frac{\partial}{\partial
x^{\mu}}\hat{\phi}\left(x\right),\hat{\phi}^{\dag}\left(y\right)\right]_{x^2=y^2}+p_{\mu}q_{\nu}\underbrace{\left[\hat{\phi}\left(x\right),\hat{\phi}^{\dag}\left(y\right)\right]_{x^2=y^2}}_{=0,~(\ref{eq:cccrkg2})}
\right\rbrace \nonumber\\&&= \frac{4i}{\left(2\pi\right)^3}
\int_{\mathbb{R}^4}\mathrm{d}^4x\,\delta\left(x_{\lambda}\,x^{\lambda}-\tau^2\right)\theta\left(x^0\right)\mathrm{e}^{ip_{\rho}\,x^{\rho}}\int_{\mathbb{R}^4}\mathrm{d}^4y\,\delta\left(y_{\lambda}\,y^{\lambda}-\tau^2\right)\theta\left(y^0\right)\mathrm{e}^{-iq_{\sigma}\,y^{\sigma}}\nonumber\\
&&\quad\times \left\lbrace q_{\sigma}\,y^{\sigma}\,
\underbrace{x^{\mu}\left[\frac{\partial}{\partial
x^{\mu}}\hat{\phi}\left(x\right),\hat{\phi}^{\dag}\left(y\right)\right]_{x^2=y^2}}_{=-ix^0\delta^3\left(\boldsymbol{x}-\boldsymbol{y}\right),~(\ref{eq:liccr})}-p_{\rho}\,x^{\rho}
\,y^{\nu}\left[\hat{\phi}\left(x\right),\frac{\partial}{\partial
y^{\nu}}\hat{\phi}^{\dag}\left(y\right)\right]_{x^2=y^2}
 \right\rbrace \nonumber\end{eqnarray}
 \begin{eqnarray}&&=
\frac{4}{\left(2\pi\right)^3}
\int_{\mathbb{R}^4}\mathrm{d}^4x\,\delta\left(x_{\lambda}\,x^{\lambda}-\tau^2\right)\theta\left(x^0\right)\mathrm{e}^{ip_{\rho}\,x^{\rho}}\int_{\mathbb{R}^4}\mathrm{d}^4y\,\delta\left(y_{\lambda}\,y^{\lambda}-\tau^2\right)\theta\left(y^0\right)\mathrm{e}^{-iq_{\sigma}\,y^{\sigma}}\nonumber\\&&\quad\times
x^0\delta^3\left(\boldsymbol{x}-\boldsymbol{y}\right) \left(
q_{\sigma}\,y^{\sigma}+p_{\rho}\,x^{\rho}
 \right)\nonumber\\
&&= 2p^0\delta^3\left(\boldsymbol{p}-\boldsymbol{q}\right).
\end{eqnarray}
$\left[\hat{b}\left(\boldsymbol{p}\right),\hat{b}^{\dag}\left(\boldsymbol{q}\right)\right]=
2p^0\delta^3\left(\boldsymbol{p}-\boldsymbol{q}\right)$ is calculated similarly.\\
For the other commutator~(\ref{eq:ccrms2}) we have
\begin{eqnarray}
&&\left[\hat{a}\left(\boldsymbol{p}\right),\hat{b}\left(\boldsymbol{q}\right)\right]\nonumber\\&&=-\frac{4}{\left(2\pi\right)^3}
\int_{\mathbb{R}^4}\mathrm{d}^4x\,\delta\left(x_{\lambda}\,x^{\lambda}-\tau^2\right)\theta\left(x^0\right)x^{\mu}\mathrm{e}^{ip_{\rho}\,x^{\rho}}\int_{\mathbb{R}^4}\mathrm{d}^4y\,\delta\left(y_{\lambda}\,y^{\lambda}-\tau^2\right)\theta\left(y^0\right)y^{\nu}\mathrm{e}^{iq_{\sigma}\,y^{\sigma}}
\nonumber\\&&\quad\times\left\lbrace
\underbrace{\left[\frac{\partial}{\partial
x^{\mu}}\hat{\phi}\left(x\right),\frac{\partial}{\partial
y^{\nu}}\hat{\phi}^{\dag}\left(y\right)\right]_{x^2=y^2}}_{=0,~(\ref{eq:cccrkg2})}-ip_{\mu}\left[\hat{\phi}\left(x\right),\frac{\partial}{\partial
y^{\nu}}\hat{\phi}^{\dag}\left(y\right)\right]_{x^2=y^2}\right.\nonumber\\
&&\qquad-\left.iq_{\nu}\left[\frac{\partial}{\partial
x^{\mu}}\hat{\phi}\left(x\right),\hat{\phi}^{\dag}\left(y\right)\right]_{x^2=y^2}+p_{\mu}q_{\nu}\underbrace{\left[\hat{\phi}\left(x\right),\hat{\phi}^{\dag}\left(y\right)\right]_{x^2=y^2}}_{=0,~(\ref{eq:cccrkg2})}
\right\rbrace \nonumber\\&&= \frac{4i}{\left(2\pi\right)^3}
\int_{\mathbb{R}^4}\mathrm{d}^4x\,\delta\left(x_{\lambda}\,x^{\lambda}-\tau^2\right)\theta\left(x^0\right)\mathrm{e}^{ip_{\rho}\,x^{\rho}}\int_{\mathbb{R}^4}\mathrm{d}^4y\,\delta\left(y_{\lambda}\,y^{\lambda}-\tau^2\right)\theta\left(y^0\right)\mathrm{e}^{iq_{\sigma}\,y^{\sigma}}\nonumber\\
&&\quad\times \left\lbrace q_{\sigma}\,y^{\sigma}\,\underbrace{
x^{\mu}\left[\frac{\partial}{\partial
x^{\mu}}\hat{\phi}\left(x\right),\hat{\phi}^{\dag}\left(y\right)\right]_{x^2=y^2}}_{=-ix^0\delta^3\left(\boldsymbol{x}-\boldsymbol{y}\right),~(\ref{eq:liccr})}+p_{\rho}\,x^{\rho}
\,y^{\nu}\left[\hat{\phi}\left(x\right),\frac{\partial}{\partial
y^{\nu}}\hat{\phi}^{\dag}\left(y\right)\right]_{x^2=y^2}
 \right\rbrace\nonumber\\&&=
-\frac{4}{\left(2\pi\right)^3}
\int_{\mathbb{R}^4}\mathrm{d}^4x\,\delta\left(x_{\lambda}\,x^{\lambda}-\tau^2\right)\theta\left(x^0\right)\mathrm{e}^{ip_{\rho}\,x^{\rho}}\int_{\mathbb{R}^4}\mathrm{d}^4y\,\delta\left(y_{\lambda}\,y^{\lambda}-\tau^2\right)\theta\left(y^0\right)\mathrm{e}^{iq_{\sigma}\,y^{\sigma}}\nonumber\\&&\quad\times
x^0\delta^3\left(\boldsymbol{x}-\boldsymbol{y}\right)
\left(p_{\rho}\,x^{\rho}- q_{\sigma}\,y^{\sigma}
\right)\nonumber\\&&=0.
\end{eqnarray}\\
$\left[\hat{b}^{\dag}\left(\boldsymbol{p}\right),\hat{a}^{\dag}\left(\boldsymbol{q}\right)\right]
=0$ can be shown in an similar way.\\
For the last commutator in~(\ref{eq:ccrms2}) we have
\begin{eqnarray}\left[\hat{a}\left(\boldsymbol{p}\right),\hat{b}^{\dag}\left(\boldsymbol{q}\right)\right]\nonumber\end{eqnarray}
 \begin{eqnarray}&&=\frac{4}{\left(2\pi\right)^3}
\int_{\mathbb{R}^4}\mathrm{d}^4x\,\delta\left(x_{\lambda}\,x^{\lambda}-\tau^2\right)\theta\left(x^0\right)x^{\mu}\mathrm{e}^{ip_{\rho}\,x^{\rho}}\int_{\mathbb{R}^4}\mathrm{d}^4y\,\delta\left(y_{\lambda}\,y^{\lambda}-\tau^2\right)\theta\left(y^0\right)y^{\nu}\mathrm{e}^{-iq_{\sigma}\,y^{\sigma}}
\nonumber\\&&\quad\times\left\lbrace
\underbrace{\left[\frac{\partial}{\partial
x^{\mu}}\hat{\phi}\left(x\right),\frac{\partial}{\partial
y^{\nu}}\hat{\phi}\left(y\right)\right]_{x^2=y^2}}_{=0,~(\ref{eq:cccrkg2})}-ip_{\mu}\underbrace{\left[\hat{\phi}\left(x\right),\frac{\partial}{\partial
y^{\nu}}\hat{\phi}\left(y\right)\right]_{x^2=y^2}}_{=0,~(\ref{eq:cccrkg2})}\right.\nonumber\\
&&\qquad+\left.iq_{\nu}\underbrace{\left[\frac{\partial}{\partial
x^{\mu}}\hat{\phi}\left(x\right),\hat{\phi}\left(y\right)\right]_{x^2=y^2}}_{=0,~(\ref{eq:cccrkg2})}+p_{\mu}q_{\nu}\underbrace{\left[\hat{\phi}\left(x\right),\hat{\phi}\left(y\right)\right]_{x^2=y^2}}_{=0,~(\ref{eq:cccrkg2})}
\right\rbrace =0.
\end{eqnarray}
$\left[\hat{b}^{\dag}\left(\boldsymbol{p}\right),\hat{a}\left(\boldsymbol{q}\right)\right]=0$ can be shown in an similar way.$\quad\bullet$\\
\\That the harmonic-oscillator commutation
relations,~(\ref{eq:ccrms}) and~(\ref{eq:ccrms2}), also  imply the
covariant canonical commutation relations,~(\ref{eq:cccrkg})
and~(\ref{eq:cccrkg2}), can be seen as follows. Again this is
shown in point form: \\
$\circ\quad$ In point form the covariant commutation
relation~(\ref{eq:cccrkg}) reads~(cf:~(\ref{eq:pfccr}))
\begin{eqnarray}
&&\int_{\mathbb{R}^4}2\mathrm{d}^4x\,\delta\left(x_{\lambda}\,x^{\lambda}-\tau^2\right)\theta\left(x^0\right)x^{\mu}\left[\hat{\phi}\left(y\right),\partial_{\mu}\hat{\phi}^{\dag}\left(x\right)\right]_{x^2=y^2=\tau^2}\nonumber\\&&=
\frac{i}{\left(2\pi\right)^3}\int_{\mathbb{R}^3}\frac{\mathrm{d}^3p}{2p_0}
\int_{\mathbb{R}^3}\frac{\mathrm{d}^3q}{2q_0}
\int_{\mathbb{R}^4}2\mathrm{d}^4x\,\delta\left(x_{\lambda}\,x^{\lambda}-\tau^2\right)\theta\left(x^0\right)x_{\lambda}\,
p^{\lambda}\nonumber\\&&\quad\times \left\lbrace \mathrm{e}^{ip_{\lambda}\, x^{\lambda}} \mathrm{e}^{-iq_{\lambda}\, y^{\lambda}}\left[\hat{a}\left(\boldsymbol{q}\right),\hat{a}^{\dag}\left(\boldsymbol{p}\right)\right]-
\mathrm{e}^{-ip_{\lambda}\, x^{\lambda}}\mathrm{e}^{-iq_{\lambda}\,y^{\lambda}}\left[\hat{a}\left(\boldsymbol{q}\right),\hat{b}\left(\boldsymbol{p}\right)\right]\right.\nonumber\\
&&\qquad\left.+\mathrm{e}^{ip_{\lambda}\,x^{\lambda}}\mathrm{e}^{iq_{\lambda}\,
y^{\lambda}}\left[\hat{b}^{\dag}\left(\boldsymbol{q}\right),\hat{a}^{\dag}\left(\boldsymbol{p}\right)\right]-\mathrm{e}^{-ip_{\lambda}\,
x^{\lambda}}\mathrm{e}^{iq_{\lambda }\,
y^{\lambda}}\left[\hat{b}^{\dag}\left(\boldsymbol{q}\right),\hat{b}\left(\boldsymbol{p}\right)\right]\right\rbrace
\nonumber\\&&=
\frac{i}{\left(2\pi\right)^3}\int_{\mathbb{R}^3}\frac{\mathrm{d}^3p}{2p_0}
\int_{\mathbb{R}^3}\mathrm{d}^3q
\int_{\mathbb{R}^4}2\mathrm{d}^4x\,\delta\left(x_{\lambda}\,x^{\lambda}-\tau^2\right)\theta\left(x^0\right)x_{\lambda}\,
p^{\lambda}\delta^3\left(\boldsymbol{p}-\boldsymbol{q}\right)\nonumber\\&&\quad\times
\left(\mathrm{e}^{ip_{\lambda}\,x^{\lambda}}
\mathrm{e}^{-iq_{\lambda}\,y^{\lambda}}+\mathrm{e}^{-ip_{\lambda}\,x^{\lambda}}\mathrm{e}^{iq_{\lambda}\,
y^{\lambda}}\right)\nonumber\\&&=
\frac{i}{\left(2\pi\right)^3}\int_{\mathbb{R}^3}\frac{\mathrm{d}^3p}{2p_0}
\int_{\mathbb{R}^4}2\mathrm{d}^4x\,\delta\left(x_{\lambda}\,x^{\lambda}-\tau^2\right)\theta\left(x^0\right)x_{\lambda}\,
p^{\lambda}\nonumber\\&&\quad\times
\left(\mathrm{e}^{ip_{\lambda}\left( x-y\right)^{\lambda}}
+\mathrm{e}^{-ip_{\lambda}\left(
x-y\right)^{\lambda}}\right)\nonumber\\&&= \frac{2i}
{\left(2\pi\right)^3}\int_{\mathbb{R}^4}\mathrm{d}^4x\,\delta\left(x_{\lambda}\,x^{\lambda}-\tau^2\right)\theta\left(x^0\right)x_{\mu}
\int_{\mathbb{R}^4}\mathrm{d}^4p\,\delta\left(p_{\lambda}\,p^{\lambda}-m^2\right)\theta\left(p^0\right)p^{\mu}\nonumber\\&&\quad\times
\left(\mathrm{e}^{ip_{\lambda}\left( x-y\right)^{\lambda}}
+\mathrm{e}^{-ip_{\lambda}\left( x-y\right)^{\lambda}}\right)
\nonumber\\&&=
i\int_{\mathbb{R}^4}\mathrm{d}^4x\,\delta\left(x_{\lambda}\,x^{\lambda}-\tau^2\right)\theta\left(x^0\right)x_{\mu}
\left(W^{\mu}\left(X,Y\right)+W^{\mu}\left(X,-Y\right)
\right)\nonumber\\&&=
i\int_{\mathbb{R}^4}\mathrm{d}^4x\,\delta\left(x_{\lambda}\,x^{\lambda}-\tau^2\right)\theta\left(x^0\right)x_{\mu}
\left(\frac{2X^{\mu}}{X_{\lambda}\,X^
{\lambda}}W\left(X,Y\right)\right.\nonumber\\&&\quad+\left.\frac{Y^{\mu}}{Y_{\lambda}\,Y^
{\lambda}}W\left(Y,Y\right)-\frac{ Y^{\mu}}{ Y_{\lambda}\,Y^
{\lambda}}W\left(Y,Y\right) \right)\nonumber\end{eqnarray}
 \begin{eqnarray}&&=
4i\int_{\mathbb{R}^4}\mathrm{d}^4x\,\delta\left(x_{\lambda}\,x^{\lambda}-\tau^2\right)\theta\left(x^0\right)
\frac{x_{\mu}\left(x+y\right)^{\mu}}{\left(x+y\right)_{\lambda}\,\left(x+y\right)^{\lambda}}\,x^0\delta^3\left(\boldsymbol{x}-\boldsymbol{y}\right)
=i,\nonumber\\
\end{eqnarray}
where we have inserted the harmonic-oscillator commutation relations~(\ref{eq:ccrms}) and~(\ref{eq:ccrms2}).
In the last step we have used the properties of $W^{\mu}$ (cf.~Appendix~\ref{app:Wmu}).\\
The other field commutators vanish, since they involve either only
vanishing harmonic-oscillator commutators~(\ref{eq:ccrms2}) or the
Pauli-Jordan functions~(\ref{eq:Delta}) vanish for spacelike
$\left(x-y\right)$ (cf.~Appendix~\ref{app:PJF}). This is, of
course, the case for $x,y$ lying on the hyperboloid
$x^2=y^2=\tau^2$.$\quad\bullet$
\section{Generators in Wigner Representation}
\subsection{Generator for Global Gauge Transformations}
\label{app:cho} We want to show that the charge operator has the
same Wigner representation in instant and point form:\\
$\circ\quad$ From~(\ref{eq:pfQ2})
we have
\begin{eqnarray}
\hat{Q}_{\mathrm{KG}}&=&2i\int_{\mathbb{R}^4}\mathrm{d}^4x\,\delta\left(x_{\lambda}\,x^{\lambda}-\tau^2\right)\theta\left(x^0\right)x^{\mu}:\left[\hat{\phi}^{\dag}\left(x\right)\left(\partial_{\mu}\hat{\phi}\left(x\right)\right)-
\left(\partial_{\mu}\hat{\phi}^{\dag}\left(x\right)\right)\hat{\phi}\left(x\right)\right]:
\nonumber\\&=&\int_{\mathbb{R}^3}
\frac{\mathrm{d}^3p}{2p_0}\int_{\mathbb{R}^3}\frac{\mathrm{d}^3q}{2q_0}\frac{2}{\left(2\pi\right)^3}\int_{\mathbb{R}^4}\mathrm{d}^4x\,\delta\left(x_{\lambda}\,x^{\lambda}-\tau^2\right)\theta\left(x^0\right)x^{\mu}\nonumber\\&&
:\left[\left(p+q\right)_{\mu}\left(\mathrm{e}^{i\left(p-q\right)_{\lambda}\, x^{\lambda}}\hat{a}^{\dag}\left(\boldsymbol{p}\right)\hat{a}\left(\boldsymbol{q}\right)-\mathrm{e}^{-i\left(p-q\right)_{\lambda}\, x^{\lambda}}\hat{b}\left(\boldsymbol{p}\right)\hat{b}^{\dag}\left(\boldsymbol{q}\right) \right)\right.\nonumber\\
&&\quad+\left.\left(p-q\right)_{\mu}\left(\mathrm{e}^{-i\left(p+q\right)_{\lambda}\,
x^{\lambda}}\hat{b}\left(\boldsymbol{p}\right)\hat{a}\left(\boldsymbol{q}\right)-\mathrm{e}^{i\left(p+q\right)_{\lambda}\,
x^{\lambda}}\hat{a}^{\dag}\left(\boldsymbol{p}\right)\hat{b}^{\dag}\left(\boldsymbol{q}\right)\right)\right]
:\nonumber\\&=& \int_{\mathbb{R}^3}
\frac{\mathrm{d}^3p}{2p_0}\int_{\mathbb{R}^3}\frac{\mathrm{d}^3q}{2q_0}:
\left[W\left(P,Q\right)\hat{a}^{\dag}\left(\boldsymbol{p}\right)\hat{a}\left(\boldsymbol{q}\right)-
W\left(P,-Q\right)\hat{b}\left(\boldsymbol{p}\right)\hat{b}^{\dag}\left(\boldsymbol{q}\right)\right.\nonumber\\
&&\quad+\left.W\left(Q,-P\right)\hat{b}\left(\boldsymbol{p}\right)\hat{a}\left(\boldsymbol{q}\right)-
W\left(Q,P\right)\hat{a}^{\dag}\left(\boldsymbol{p}\right)\hat{b}^{\dag}\left(\boldsymbol{q}\right)\right]:
\nonumber\\&=&\int_{\mathbb{R}^3}
\frac{\mathrm{d}^3p}{2p_0}\left(\hat{a}^{\dag}\left(\boldsymbol{p}\right)\hat{a}\left(\boldsymbol{p}\right)-\hat{b}^{\dag}\left(\boldsymbol{p}\right)\hat{b}\left(\boldsymbol{p}\right)\right)
,\end{eqnarray}
 where we have used the properties of the $W$ distribution~(cf.~Appendix~\ref{app:W}).$\quad\bullet$
\subsection{Translation Generator}
\label{app:Pmu}
We want to show that the four-momentum operator is the same in instant and point form.\\
$\circ\quad$
From~(\ref{eq:pfP}) we have
\begin{eqnarray}
\hat{P}_{\mathrm{KG}}^{\mu}&=&\int_{\mathbb{R}^4}2\mathrm{d}^4x\,
\delta\left(x_{\lambda}\,x^{\lambda}-\tau^2\right)\theta\left(x^0\right)x_{\nu}\hat{\mathcal{T}}_{\mathrm{KG}}^{\mu\nu}\left(x\right)\nonumber\end{eqnarray}
 \begin{eqnarray}&=&\frac{2}{\left(2\pi\right)^3}\int_{\mathbb{R}^4}\mathrm{d}^4x\,
\delta\left(x_{\lambda}\,x^{\lambda}-\tau^2\right)\theta\left(x^0\right)\int_{\mathbb{R}^3}\frac{\mathrm{d}^3p}{2p^0}\int_{\mathbb{R}^3}\frac{\mathrm{d}^3q}{2q^0}\nonumber\\
&&\times\left\lbrace \left[ p^{\mu} \,x_{\lambda}
\,q^{\lambda}+p_{\lambda}\, x^{\lambda}\, q^{\mu}-x^{\mu}\left( q_{\lambda}
\,p^{\lambda}-m^2\right) \right]\right.\nonumber\\&&\qquad\times\left. \left( \mathrm{e}^{i\left(
p-q\right)_{\lambda} \,x^{\lambda} }
\hat{a}^{\dag}\left(\boldsymbol{p}\right)\hat{a}\left(\boldsymbol{q}\right)+\mathrm{e}^{-i\left(
p-q\right)_{\lambda}\, x^{\lambda} }
\hat{b}^{\dag}\left(\boldsymbol{q}\right)\hat{b}\left(\boldsymbol{p}\right)\right)\right.\nonumber\\
 &&\quad-\left.\left[ p^{\mu}\, x_{\lambda}\, q^{\lambda}+p_{\lambda} \,x^{\lambda}\,
q^{\mu}-x^{\mu}\left( q_{\lambda}\, p^{\lambda}+m^2\right) \right]\right.\nonumber\\&&\qquad\times\left.
\left( \mathrm{e}^{i\left( p+q\right)_{\lambda}\, x^{\lambda} }
\hat{a}^{\dag}\left(\boldsymbol{p}\right)\hat{b}^{\dag}\left(\boldsymbol{q}\right)+\mathrm{e}^{-i\left(
p+q\right)_{\lambda}\, x^{\lambda} }
\hat{b}\left(\boldsymbol{p}\right)\hat{a}\left(\boldsymbol{q}\right)\right)
\right\rbrace.\nonumber\\
\end{eqnarray} Rewriting the square
brackets in terms of $P=p+q$ and $Q=p-q$ gives
\begin{eqnarray}p^{\mu} \,x_{\lambda}
\,q^{\lambda}+p_{\lambda}\, x^{\lambda}\, q^{\mu}-x^{\mu}\left( q_{\lambda}
\,p^{\lambda}-m^2\right) &=&\frac{1}{2}\left[ P^{\mu}\,x_{\lambda}\,
P^{\lambda}-Q^{\mu}\,x_{\lambda}\,
Q^{\lambda}+x^{\mu}\,Q_{\lambda}\,Q^{\lambda}\right],\nonumber\\
\\p^{\mu}\,
x_{\lambda}\, q^{\lambda}+p_{\lambda} \,x^{\lambda}\, q^{\mu}-x^{\mu}\left(
q_{\lambda}\, p^{\lambda}+m^2\right) &=&\frac{1}{2}\left[
P^{\mu}\,x_{\lambda}\, P^{\lambda}-Q^{\mu}\,x_{\lambda}\,
Q^{\lambda}-x^{\mu}\,P_{\lambda}\,P^{\lambda}\right].\nonumber\\
\end{eqnarray} Interchanging position and momentum integrations
yields
\begin{eqnarray}&&\frac{1}{2}\int_{\mathbb{R}^3}\frac{\mathrm{d}^3p}{2p^0}\int_{\mathbb{R}^3}\frac{\mathrm{d}^3q}{2q^0}\frac{2}{\left(2\pi\right)^3}\int_{\mathbb{R}^4}\mathrm{d}^4x\,
\delta\left(x_{\lambda}\,x^{\lambda}-\tau^2\right)\theta\left(x^0\right)\nonumber\\
&&\quad\times\left\lbrace \left[ P^{\mu}\,x_{\lambda}\,
P^{\lambda}-Q^{\mu}\,x_{\lambda}\,
Q^{\lambda}+x^{\mu}\,Q_{\lambda}\,Q^{\lambda}\right]\left(
\mathrm{e}^{iQ_{\lambda} \,x^{\lambda} }
\hat{a}^{\dag}\left(\boldsymbol{p}\right)\hat{a}\left(\boldsymbol{q}\right)+\mathrm{e}^{-iQ_{\lambda}\,
x^{\lambda} }
\hat{b}^{\dag}\left(\boldsymbol{q}\right)\hat{b}\left(\boldsymbol{p}\right)\right)-\right.\nonumber\\
 &&\qquad-\left. \left[ P^{\mu}\,x_{\lambda}\, P^{\lambda}-Q^{\mu}\,x_{\lambda}\,
Q^{\lambda}-x^{\mu}\,P_{\lambda}\,P^{\lambda}\right] \left(
\mathrm{e}^{-iP_{\lambda}\, x^{\lambda} }
\hat{a}^{\dag}\left(\boldsymbol{p}\right)\hat{b}^{\dag}\left(\boldsymbol{q}\right)+\mathrm{e}^{iP_{\lambda}\,
x^{\lambda} }
\hat{b}\left(\boldsymbol{p}\right)\hat{a}\left(\boldsymbol{q}\right)\right)
\right\rbrace \nonumber\\&&=
\frac{1}{2}\int_{\mathbb{R}^3}\frac{\mathrm{d}^3p}{2p^0}\int_{\mathbb{R}^3}\frac{\mathrm{d}^3q}{2q^0}\nonumber\\
&&\quad\times\left\lbrace
\left[P^{\mu}\,W\left(P,Q\right)-Q^{\mu}\,Q^{\lambda}\,W_{\lambda}\left(P,Q\right)+Q_{\lambda}\,Q^{\lambda}\,W^{\mu}\left(P,Q\right)\right]\hat{a}^{\dag}\left(\boldsymbol{p}\right)\hat{a}\left(\boldsymbol{q}\right)\right.\nonumber\\
&&\qquad+\left.\left[P^{\mu}\,W\left(P,-Q\right)-Q^{\mu}\,Q^{\lambda}\,W_{\lambda}\left(P,-Q\right)+Q_{\lambda}\,Q^{\lambda}\,W^{\mu}\left(P,-Q\right)\right]\hat{b}^{\dag}\left(\boldsymbol{q}\right)\hat{b}\left(\boldsymbol{p}\right)\right.\nonumber\\
&&\qquad-\left.
\left[P^{\mu}\,P^{\lambda}\,W_{\lambda}\left(Q,-P\right)-Q^{\mu}\,W\left(Q,-P\right)-P_{\lambda}\,P^{\lambda}\,W^{\mu}\left(Q,-P\right)\right]\hat{a}^{\dag}\left(\boldsymbol{p}\right)\hat{b}^{\dag}\left(\boldsymbol{q}\right)
\right.\nonumber\\
&&\qquad-\left.
\left[P^{\mu}\,P^{\lambda}\,W_{\lambda}\left(Q,P\right)-Q^{\mu}\,W\left(Q,P\right)-P_{\lambda}\,P^{\lambda}\,W^{\mu}\left(Q,P\right)
\right]\hat{b}\left(\boldsymbol{p}\right)\hat{a}\left(\boldsymbol{q}\right)
\right\rbrace,
\end{eqnarray}
where we have expressed the $x$-integrals by the integral
representation of the $W$-distribution~(\ref{eq:irepW}) and its
derivative~(\ref{eq:irepWmu}). It can be easily seen that
$P^{\mu}\,W\left(P,Q\right)\hat{a}^{\dag}\left(\boldsymbol{p}\right)\hat{a}\left(\boldsymbol{q}\right)
$ and
$P^{\mu}\,W\left(P,-Q\right)\hat{b}^{\dag}\left(\boldsymbol{q}\right)\hat{b}\left(\boldsymbol{p}\right)$
are the only surviving terms. For the
$\hat{a}^{\dag}\hat{a}$-contribution we have
\begin{eqnarray}&&
P^{\mu}\,W\left(P,Q\right)-Q^{\mu}\,Q^{\lambda}\,W_{\lambda}\left(P,Q\right)+Q_{\lambda}\,Q^{\lambda}\,W^{\mu}\left(P,Q\right)\nonumber\\&&=
P^{\mu}\,W\left(P,Q\right)-Q^{\mu}\,Q^{\lambda}\,\frac{P_{\lambda}}{P_{\sigma}\,P^{\sigma}}W\left(P,Q\right)-
Q^{\mu}\,Q^{\lambda}\,\frac{Q_{\lambda}}{Q_{\sigma}\,Q^{\sigma}}W\left(Q,Q\right)\nonumber\\&&\quad+Q_{\lambda}\,Q^{\lambda}\,\frac{P^{\mu}}{P_{\sigma}\,P^{\sigma}}W\left(P,Q\right)+
Q_{\lambda}\,Q^{\lambda}\,\frac{Q^{\mu}}{Q_{\sigma}\,Q^{\sigma}}W\left(Q,Q\right)\nonumber\\&&=P^{\mu}\,W\left(P,Q\right)
,
\end{eqnarray}
since $P_{\lambda}\,Q^{\lambda}=0$ and
$Q_{\lambda}\,Q^{\lambda}\,W\left(P,Q\right)=-\tilde{\boldsymbol{Q}}^2\,\tilde{P}^0\,\delta^3\left(\tilde{\boldsymbol{Q}}\right)=0.$
\\The
$\hat{b}^{\dag}\hat{b}$-contribution is calculated similarly.\\
For the $\hat{a}^{\dag}\hat{b}^{\dag}$-contribution we have
\begin{eqnarray}&&
P^{\mu}\,P^{\lambda}\,W_{\lambda}\left(Q,-P\right)-Q^{\mu}\,W\left(Q,-P\right)-P_{\lambda}\,P^{\lambda}\,W^{\mu}\left(Q,-P\right)\nonumber\\&&=
P^{\mu}\,P^{\lambda}\,\frac{P_{\lambda}}{P_{\sigma}\,P^{\sigma}}W\left(P,-P\right)-P_{\lambda}\,P^{\lambda}\,\frac{P^{\mu}}{P_{\sigma}\,P^{\sigma}}W\left(P,-P\right)
\nonumber\\&&=0,
\end{eqnarray} since
$W\left(Q,-P\right)=0$.\\ The $\hat{b}\,\hat{a}$-contribution
vanishes similarly.
\\
Therefore, we finally have
\begin{eqnarray} \hat{P}_{\mathrm{KG}}^{\mu}&=&
\frac{1}{2}\int_{\mathbb{R}^3}\frac{\mathrm{d}^3p}{2p^0}\int_{\mathbb{R}^3}\frac{\mathrm{d}^3q}{2q^0}P^{\mu}
\left[W\left(P,Q\right)\hat{a}^{\dag}\left(\boldsymbol{p}\right)\hat{a}\left(\boldsymbol{q}\right)+W\left(P,-Q\right)\hat{b}^{\dag}\left(\boldsymbol{p}\right)\hat{b}\left(\boldsymbol{q}\right)\right]\nonumber\\&=&
\frac{1}{2}\int_{\mathbb{R}^3}\frac{\mathrm{d}^3p}{2p^0}\int_{\mathbb{R}^3}\frac{\mathrm{d}^3q}{2q^0}P^{\mu}P^{0}\delta^3\left(\boldsymbol{Q}\right)
\left[\hat{a}^{\dag}\left(\boldsymbol{p}\right)\hat{a}\left(\boldsymbol{q}\right)+\hat{b}^{\dag}\left(\boldsymbol{p}\right)\hat{b}\left(\boldsymbol{q}\right)\right]
\nonumber\\&=&\int_{\mathbb{R}^3}\frac{\mathrm{d}^3p}{2p_0}p^{\mu}\left(\hat
{a}^{\dag}\left(\boldsymbol{p}\right)\hat{a}\left(\boldsymbol{p}\right)+\hat{b}^{\dag}\left(\boldsymbol{p}\right)\hat{b}\left(\boldsymbol{p}\right)\right)
.\quad\bullet\end{eqnarray}
\chapter{Dirac Fields}
\label{app:s12f}
\section{Invariant Scalar Product}
\label{app:pfspd}
For the scalar product between positive frequency modes~(\ref{eq:spbpfm}) we have
\begin{eqnarray}
\left(\psi_{\rho,\boldsymbol{p}},\psi_{\sigma,\boldsymbol{q}}\right)_{\Sigma_{\tau}}&=&
\frac{2}{\left(2\pi\right)^3}\int_{\mathbb{R}^4}\mathrm{d}^4x\,
\delta\left(x_{\lambda}\,x^{\lambda}-\tau^2\right)\theta\left(x^0\right)x^{\mu}\mathrm{e}^{ix_{\lambda}\left(p-q\right)^{\lambda}}\bar{u}_{\rho}\left(\boldsymbol{p}\right)\gamma_{\mu}u_{\sigma}\left(\boldsymbol{q}\right)
\nonumber\\&=&
W^{\mu}\left(P,Q\right)\bar{u}_{\rho}\left(\boldsymbol{p}\right)\gamma_{\mu}u_{\sigma}\left(\boldsymbol{q}\right)
\nonumber\\&=&\frac{W\left(P,Q\right)}{P_{\lambda}\,P^{\lambda}}\bar{u}_{\rho}\left(\boldsymbol{p}\right)\gamma_{\mu}\left(p+q\right)^{\mu}u_{\sigma}\left(\boldsymbol{q}\right)+
\frac{W\left(Q,Q\right)}{Q_{\lambda}\,Q^{\lambda}}
\underbrace{\bar{u}_{\rho}\left(\boldsymbol{p}\right)\gamma_{\mu}\left(p-q\right)^{\mu}u_{\sigma}\left(\boldsymbol{q}\right)}_{=\left(m-m\right)
\bar{u}_{\rho}\left(\boldsymbol{p}\right)u_{\sigma}\left(\boldsymbol{q}\right)=0}
\nonumber\\&=&\frac{2p^0\delta^3\left(\boldsymbol{p}-\boldsymbol{q}\right)}{4p_{\lambda}\,p^{\lambda}}2m\,\bar{u}_{\rho}\left(\boldsymbol{p}\right)u_{\sigma}\left(\boldsymbol{p}\right)=
2p^0\delta_{\rho\sigma}\delta^3\left(\boldsymbol{p}-\boldsymbol{q}\right),\end{eqnarray}
where we have used the Dirac equations for
$\bar{u}_{\rho}\left(\boldsymbol{p}\right)$~(\ref{eq:amde1})
and $u_{\sigma}\left(\boldsymbol{q}\right)$~(\ref{eq:mde1}). The
scalar product between negative frequency modes gives the same
result. For the mixed scalar product~(\ref{eq:spbmfm}) we have
\begin{eqnarray}
\left(\psi_{\rho,\boldsymbol{p}},\chi_{\sigma,\boldsymbol{q}}\right)_{\Sigma_{\tau}}&=&\frac{2}{\left(2\pi\right)^3}\int_{\mathbb{R}^4}\mathrm{d}^4x\,
\delta\left(x_{\lambda}\,x^{\lambda}-\tau^2\right)\theta\left(x^0\right)x^{\mu}\mathrm{e}^{ix_{\lambda}\left(p+q\right)^{\lambda}}\bar{u}_{\rho}\left(\boldsymbol{p}\right)\gamma_{\mu}v_{\sigma}\left(\boldsymbol{q}\right)
\nonumber\\&=&
W^{\mu}\left(Q,P\right)\bar{u}_{\rho}\left(\boldsymbol{p}\right)\gamma_{\mu}v_{\sigma}\left(\boldsymbol{q}\right)=
\frac{W\left(P,P\right)}{P_{\lambda}\,P^{\lambda}}\bar{u}_{\rho}\left(\boldsymbol{p}\right)\gamma_{\mu}\left(p+q\right)^{\mu}v_{\sigma}\left(\boldsymbol{q}\right)
\nonumber\\&=&\frac{W\left(P,P\right)}{P_{\lambda}\,P^{\lambda}}\left(m-m\right)\bar{u}_{\rho}\left(\boldsymbol{p}\right)v_{\sigma}\left(\boldsymbol{q}\right)=0,
\end{eqnarray} where we have used the Dirac equations for
$\bar{u}_{\rho}\left(\boldsymbol{p}\right)$~(\ref{eq:amde1}) and
$v_{\sigma}\left(\boldsymbol{q}\right)$~(\ref{eq:mde2}). The other mixed scalar
product vanishes in a similar way.
\section{Covariant Canonical Anticommutation Relations}
\label{app:ccacr} We use hyperbolic coordinates
$(\tau,\xi\equiv\mathrm{cosh}\beta,\vartheta,\varphi)$~(\ref{eq:xhc}) with the hypersurface element~(\ref{eq:hse}).\\
$\circ\quad$ The Lagrangian operator for free Dirac fields in
hyperbolic coordinates reads
\begin{eqnarray}
\hat{L}_{\mathrm{D}}
&=&\int_{\Sigma_{\tau}}\tau^3\sqrt{\xi^2-1}\mathrm{d}\xi\,
\mathrm{d}\mathrm{cos}\vartheta \,\mathrm{d}\varphi\left\lbrace
i\hat{\bar{\psi}}\,\frac{\gamma^{\lambda}\,
x_{\lambda}}{\tau}\left[\frac{\partial}{\partial\tau}-\frac{\sigma^{\mu\nu}\hat{M}_{\mu\nu}}{2\tau}\right]\hat{\psi}-
m\hat{\bar{\psi}}\hat{\psi} \right\rbrace
\nonumber\\&=&\int_{\Sigma_{\tau}} \mathrm{d}\Sigma_{\tau}
\hat{\mathcal{L}}_{\mathrm{D}}\left(\tau,\xi,\vartheta,\varphi\right).\end{eqnarray}
The equal-$\tau$ canonical
anticommutation relations are given by~\cite{Gromes:1974yu}
\begin{eqnarray}
&&\left\lbrace
\hat{\psi}_{\alpha}\left(y\right),\hat{\bar{\psi}}_{\beta}\left(x\right)\right\rbrace_{\tau\left(x\right)=\tau\left(y\right)=\tau}\nonumber\\&&=
 \frac{\left[\gamma^{\lambda}\, x_{\lambda}\right]_{\alpha\beta}}{\tau^4\sqrt{\left(\xi\left(x\right)\right)^2-1}}\,\delta\left(\xi\left(x\right)-\xi\left(y\right)\right)
\delta\left(\mathrm{cos}\,\vartheta
\left(x\right)-\mathrm{cos}\,\vartheta \left(y\right)\right)
\delta\left(\varphi\left(x\right)-\varphi\left(y\right)\right),\nonumber\\\label{eq:anticom}
\end{eqnarray} such
 that the covariant anticommutator relation~(\ref{eq:ccar}) is satisfied,
\begin{eqnarray}&&
i\int_{\Sigma}\mathrm{d}\Sigma^{\mu}\left(x\right)\left[\gamma_{\mu}\right]^{~\beta}_{\gamma}\left\lbrace
\hat{\psi}_{\alpha}\left(y\right),\hat{\bar{\psi}}_{\beta}\left(x\right)\right\rbrace_{x,y\in\Sigma}\nonumber\\&&=
i\int_{\Sigma_{\tau}}\tau^2\sqrt{\left(\xi\left(x\right)\right)^2-1}\,\mathrm{d}\xi\left(x\right)
\,\mathrm{d}\mathrm{cos}\,\vartheta\left(x\right)\,
\mathrm{d}\varphi\left(x\right)\,\left[\gamma_{\mu}\,
x^{\mu}\right]^{~\beta}_{\gamma}\left\lbrace\hat{\psi}_{\alpha}\left(y\right),\hat{\bar{\psi}}_{\beta}\left(x\right)\right\rbrace_{\tau\left(x\right)=\tau\left(y\right)=\tau}\nonumber\\&&=
i\delta_{\gamma\alpha}.
\end{eqnarray}
With relation~(\ref{eq:userel}) we see, that the
anticommutators~(\ref{eq:anticom}) and~(\ref{eq:licacr}) are
essentially the same.$\quad\bullet$
\section{Anticommutation Relations in Momentum Space}
\label{app:cacrms} We want to show explicitly in point form, that
the covariant canonical anticommutation relation,~(\ref{eq:ccar2}) and~(\ref{eq:licacr}), imply the anticommutation relations in
momentum space,~(\ref{eq:acrms}) and~(\ref{eq:acrms2}).\\
$\circ\quad$
 For the anticommutators~(\ref{eq:acrms}) on the hyperboloid we have
\begin{eqnarray}
\left\lbrace \hat{c}_{\rho}\left(\boldsymbol{p}\right),
\hat{c}^{\dag}_{\sigma}\left(\boldsymbol{q}\right)\right\rbrace&=&
4\int_{\mathbb{R}^4}\mathrm{d}^4x\,\delta\left(x_{\lambda}\,x^{\lambda}-\tau^2\right)\theta\left(x^0\right)x^{\mu}
\int_{\mathbb{R}^4}\mathrm{d}^4y\,\delta\left(y_{\lambda}\,y^{\lambda}-\tau^2\right)\theta\left(y^0\right)y^{\nu}\nonumber\\&&\times
\left\lbrace\left[\bar{\psi}_{\rho,\boldsymbol{p}}\left(x\right)\gamma_{\mu}\right]^{\alpha}
\left[\hat{\psi}\left(x\right)\right]_{\alpha},\left[\hat{\bar{\psi}}\left(y\right)\gamma_{\nu}\right]^{\beta}\left[\psi_{\sigma,\boldsymbol{q}}\left(y\right)\right]_{\beta}\right\rbrace
\nonumber\\&=&
\frac{4}{\left(2\pi\right)^3}\int_{\mathbb{R}^4}\mathrm{d}^4x\,\delta\left(x_{\lambda}\,x^{\lambda}-\tau^2\right)\theta\left(x^0\right)x^{\mu}
\int_{\mathbb{R}^4}\mathrm{d}^4y\,\delta\left(y_{\lambda}\,y^{\lambda}-\tau^2\right)\theta\left(y^0\right)\nonumber\\&&\times
\mathrm{e}^{ip_{\rho}\,x^{\rho}}\mathrm{e}^{-iq_{\lambda}\,y^{\lambda}}\left[\bar{u}_{\rho}\left(\boldsymbol{p}\right)\gamma_{\mu}\right]^{\alpha}\left[u_{\sigma}\left(\boldsymbol{q}\right)\right]_{\beta}\underbrace{y^{\nu}
\left\lbrace\hat{\psi}_{\alpha}\left(x\right),
\left[\hat{\bar{\psi}}\left(y\right)\gamma_{\nu}\right]^{\beta}\right\rbrace}_{=x^0\delta_{\alpha}^{\beta}\delta^3\left(\boldsymbol{x}-\boldsymbol{y}\right),~(\ref{eq:licacr})}\nonumber\\&=&
\frac{2}{\left(2\pi\right)^3}\int_{\mathbb{R}^4}\mathrm{d}^4x\,\delta\left(x_{\lambda}\,x^{\lambda}-\tau^2\right)\theta\left(x^0\right)x^{\mu}\mathrm{e}^{i\left(p-q\right)_{\lambda}\,
x^{\lambda}}\bar{u}_{\rho}\left(\boldsymbol{p}\right)\gamma_{\mu}u_{\sigma}\left(\boldsymbol{q}\right)\nonumber\end{eqnarray}\\\begin{eqnarray}&=&W^{\mu}\left(P,Q\right)\bar{u}_{\rho}\left(\boldsymbol{p}\right)\gamma_{\mu}u_{\sigma}\left(\boldsymbol{q}\right)=
\frac{2\left(p+q\right)^{\mu}\,p_{\mu}}{\left(p+q\right)_{\lambda}\,\left(p+q\right)^{\lambda}}
\left(p+q\right)^0\delta_{\rho\sigma}\delta^3\left(\boldsymbol{p}-\boldsymbol{q}\right)\nonumber\\&&+\frac{1}{Q_{\lambda}\,Q^{\lambda}}
\underbrace{\bar{u}_
{\rho}\left(\boldsymbol{p}\right)\gamma_{\mu}\left(p-q\right)^{\mu}u_{\sigma}\left(\boldsymbol{q}\right)}_{=0,~(\ref{eq:mde1}),~(\ref{eq:amde1})}W\left(Q,Q\right)\nonumber\\&=&
2p^0\delta_{\rho\sigma}\delta^3\left(\boldsymbol{p}-\boldsymbol{q}\right),
\end{eqnarray} where we have used the properties of $W^{\mu}$ (cf.~Appendix~\ref{app:Wmu}).
$ \left\lbrace \hat{d}_{\rho}\left(\boldsymbol{p}\right),
\hat{d}^{\dag}_{\sigma}\left(\boldsymbol{q}\right)\right\rbrace=2p^0\delta_{\rho\sigma}\delta^3\left(\boldsymbol{p}-\boldsymbol{q}\right)$ is calculated similarly.\\
For the anticommutators~(\ref{eq:acrms2}) we have
\begin{eqnarray}
\left\lbrace \hat{c}_{\rho}\left(\boldsymbol{p}\right),
\hat{d}_{\sigma}\left(\boldsymbol{q}\right)\right\rbrace&=&
4\int_{\mathbb{R}^4}\mathrm{d}^4x\,\delta\left(x_{\lambda}\,x^{\lambda}-\tau^2\right)\theta\left(x^0\right)x^{\mu}
\int_{\mathbb{R}^4}\mathrm{d}^4y\,\delta\left(y_{\lambda}\,y^{\lambda}-\tau^2\right)\theta\left(y^0\right)y^{\nu}\nonumber\\&&\times
\left\lbrace\left[\bar{\psi}_{\rho,\boldsymbol{p}}\left(x\right)\gamma_{\mu}\right]^{\alpha}
\left[\hat{\psi}\left(x\right)\right]_{\alpha},\left[\hat{\bar{\psi}}\left(y\right)\gamma_{\nu}\right]^{\beta}\left[\chi_{\sigma,\boldsymbol{q}}\left(y\right)\right]_{\beta}\right\rbrace
\nonumber\\&=&
\frac{4}{\left(2\pi\right)^3}\int_{\mathbb{R}^4}\mathrm{d}^4x\,\delta\left(x_{\lambda}\,x^{\lambda}-\tau^2\right)\theta\left(x^0\right)x^{\mu}
\int_{\mathbb{R}^4}\mathrm{d}^4y\,\delta\left(y_{\lambda}\,y^{\lambda}-\tau^2\right)\theta\left(y^0\right)\nonumber\\&&\times
\mathrm{e}^{ip_{\rho}\,x^{\rho}}\mathrm{e}^{iq_{\lambda}\,y^{\lambda}}\left[\bar{u}_{\rho}\left(\boldsymbol{p}\right)\gamma_{\mu}\right]^{\alpha}\left[v_{\sigma}\right]_{\beta}\left(\boldsymbol{q}\right)\underbrace{y^{\nu}
\left\lbrace\hat{\psi}_{\alpha}\left(x\right),
\left[\hat{\bar{\psi}}\left(y\right)\gamma_{\nu}\right]^{\beta}\right\rbrace}_{=x^0\delta_{\alpha}^{\beta}\delta^3\left(\boldsymbol{x}-\boldsymbol{y}\right),~(\ref{eq:licacr})}\nonumber\\&=&
\frac{2}{\left(2\pi\right)^3}\int_{\mathbb{R}^4}\mathrm{d}^4x\,\delta\left(x_{\lambda}\,x^{\lambda}-\tau^2\right)\theta\left(x^0\right)x^{\mu}\mathrm{e}^{i\left(p+q\right)_{\lambda}\,
x^{\lambda}}\bar{u}_{\rho}\left(\boldsymbol{p}\right)\gamma_{\mu}v_{\sigma}\left(\boldsymbol{q}\right)\nonumber\\&=&W^{\mu}\left(Q,P\right)\bar{u}_{\rho}\left(\boldsymbol{p}\right)\gamma_{\mu}u_{\sigma}\left(\boldsymbol{q}\right)
\nonumber\\&=&
\frac{1}{P_{\lambda}\,P^{\lambda}}\bar{u}_{\rho}\left(\boldsymbol{p}\right)\gamma_{\mu}\left(p^{\mu}+q^{\mu}\right)v_{\sigma}\left(\boldsymbol{q}\right)W\left(P,P\right)\nonumber\\&=&\frac{1}{P_{\lambda}\,P^{\lambda}}\left(m-m\right)\bar{u}_{\rho}\left(\boldsymbol{p}\right)v_{\sigma}\left(\boldsymbol{q}\right)W\left(P,P\right)=0,
\end{eqnarray} where we have used~(\ref{eq:amde1}) and~(\ref{eq:mde2}).
$ \left\lbrace \hat{d}^{\dag}_{\rho}\left(\boldsymbol{p}\right),
\hat{c}^{\dag}_{\sigma}\left(\boldsymbol{q}\right)\right\rbrace$ can be shown to vanish in a similar way.\\
For the last anticommutators in~(\ref{eq:acrms2}) we have
\begin{eqnarray}
\left\lbrace
\hat{c}_{\rho}\left(\boldsymbol{p}\right),\hat{d}^{\dag}_{\sigma}\left(\boldsymbol{q}\right)\right\rbrace&=&
4\int_{\mathbb{R}^4}\mathrm{d}^4x\,\delta\left(x_{\lambda}\,x^{\lambda}-\tau^2\right)\theta\left(x^0\right)x^{\mu}
\int_{\mathbb{R}^4}\mathrm{d}^4y\,\delta\left(y_{\lambda}\,y^{\lambda}-\tau^2\right)\theta\left(y^0\right)y^{\nu}\nonumber\\&&\times
\left\lbrace\left[\bar{\psi}_{\rho,\boldsymbol{p}}\left(x\right)\gamma_{\mu}\right]_{\alpha}
\left[\hat{\psi}\left(x\right)\right]_{\alpha},\left[\bar{\chi}_{\sigma,\boldsymbol{q}}\left(y\right)\gamma_{\nu}\right]_{\beta}\left[\hat{\psi}\left(y\right)\right]_{\beta}\right\rbrace
\nonumber\\&=&
\frac{4}{\left(2\pi\right)^3}\int_{\mathbb{R}^4}\mathrm{d}^4x\,\delta\left(x_{\lambda}\,x^{\lambda}-\tau^2\right)\theta\left(x^0\right)x^{\mu}
\int_{\mathbb{R}^4}\mathrm{d}^4y\,\delta\left(y_{\lambda}\,y^{\lambda}-\tau^2\right)\theta\left(y^0\right)y^{\nu}\nonumber\\&&\times
\mathrm{e}^{ip_{\rho}\,x^{\rho}}\mathrm{e}^{-iq_{\lambda}\,y^{\lambda}}\left[\bar{u}_{\rho}\left(\boldsymbol{p}\right)\gamma_{\mu}\right]_{\alpha}\left[\bar{v}_{\sigma}\left(\boldsymbol{q}\right)\gamma_{\nu}\right]_{\beta}\underbrace{
\left\lbrace\hat{\psi}_{\alpha}\left(x\right),
\hat{\psi}_{\beta}\left(y\right)\right\rbrace}_{=0,~(\ref{eq:ccar2})}\nonumber\\&=&0\nonumber.
\end{eqnarray}
Similarly, $\left\lbrace\hat{d}^{\dag}_{\rho}\left(\boldsymbol{p}\right),
\hat{c}_{\sigma}\left(\boldsymbol{q}\right)\right\rbrace$ can be shown to vanish.$\quad\bullet$\\
\\That the harmonic-oscillator anticommutation
relations,~(\ref{eq:acrms}) and~(\ref{eq:acrms2}), imply the
covariant canonical anticommutation relations,~(\ref{eq:ccar})
and~(\ref{eq:ccar2}), can be seen as follows. Before that, we
introduce the projector for positive frequencies being
\begin{eqnarray}
\sum_{\rho=\pm\frac12}\left[u_{\rho}\right]_{\alpha}\left(\boldsymbol{p}\right)\left[\bar{u}_{\rho}\right]_{\beta}\left(\boldsymbol{p}\right)
=\left[\gamma^{\mu}\,p_{\mu}+m\right]_{\alpha\beta}
 \label{eq:proj},\end{eqnarray}
and the projector for negative frequencies as
\begin{eqnarray}
-\sum_{\rho=\pm\frac12}\left[v_{\rho}\right]_{\alpha}\left(\boldsymbol{p}\right)\left[\bar{v}_{\rho}\right]_{\beta}\left(\boldsymbol{p}\right)
=\left[-\gamma^{\mu}\,p_{\mu}+m\right]_{\alpha\beta}
 \label{eq:proj2}.\end{eqnarray}\\
$\circ\quad$ In point form the covariant
anticommutation relations are
\begin{eqnarray}&&i\int_{\mathbb{R}^4}2\mathrm{d}^4x\,\delta\left(x_{\lambda}\,x^{\lambda}-\tau^2\right)\theta\left(x^0\right)x^{\mu}\left\lbrace \hat{\psi}_{\alpha}\left(y\right),\left[\hat{\bar{\psi}}\left(x\right)\gamma_{\mu}\right]_{\beta}\right\rbrace_{x^2=y^2=\tau^2}\nonumber\\&&=
i\int_{\mathbb{R}^4}2\mathrm{d}^4x\,\delta\left(x_{\lambda}\,x^{\lambda}-\tau^2\right)\theta\left(x^0\right)x^{\mu}\sum_{\rho,\sigma=\pm\frac12}
\frac{1}{\left(2\pi\right)^{3}}\int_{\mathbb{R}^3}\frac{\mathrm{d}^3p}{2p^0}
\int_{\mathbb{R}^3}\frac{\mathrm{d}^3q}{2q^0}
\nonumber\\&&\quad\times \left[ \mathrm{e}^{-iq_{\rho}\, y^{\rho}}
\mathrm{e}^{ip_{\nu}\, x^{\nu}}\left[
u_{\sigma}\left(\boldsymbol{q}\right)\right]_{\alpha}\left[\bar{u}_{\rho}\left(\boldsymbol{p}\right)\gamma_{\mu}\right]_{\beta}
\left\lbrace
\hat{c}_{\sigma}\left(\boldsymbol{q}\right),\hat{c}_{\rho}^{\dag}\left(\boldsymbol{p}\right)
\right\rbrace \right.
\nonumber\\
&&\qquad+\left.\mathrm{e}^{iq_{\rho}\, y^{\rho}}
\mathrm{e}^{-ip_{\nu}\, x^{\nu}}\left[
v_{\sigma}\left(\boldsymbol{q}\right)\right]_{\alpha}\left[\bar{v}_{\rho}\left(\boldsymbol{p}\right)\gamma_{\mu}\right]_{\beta}
\left\lbrace
\hat{d}^{\dag}_{\sigma}\left(\boldsymbol{q}\right),\hat{d}_{\rho}\left(\boldsymbol{p}\right)
\right\rbrace\right.
\nonumber\\
&&\qquad+\left.
 \mathrm{e}^{-iq_{\rho}\, y^{\rho}} \mathrm{e}^{-ip_{\nu}\, x^{\nu}}\left[ u_{\sigma}\left(\boldsymbol{q}\right)\right]_{\alpha}\left[\bar{v}_{\rho}\left(\boldsymbol{p}\right)\gamma_{\mu}\right]_{\beta}
\left\lbrace
\hat{c}_{\sigma}\left(\boldsymbol{q}\right),\hat{d}_{\rho}\left(\boldsymbol{p}\right)
\right\rbrace \right.
\nonumber\\
&&\qquad+\left.\mathrm{e}^{iq_{\rho}\, y^{\rho}}
\mathrm{e}^{ip_{\nu}\, x^{\nu}}\left[
v_{\sigma}\left(\boldsymbol{q}\right)\right]_{\alpha}\left[\bar{u}_{\rho}\left(\boldsymbol{p}\right)\gamma_{\mu}\right]_{\beta}
\left\lbrace
\hat{d}^{\dag}_{\sigma}\left(\boldsymbol{q}\right),\hat{c}_{\rho}^{\dag}\left(\boldsymbol{p}\right)
\right\rbrace \right]\nonumber\\&&=
i\int_{\mathbb{R}^4}2\mathrm{d}^4x\,\delta\left(x_{\lambda}\,x^{\lambda}-\tau^2\right)\theta\left(x^0\right)x^{\mu}
\frac{1}{\left(2\pi\right)^{3}}\int_{\mathbb{R}^4}\mathrm{d}^4p\,\delta\left(p_{\lambda}\,
p^{\lambda}-m^2\right)\theta\left(p^0\right)
\nonumber\\&&\quad\times \left[
\mathrm{e}^{ip_{\lambda}\left(x-y\right)^{\lambda}}
\left[u_{\rho}\left(\boldsymbol{p}\right)\right]_{\alpha}\left[\bar{u}_{\rho}\left(\boldsymbol{p}\right)\gamma_{\mu}\right]_{\beta}
+\mathrm{e}^{-ip_{\lambda}\left(x-y\right)^{\lambda}}\left[
v_{\rho}\left(\boldsymbol{p}\right)\right]_{\alpha}\left[\bar{v}_{\rho}\left(\boldsymbol{p}\right)\gamma_{\mu}\right]_{\beta}
\right]\nonumber\\&&=
i\int_{\mathbb{R}^4}2\mathrm{d}^4x\,\delta\left(x_{\lambda}\,x^{\lambda}-\tau^2\right)\theta\left(x^0\right)
\frac{1}{\left(2\pi\right)^{3}}\int_{\mathbb{R}^4}\,\mathrm{d}^4p\,\delta\left(p_{\lambda}\,
p^{\lambda}-m^2\right)\theta\left(p^0\right)
\nonumber\\&&\quad\times \left[
\mathrm{e}^{ip_{\lambda}\left(x-y\right)^{\lambda}}\underbrace{\left[u_{\rho}\left(\boldsymbol{p}\right)\right]_{\alpha}\left[\bar{u}_{\rho}\left(\boldsymbol{p}\right)\right]_{\gamma}}_{=\left[p_{\nu}\,\gamma^{\nu}\right]_{\alpha\gamma}+m\delta_{\alpha\gamma},\text{~(\ref{eq:proj})}}\left[x_{\mu}\,\gamma^{\mu}\right]^{\gamma}_{~\beta}
\right.\nonumber\\&&\qquad+\left.\mathrm{e}^{-ip_{\lambda}\left(x-y\right)^{\lambda}}\underbrace{\left[v_{\rho}\left(\boldsymbol{p}\right)\right]_{\alpha}\left[\bar{v}_{\rho}\left(\boldsymbol{p}\right)\right]_{\gamma}}_{=\left[p_{\nu}\,\gamma^{\nu}\right]_{\alpha\gamma}-m\delta_{\alpha\gamma},\text{~(\ref{eq:proj2})}}\left[x_{\mu}\,\gamma^{\mu}\right]^{\gamma}_{~\beta}\right]\nonumber\\&&=
i\int_{\mathbb{R}^4}2\mathrm{d}^4x\,\delta\left(x_{\lambda}\,x^{\lambda}-\tau^2\right)\theta\left(x^0\right)
\frac{1}{\left(2\pi\right)^{3}}\int_{\mathbb{R}^4}\mathrm{d}^4p\,\delta\left(p_{\lambda}\,
p^{\lambda}-m^2\right)\theta\left(p^0\right)
\nonumber\\&&\quad\times \left[p_{\nu}\, x^{\nu}
\delta_{\alpha\beta}
\left(\mathrm{e}^{ip_{\lambda}\left(x-y\right)^{\lambda}}+\mathrm{e}^{-ip_{\lambda}\left(x-y\right)^{\lambda}}\right)+
m\left[x_{\mu}\,\gamma^{\mu}\right]_{\alpha\beta}\left(\mathrm{e}^{ip_{\lambda}\left(x-y\right)^{\lambda}}-
\mathrm{e}^{-ip_{\lambda}\left(x-y\right)^{\lambda}}\right)\right]\nonumber\end{eqnarray}\\\begin{eqnarray}&&=
\int_{\mathbb{R}^4}\mathrm{d}^4x\,\delta\left(x_{\lambda}\,x^{\lambda}-\tau^2\right)\theta\left(x^0\right)
\nonumber\\&&\quad\times \left\lbrace i\delta_{\alpha\beta}
\,x^{\nu}\left[\ W_{\nu}\left(X,Y\right)+
W_{\nu}\left(X,-Y\right)\right]+2m
\left[x_{\mu}\,\gamma^{\mu}\right]_{\alpha\beta}\underbrace{\Delta\left(x-y\right)\left\vert_{
x^2=y^2=\tau^2}\right.}_{=0,~(\ref{eq:Delta=0})}\right\rbrace\nonumber\\&&=
i\delta_{\alpha\beta}\int_{\mathbb{R}^4}\mathrm{d}^4x\,\delta\left(x_{\lambda}\,x^{\lambda}-\tau^2\right)\theta\left(x^0\right)\frac{2x^{\nu}\left(x+y\right)_{\nu}}{\left(x+y\right)_{\lambda}\,\left(x+y\right)^{\lambda}}\left(x+y\right)^0
\delta^3\left(\boldsymbol{x}-\boldsymbol{y}\right)\nonumber\\
&&=i\delta_{\alpha\beta}, \end{eqnarray} where we have
used~(\ref{eq:amugamma}) for c-numbers $p,\,x$, i.e.
$p_{\mu}\,\gamma^{\mu}\,x_{\nu}\,\gamma^{\nu}=p_{\mu}\,x^{\mu}$.
\section{Generators in Wigner Representation}
\subsection{Generator for Global Gauge Transformations}
\label{app:ggt} We want to find the Wigner representation of $\hat{Q}_{\mathrm{D}}$ calculated in point form:\\
$\circ\quad$ From~(\ref{eq:pfQ}) we have
\begin{eqnarray}
\hat{Q}_{\mathrm{D}}&=&\int_{\mathbb{R}^4}2\mathrm{d}^4x\,\delta\left(x_{\lambda}\,x^{\lambda}-\tau^2\right)\theta\left(x^0\right)x^{\mu}:\hat{\bar{\psi}}\left(x\right)\gamma_{\mu}\hat{\psi}\left(x\right):
\nonumber\\&=&\sum_{\rho,\sigma=\pm\frac12}\int_{\mathbb{R}^3}
\frac{\mathrm{d}^3p}{2p_0}\int_{\mathbb{R}^3}\frac{\mathrm{d}^3q}{2q_0}\frac{2}{\left(2\pi\right)^3}\int_{\mathbb{R}^4}\mathrm{d}^4x\,\delta\left(x_{\lambda}\,x^{\lambda}-\tau^2\right)\theta\left(x^0\right)x^{\mu}\nonumber\\&&
\times:\left(\mathrm{e}^{i\left(p-q\right)_{\nu}\,
x^{\nu}}\bar{u}_{\rho}\left(\boldsymbol{p}\right)\gamma_{\mu}u_{\sigma}\left(\boldsymbol{q}\right)\hat{c}_{\rho}^{\dag}\left(\boldsymbol{p}\right)\hat{c}_{\sigma}\left(\boldsymbol{q}\right)\right.
\nonumber\\&&\quad+ \left.\mathrm{e}^{-i\left(p-q\right)_{\nu}\,
x^{\nu}}\bar{v}_{\rho}\left(\boldsymbol{p}\right)\gamma_{\mu}v_{\sigma}\left(\boldsymbol{q}\right)\hat{d}_{\rho}\left(\boldsymbol{p}\right)\hat{d}^{\dag}_{\sigma}\left(\boldsymbol{q}\right)\right.\nonumber\\&&\quad+\left.\mathrm{e}^{i\left(p+q\right)_{\nu}\,
x^{\nu}}\bar{u}_{\rho}\left(\boldsymbol{p}\right)\gamma_{\mu}v_{\sigma}\left(\boldsymbol{q}\right)\hat{c}_{\rho}^{\dag}\left(\boldsymbol{p}\right)\hat{d}^{\dag}_{\sigma}\left(\boldsymbol{q}\right)\right.
\nonumber\\&&\quad+ \left.\mathrm{e}^{-i\left(p+q\right)_{\nu}\,
x^{\nu}}\bar{v}_{\rho}\left(\boldsymbol{p}\right)\gamma_{\mu}u_{\sigma}\left(\boldsymbol{q}\right)\hat{d}_{\rho}\left(\boldsymbol{p}\right)\hat{c}_{\sigma}\left(\boldsymbol{q}\right)\right):
\nonumber\\&=&\sum_{\rho,\sigma=\pm\frac12} \int_{\mathbb{R}^3}
\frac{\mathrm{d}^3p}{2p_0}\int_{\mathbb{R}^3}\frac{\mathrm{d}^3q}{2q_0}
\nonumber\\&&\times:\left[W^{\mu}\left(P,Q\right)\left(\bar{u}_{\rho}\left(\boldsymbol{p}\right)\gamma_{\mu}u_{\sigma}\left(\boldsymbol{q}\right)\hat{c}_{\rho}^{\dag}\left(\boldsymbol{p}\right)\hat{c}_{\sigma}\left(\boldsymbol{q}\right)+
\bar{v}_{\rho}\left(\boldsymbol{p}\right)\gamma_{\mu}v_{\sigma}\left(\boldsymbol{q}\right)\hat{d}_{\rho}\left(\boldsymbol{p}\right)\hat{d}^{\dag}_{\sigma}\left(\boldsymbol{q}\right)\right)\right.
\nonumber\\&&\quad+
\left.W^{\mu}\left(Q,P\right)\left(\bar{u}_{\rho}\left(\boldsymbol{p}\right)\gamma_{\mu}v_{\sigma}\left(\boldsymbol{q}\right)\hat{c}_{\rho}^{\dag}\left(\boldsymbol{p}\right)\hat{d}^{\dag}_{\sigma}\left(\boldsymbol{q}\right)+
\bar{v}_{\rho}\left(\boldsymbol{p}\right)\gamma_{\mu}u_{\sigma}\left(\boldsymbol{q}\right)\hat{d}_{\rho}\left(\boldsymbol{p}\right)\hat{c}_{\sigma}\left(\boldsymbol{q}\right)\right)
\right]: \nonumber\end{eqnarray}\\\begin{eqnarray}&=&
\sum_{\rho,\sigma=\pm\frac12} \int_{\mathbb{R}^3}
\frac{\mathrm{d}^3p}{2p_0}\int_{\mathbb{R}^3}\frac{\mathrm{d}^3q}{2q_0}
\nonumber\\&&\times:\left[
\frac{\left(p+q\right)^{\mu}\left(p+q\right)^0}{\left(p+q\right)_{\lambda}\,\left(p+q\right)^{\lambda}}\,\delta^3\left(\boldsymbol{p}-\boldsymbol{q}\right)\right.\nonumber\\&&\qquad\times\left.\left(\bar{u}_{\rho}\left(\boldsymbol{p}\right)\gamma_{\mu}u_{\sigma}\left(\boldsymbol{q}\right)\hat{c}_{\rho}^{\dag}\left(\boldsymbol{p}\right)\hat{c}_{\sigma}\left(\boldsymbol{q}\right)+
\bar{v}_{\rho}\left(\boldsymbol{p}\right)\gamma_{\mu}v_{\sigma}\left(\boldsymbol{q}\right)\hat{d}_{\rho}\left(\boldsymbol{p}\right)\hat{d}^{\dag}_{\sigma}\left(\boldsymbol{q}\right)\right)
\right. \nonumber\\&&\quad+ \left.
\frac{\left(p-q\right)^{\mu}}{Q_{\lambda}\,Q^{\lambda}}W\left(Q,Q\right)\right.\nonumber\\&&\qquad\times\left.\left(\bar{u}_{\rho}\left(\boldsymbol{p}\right)\gamma_{\mu}u_{\sigma}\left(\boldsymbol{q}\right)\hat{c}_{\rho}^{\dag}\left(\boldsymbol{p}\right)\hat{c}_{\sigma}\left(\boldsymbol{q}\right)+
\bar{v}_{\rho}\left(\boldsymbol{p}\right)\gamma_{\mu}v_{\sigma}\left(\boldsymbol{q}\right)\hat{d}_{\rho}\left(\boldsymbol{p}\right)\hat{d}^{\dag}_{\sigma}\left(\boldsymbol{q}\right)\right)
\right. \nonumber\\&&\quad+ \left.
\frac{\left(p-q\right)^{\mu}}{P_{\lambda}\,P^{\lambda}}W\left(P,P\right)\right.\nonumber\\&&\qquad\times\left.\left(\bar{u}_{\rho}\left(\boldsymbol{p}\right)\gamma_{\mu}v_{\sigma}\left(\boldsymbol{q}\right)\hat{c}_{\rho}^{\dag}\left(\boldsymbol{p}\right)\hat{d}^{\dag}_{\sigma}\left(\boldsymbol{q}\right)+
\bar{v}_{\rho}\left(\boldsymbol{p}\right)\gamma_{\mu}u_{\sigma}\left(\boldsymbol{q}\right)\hat{d}_{\rho}\left(\boldsymbol{p}\right)\hat{c}_{\sigma}\left(\boldsymbol{q}\right)\right)
\right]:\nonumber\\&=& \sum_{\rho=\pm\frac12}
 \int_{\mathbb{R}^3}
\frac{\mathrm{d}^3p}{2p_0}\left(\hat{c}_{\rho}^{\dag}\left(\boldsymbol{p}\right)\hat{c}_{\rho}\left(\boldsymbol{p}\right)-\hat{d}^{\dag}_{\rho}\left(\boldsymbol{p}\right)\hat{d}_{\rho}\left(\boldsymbol{p}\right)\right)
,
\end{eqnarray} where we have used the Dirac equations of the
spinors,~(\ref{eq:mde1}),~(\ref{eq:mde2}),~(\ref{eq:amde1})
and~(\ref{eq:amde2}).$\quad\bullet$
\subsection{Translation Generator}
\label{app:PmuD}
We want to calculate $\hat{P}_{\mathrm{D}}^{\mu}$ in Wigner basis:\\
$\circ\quad$From~(\ref{eq:PmuD}) we have
\begin{eqnarray}\hat{P}_{\mathrm{D}}^{\mu}&=&\frac{i}{2}
\int_{\mathbb{R}^4}2\mathrm{d}^4x\,\delta\left(x_{\lambda}\,x^{\lambda}-\tau^2\right)\theta\left(x^0\right)x^{\nu}\nonumber\\
&&\times:\left[\hat{\bar{\psi}}\left(x\right)\gamma_{\nu}\left(\partial^{\mu}\hat{\psi}\left(x\right)\right)-
\left(\partial^{\mu}\hat{\bar{\psi}}\left(x\right)\right)\gamma_{\nu}\hat{\psi}\left(x\right)
\right]:\nonumber\\ &=&
\frac{1}{2\left(2\pi\right)^3}\sum_{\rho,\sigma=\pm\frac12}\int_{\mathbb{R}^3}\frac{\mathrm{d}^3p}{2p_0}\int_{\mathbb{R}^3}\frac{\mathrm{d}^3q}{2q_0}
\int_{\mathbb{R}^4}2\mathrm{d}^4x\,\delta\left(x_{\lambda}\,x^{\lambda}-\tau^2\right)\theta\left(x^0\right)x^{\nu}\nonumber\\
&&\times :\left[
\left(p+q\right)^{\mu}\left(\mathrm{e}^{+ix_{\lambda}\left(p-q\right)^{\lambda}}\bar{u}_{\rho}\left(\boldsymbol{p}\right)\gamma_{\nu}u_{\sigma}\left(\boldsymbol{q}\right)\hat{c}_{\rho}^{\dag}\left(\boldsymbol{p}\right)\hat{c}_{\sigma}\left(\boldsymbol{q}\right)\right.\right.\nonumber\\&&\qquad-\left.\left.\mathrm{e}^{-ix_{\lambda}\,\left(p-q\right)^{\lambda}}\bar{v}_{\rho}\left(\boldsymbol{p}\right)\gamma_{\nu}v_{\sigma}\left(\boldsymbol{q}\right)\hat{d}_{\rho}\left(\boldsymbol{p}\right)\hat{d}_{\sigma}^{\dag}\left(\boldsymbol{q}\right)\right)\right.\nonumber
\\
&&\quad\left.+\left(p-q\right)^{\mu}\left(\mathrm{e}^{ix_{\lambda}\,\left(p+q\right)^{\lambda}}\bar{u}_{\rho}\left(\boldsymbol{p}\right)\gamma_{\nu}v_{\sigma}\left(\boldsymbol{q}\right)\hat{c}_{\rho}^{\dag}\left(\boldsymbol{p}\right)\hat{d}_{\sigma}^{\dag}\left(\boldsymbol{q}\right)\right.\right.\nonumber\\&&\qquad-\left.\left.\mathrm{e}^{-ix_{\lambda}\left(p+q\right)^{\lambda}}\bar{v}_{\rho}\left(\boldsymbol{p}\right)\gamma_{\nu}u_{\sigma}\left(\boldsymbol{q}\right)\hat{d}_{\rho}\left(\boldsymbol{p}\right)\hat{c}_{\sigma}\left(\boldsymbol{q}\right)\right)\right]\nonumber
\nonumber\end{eqnarray}\\\begin{eqnarray}&=&
\frac{1}{2}\sum_{\rho,\sigma=\pm\frac12}\int_{\mathbb{R}^3}\frac{\mathrm{d}^3p}{2p_0}\int_{\mathbb{R}^3}\frac{\mathrm{d}^3q}{2q_0}\nonumber\\
&&\times :\left[
P^{\mu}\left(W^{\nu}\left(P,Q\right)\bar{u}_{\rho}\left(\boldsymbol{p}\right)\gamma_{\nu}u_{\sigma}\left(\boldsymbol{q}\right)\hat{c}_{\rho}^{\dag}\left(\boldsymbol{p}\right)\hat{c}_{\sigma}\left(\boldsymbol{q}\right)\right.\right.\nonumber\\&&\qquad-\left.\left.W^{\nu}\left(P,-Q\right)\bar{v}_{\rho}\left(\boldsymbol{p}\right)\gamma_{\nu}v_{\sigma}\left(\boldsymbol{q}\right)\hat{d}_{\rho}\left(\boldsymbol{p}\right)\hat{d}_{\sigma}^{\dag}\left(\boldsymbol{q}\right)\right)\right.\nonumber
\\
&&\quad\left.+Q^{\mu}\left(W^{\nu}\left(Q,P\right)\bar{u}_{\rho}\left(\boldsymbol{p}\right)\gamma_{\nu}v_{\sigma}\left(\boldsymbol{q}\right)\hat{c}_{\rho}^{\dag}\left(\boldsymbol{p}\right)\hat{d}_{\sigma}^{\dag}\left(\boldsymbol{q}\right)\right.\right.\nonumber\\&&\qquad-\left.\left.W^{\nu}\left(Q,-P\right)\bar{v}_{\rho}\left(\boldsymbol{p}\right)\gamma_{\nu}u_{\sigma}\left(\boldsymbol{q}\right)\hat{d}_{\rho}\left(\boldsymbol{p}\right)\hat{c}_{\sigma}\left(\boldsymbol{q}\right)\right)\right]:\nonumber
\nonumber\\&=&
\frac{1}{2}\sum_{\rho,\sigma=\pm\frac12}\int_{\mathbb{R}^3}\frac{\mathrm{d}^3p}{2p_0}\int_{\mathbb{R}^3}\frac{\mathrm{d}^3q}{2q_0}\nonumber\\
&&\times :\left[
P^{\mu}\left(\frac{\left(p+q\right)^{\nu}\,\left(p+q\right)^0}{\left(p+q\right)_{\lambda}\,\left(p+q\right)^{\lambda}}
\delta^3\left(\boldsymbol{p}-\boldsymbol{q}\right)\bar{u}_{\rho}\left(\boldsymbol{p}\right)\gamma_{\nu}
u_{\sigma}\left(\boldsymbol{q}\right)\hat{c}_{\rho}^{\dag}\left(\boldsymbol{p}\right)
\hat{c}_{\sigma}\left(\boldsymbol{q}\right)\right.\right.\nonumber\\
&&\quad+\left.\left.
\frac{\left(p-q\right)^{\nu}}{Q_{\lambda}\,Q^{\lambda}}W\left(Q,Q\right)\bar{u}_{\rho}\left(\boldsymbol{p}\right)\gamma_{\nu}u_{\sigma}\left(\boldsymbol{q}\right)\hat{c}_{\rho}^{\dag}\left(\boldsymbol{p}\right)\hat{c}_{\sigma}\left(\boldsymbol{q}\right)\right.\right.\nonumber\\
&&\quad-\left.\left.
\frac{\left(p+q\right)^{\nu}\,\left(p+q\right)^0}{\left(p+q\right)_{\lambda}\,\left(p+q\right)^{\lambda}}
\delta^3\left(\boldsymbol{p}-\boldsymbol{q}\right)\bar{v}_{\rho}\left(\boldsymbol{p}\right)\gamma_{\nu}v_{\sigma}\left(\boldsymbol{q}\right)\hat{d}_{\rho}\left(\boldsymbol{p}\right)\hat{d}_{\sigma}^{\dag}\left(\boldsymbol{q}\right)\right.\right.\nonumber
\\
&&\quad-\left.\left.\frac{\left(p-q\right)^{\nu}}{Q_{\lambda}\,Q^{\lambda}}
W\left(Q,-Q\right)\bar{v}_{\rho}\left(\boldsymbol{p}\right)\gamma_{\nu}v_{\sigma}\left(\boldsymbol{q}\right)\hat{d}_{\rho}\left(\boldsymbol{p}\right)\hat{d}_{\sigma}^{\dag}\left(\boldsymbol{q}\right)\right)\right.\nonumber
\\&&\quad
\left.+Q^{\mu}\left(\frac{\left(p+q\right)^{\nu}}{P_{\lambda}\,P^{\lambda}}W\left(P,P\right)\bar{u}_{\rho}\left(\boldsymbol{p}\right)\gamma_{\nu}v_{\sigma}\left(\boldsymbol{q}\right)\hat{c}_{\rho}^{\dag}\left(\boldsymbol{p}\right)\hat{d}_{\sigma}^{\dag}\left(\boldsymbol{q}\right)\right.\right.\nonumber\\
&&\quad-\left.\left.\frac{\left(p+q\right)^{\nu}}{P_{\lambda}\,P^{\lambda}}W\left(P,-P\right)\bar{v}_{\rho}\left(\boldsymbol{p}\right)\gamma_{\nu}u_{\sigma}\left(\boldsymbol{q}\right)\hat{d}_{\rho}\left(\boldsymbol{p}\right)\hat{c}_{\sigma}\left(\boldsymbol{q}\right)\right)\right]:\nonumber
\nonumber\\&=& \sum_{\rho=\pm\frac12}
\int_{\mathbb{R}^3}\frac{\mathrm{d}^3p}{2p^0}p^{\mu}\left(\hat{c}_{\rho}^{\dag}\left(\boldsymbol{p}\right)\hat{c}_{\rho}\left(\boldsymbol{p}\right)+\hat{d}_{\rho}^{\dag}\left(\boldsymbol{p}\right)\hat{d}_{\rho}\left(\boldsymbol{p}\right)\right),
\end{eqnarray} where we have used~(\ref{eq:ugammau}) and the Dirac equations in momentum space for the
spinors,~(\ref{eq:mde1}),~(\ref{eq:mde2}),~(\ref{eq:amde1})
and~(\ref{eq:amde2}).$\quad\bullet$
\chapter{Dyson Expansion}
\label{app:st} The objective of this appendix is to show order by
order that the Dyson expansions for the S operator in the usual
instant-form~(\ref{eq:Des2}) and in the generalized point-form
formulation~(\ref{eq:Des1}) are equivalent.
\section{First Order}
We initially show that the second term of~(\ref{eq:Des1}) is
equivalent to
the second term of~(\ref{eq:Des2}).\\
 $\circ\quad$
Since $k$ is a timelike vector, it can be written as a Lorentz-boosted unit vector in $x^0$-direction,
\begin{eqnarray}
k=\Lambda\left(v\right)\tilde{k},\quad\text{with}\quad\tilde{k}=\left(%
\begin{array}{c}
  1 \\
  0 \\
0\\0
\end{array}
\right).%
\end{eqnarray}
with \lq\lq\,\,$\widetilde{\,}$\,\,\rq\rq\,denoting \lq\lq boosted with
$\Lambda^{-1}\left(v\right)$\rq\rq. To simplify notation in the
following, we will write $\Lambda\left(v\right)$ as $\Lambda$. The
second term of~(\ref{eq:Des1}) is
\begin{eqnarray}&&2\int_{-\infty}^{\infty}\mathrm{d}s\, k_{\mu} \int_{\mathbb{R}^4}\mathrm{d}^4x\,\delta\left(x_{\lambda}\,
x^{\lambda}-\tau^2\right)\theta\left(x^0\right)x^{\mu}:\hat{\mathcal
{L}}_{\mathrm{int}}\left(x+s\,k+a\right):\nonumber\\&=&\int_{-\infty}^{\infty}\mathrm{d}s
\int_{\mathbb{R}^4}\mathrm{d}^4\tilde{x}\,\delta\left(\tilde{x}_{\lambda}\,\tilde{x}^{\lambda}-\tau^2\right)
\theta\left(\tilde{x}^0\right)\tilde{x}^{0}:\hat{\mathcal
{L}}_{\mathrm{int}}\left(\Lambda\left(\tilde{x}+s\,\tilde{k}+\tilde{a}\right)\right):\nonumber\\&=&
\int_{-\infty}^{\infty}\mathrm{d}s\int_{\mathbb{R}^3}\mathrm{d}^3\tilde{x}:\hat{\mathcal
{L}}_{\mathrm{int}}\left(\Lambda\left(\tilde{x}+s\,\tilde{k}+\tilde{a}\right)\right):,
\end{eqnarray}
where we have used Lorentz invariance of the hypersurface element.
We define now a new variable $z$ as
\begin{eqnarray}\label{eq:z}
z:=\left(%
\begin{array}{c}
  \sqrt{\boldsymbol{\tilde{x}}^2+\tau^2} +s+\tilde{a}^0\\
   \boldsymbol{\tilde{x}}+\tilde{\boldsymbol{a}}\\
\end{array}
\right).
\end{eqnarray} The invariant volume element transforms as
\begin{eqnarray}
\mathrm{d}^4z=
\left\vert\frac{\partial\left(z^0,\boldsymbol{z}\right)}{\partial\left(s,\boldsymbol{\tilde{x}}\right)}
\right\vert \mathrm{d}s\, \mathrm{d}^3\tilde{x},
\end{eqnarray} with the Jacobian determinant being
\begin{eqnarray}
\frac{\partial\left(z^0,\boldsymbol{z}\right)}{\partial\left(s,\boldsymbol{\tilde{x}}\right)}=\left\vert%
\begin{array}{cccc}
1&\frac{\tilde{x}^1}{\sqrt{\boldsymbol{\tilde{x}}^2+\tau^2}}&\frac{\tilde{x}^2}{\sqrt{\boldsymbol{\tilde{x}}^2+\tau^2}}&
\frac{\tilde{x}^3}{\sqrt{\boldsymbol{\tilde{x}}^2+\tau^2}}\\
  0&1&0&0\\
0&0&1&0\\
0&0&0&1\\
\end{array}
\right\vert%
=1.
\end{eqnarray}
Thus, we obtain \begin{eqnarray}
\int_{-\infty}^{\infty}\mathrm{d}s\int_{\mathbb{R}^3}\mathrm{d}^3\tilde{x}:\hat{\mathcal
{L}}_{\mathrm{int}}\left(\Lambda\,\left(\tilde{x}+s\,\tilde{k}+\tilde{a}\right)\right):=\int_{\mathbb{R}^4}\mathrm{d}^4z:\hat{\mathcal
{L}}_{\mathrm{int}}\left(\Lambda\,
z\right):=\int_{\mathbb{R}^4}\mathrm{d}^4z:\hat{\mathcal
{L}}_{\mathrm{int}}\left(z\right):.\nonumber\\\end{eqnarray} Here,
we have again used Lorentz invariance of the volume element and
that $\hat{\mathcal {L}}_{\mathrm{int}}$ transforms like a Lorentz
scalar~(\ref{eq:scalarfield}). This result is equivalent to the
first order instant-form Dyson expansion of the S
operator~(\ref{eq:Des2}).$\quad\bullet$
\section{Second Order}
$\circ\quad$ For higher orders we have to take $s$-ordering into
account. The second order contribution to the S
operator~(\ref{eq:Des1}) reads
\begin{eqnarray}
&&2^2\int_{-\infty}^{\infty}\mathrm{d}s_1\,
k_{\mu}\int_{-\infty}^{\infty}\mathrm{d}s_2\,
k_{\nu}\,\int_{\mathbb{R}^4}\mathrm{d}^4x_1\,\delta\left(x_{1\lambda}\,x_1^{\lambda}-\tau^2\right)\theta\left(x_1^0\right)x_1^{\mu}\nonumber\\&&\times\int_{\mathbb{R}^4}\mathrm{d}^4x_2\,\delta\left(x_{2\lambda}\,x_2^{\lambda}-\tau^2\right)\theta\left(x_2^0\right)x_2^{\nu}\,\mathcal{S}\left[:\hat{\mathcal
{L}}_{\mathrm{int}}\left(x_1+s_1\,k+a\right)::\hat{\mathcal
{L}}_{\mathrm{int}}\left(x_2+s_2\,k+a\right):\right]
\nonumber\\&&= 2^2\int_{-\infty}^{\infty}\mathrm{d}s_1\,
k_{\mu}\int_{-\infty}^{\infty}\mathrm{d}s_2\,
k_{\nu}\,\nonumber\\&&\quad\times\int_{\mathbb{R}^4}\mathrm{d}^4x_1\,\delta\left(x_{1\lambda}\,x_1^{\lambda}-\tau^2\right)\theta\left(x_1^0\right)x_1^{\mu}\int_{\mathbb{R}^4}\mathrm{d}^4x_2\,\delta\left(x_{2\lambda}\,x_2^{\lambda}-\tau^2\right)\theta\left(x_2^0\right)x_2^{\nu}\nonumber\\&&\quad\times
\left[\theta\left(s_1-s_2\right):\hat{\mathcal
{L}}_{\mathrm{int}}\left(x_1+s_1\,k+a\right)::\hat{\mathcal
{L}}_{\mathrm{int}}\left(x_2+s_2\,k+a\right):\right.\nonumber\\&&\qquad+\left.\theta\left(s_2-s_1\right):\hat{\mathcal
{L}}_{\mathrm{int}}\left(x_2+s_2\,k+a\right)::\hat{\mathcal
{L}}_{\mathrm{int}}\left(x_1+s_1\,k+a\right):\right]\nonumber\\&&=
\int_{-\infty}^{\infty}\mathrm{d}s_1\int_{-\infty}^{\infty}\mathrm{d}s_2\int_{\mathbb{R}^3}\mathrm{d}^3\tilde{x}_1
\int_{\mathbb{R}^3}\mathrm{d}^3\tilde{x}_2\nonumber\\&&\quad\times\left[\theta\left(s_1-s_2\right):\hat{\mathcal
{L}}_{\mathrm{int}}\left(\Lambda\,\left(\tilde{x}_1+s_1\,\tilde{k}+\tilde{a}\right)\right)::\hat{\mathcal
{L}}_{\mathrm{int}}\left(\Lambda\,\left(\tilde{x}_2+s_2\,\tilde{k}+\tilde{a}\right)\right):\right.\nonumber\\&&\qquad+\left.\theta\left(s_2-s_1\right):\hat{\mathcal
{L}}_{\mathrm{int}}\left(\Lambda\,\left(\tilde{x}_2+s_2\,\tilde{k}+\tilde{a}\right)\right)::\hat{\mathcal{L}}_{\mathrm{int}}\left(\Lambda\,\left(\tilde{x}_1+s_1\,\tilde{k}+\tilde{a}\right)\right):\right]\label{eq:2ndo}
\end{eqnarray}
where we have used Lorentz invariance as before. Introducing again
new variables $z_1,z_2$ given by the transformation~(\ref{eq:z})
with $ (s_i,\boldsymbol{x}_i)\longrightarrow z_i,\,i=1,2$ and with
the abbreviation
$d\left(\boldsymbol{z}_1,\boldsymbol{z}_2\right):=
\sqrt{\left(\boldsymbol{z}_1-\tilde{\boldsymbol{a}}\right)^2+\tau^2}-\sqrt{\left(\boldsymbol{z}_2-\tilde{\boldsymbol{a}}\right)^2+\tau^2}$
we obtain for~(\ref{eq:2ndo})
\begin{eqnarray}&&
\int_{\mathbb{R}^4}\mathrm{d}^4z_1
\int_{\mathbb{R}^4}\mathrm{d}^4z_2\left[\theta\left(z^0_1-z_2^0-d\left(\boldsymbol{z}_1,\boldsymbol{z}_2\right)\right):\hat{\mathcal
{L}}_{\mathrm{int}}\left(\Lambda\,z_1\right)::\hat{\mathcal
{L}}_{\mathrm{int}}\left(\Lambda\,z_2\right):\right.\nonumber\\&&\quad+\left.\theta\left(z_2^0-z^0_1+d\left(\boldsymbol{z}_1,\boldsymbol{z}_2\right)\right):\hat{\mathcal
{L}}_{\mathrm{int}}\left(\Lambda\,z_2\right)::\hat{\mathcal
{L}}_{\mathrm{int}}\left(\Lambda\,z_1\right):\right]
\nonumber\end{eqnarray}\\\begin{eqnarray}&&=
\int_{\mathbb{R}^4}\mathrm{d}^4z_1
\int_{\mathbb{R}^3}\mathrm{d}^3z_2
\left[\int_{-\infty}^{z^0_1-d\left(\boldsymbol{z}_1,\boldsymbol{z}_2\right)}\mathrm{d}z_2^0:\hat{\mathcal
{L}}_{\mathrm{int}}\left(\Lambda\,z_1\right)::\hat{\mathcal
{L}}_{\mathrm{int}}\left(\Lambda\,z_2\right):\right.\nonumber\\&&\quad+\left.\int_{z^0_1-d\left(\boldsymbol{z}_1,\boldsymbol{z}_2\right)}^{\infty}\mathrm{d}z_2^0:\hat{\mathcal
{L}}_{\mathrm{int}}\left(\Lambda\,z_2\right)::\hat{\mathcal
{L}}_{\mathrm{int}}\left(\Lambda\,z_1\right):\right]\nonumber\\&&=
\int_{\mathbb{R}^4}\mathrm{d}^4z_1
\int_{\mathbb{R}^3}\mathrm{d}^3z_2
\left[\int_{-\infty}^{z^0_1}\mathrm{d}z_2^0:\hat{\mathcal
{L}}_{\mathrm{int}}\left(\Lambda\,z_1\right)::\hat{\mathcal
{L}}_{\mathrm{int}}\left(\Lambda\,z_2\right):\right.\nonumber\\&&\quad+\left.\int_{z^0_1}^{\infty}\mathrm{d}z_2^0:\hat{\mathcal
{L}}_{\mathrm{int}}\left(\Lambda\,z_2\right)::\hat{\mathcal
{L}}_{\mathrm{int}}\left(\Lambda\,z_1\right):+\int_{z^0_1}^{z^0_1-d\left(\boldsymbol{z}_1,
\boldsymbol{z}_2\right)}\mathrm{d}z_2^0:\hat{\mathcal
{L}}_{\mathrm{int}}\left(\Lambda\,z_1\right)::\hat{\mathcal
{L}}_{\mathrm{int}}\left(\Lambda\,z_2\right):\right.\nonumber\\&&\quad+\left.\int_{z^0_1-d\left(\boldsymbol{z}_1,\boldsymbol{z}_2\right)}^{z^0_1}
\mathrm{d}z_2^0:\hat{\mathcal
{L}}_{\mathrm{int}}\left(\Lambda\,z_2\right)::\hat{\mathcal
{L}}_{\mathrm{int}}\left(\Lambda\,z_1\right):\right].\nonumber\\\label{eq:2ndo1}
\end{eqnarray}
The last 2 terms cancel out as follows: At the integration limit
$z_2^0=z^0_1$ of the $z_2^0$-integration we see that $z_1$ and
$z_2$ are spacelike separated,
\begin{eqnarray}
\left(z_1-z_2\right)_{\lambda}\,\left(z_1-z_2\right)^{\lambda}\left\vert_{z_2^0=z^0_1}\right.=-\left(\boldsymbol{z}_1-\boldsymbol{z}_2\right)^2<0.\label{eq:cancelout2}
\end{eqnarray}
This holds as well for the other integration limit
$z_2^0=z^0_1-d\left(\boldsymbol{z}_1,\boldsymbol{z}_2\right)$,
which can be seen as follows: We have to show that
\begin{eqnarray}
\left(z_1-z_2\right)^2\left\vert_{z_2^0=z^0_1-d\left(\boldsymbol{z}_1,\boldsymbol{z}_2\right)}\right.&<&0;\nonumber\\
d^2\left(\boldsymbol{z}_1,\boldsymbol{z}_2\right)-\left(\boldsymbol{z}_1-\boldsymbol{z}_2\right)^2=&&\nonumber\\
2\tau^2-2\sqrt{\left(\boldsymbol{z}_1-\tilde{\boldsymbol{a}}\right)^2+\tau^2}\sqrt{\left(\boldsymbol{z}_2-\tilde{\boldsymbol{a}}\right)^2+\tau^2}+
2\left(\boldsymbol{z}_1-\tilde{\boldsymbol{a}}\right)\cdot\left(\boldsymbol{z}_2-\tilde{\boldsymbol{a}}\right)&<&0.\nonumber\\
\end{eqnarray}
By bringing the square root on the other side we have on both sides positive values, thus squaring gives
\begin{eqnarray}
&&\tau^4+2\tau^2\left(\boldsymbol{z}_1-\tilde{\boldsymbol{a}}\right)\cdot\left(\boldsymbol{z}_2-\tilde{\boldsymbol{a}}\right)+
\left(\boldsymbol{z}_1-\tilde{\boldsymbol{a}}\right)^2\left(\boldsymbol{z}_2-\tilde{\boldsymbol{a}}\right)^2\mathrm{cos}^2\left(\boldsymbol{z}_1,\boldsymbol{z}_2\right)\nonumber\\&&<
\left(\boldsymbol{z}_1-\tilde{\boldsymbol{a}}\right)^2\left(\boldsymbol{z}_2-\tilde{\boldsymbol{a}}\right)^2+\tau^4+
\tau^2\left[\left(\boldsymbol{z}_1-\tilde{\boldsymbol{a}}\right)^2+\left(\boldsymbol{z}_2-\tilde{\boldsymbol{a}}\right)^2\right]\nonumber\\
&&\Longrightarrow
\left(\boldsymbol{z}_1-\tilde{\boldsymbol{a}}\right)^2\left(\boldsymbol{z}_2-\tilde{\boldsymbol{a}}\right)^2\left(\mathrm{cos}^2\left(\boldsymbol{z}_1,\boldsymbol{z}_2\right)-1\right)<
\tau^2\left(\boldsymbol{z}_1-\boldsymbol{z}_2\right)^2.
\end{eqnarray} This is always satisfied.
What is left to show is that $z_1$ and $z_2$ are spacelike
separated also between these integration limits. The function
\begin{eqnarray}
f\left(z_2^0\right)=\left(z_1-z_2\right)_{\lambda}\,\left(z_1-z_2\right)^{\lambda}=\left(z^0_1-z_2^0\right)^2-\left(\boldsymbol{z}_1-\boldsymbol{z}_2\right)^2
\end{eqnarray} with fixed
$z^0_1,\boldsymbol{z}_1,\boldsymbol{z}_2$
has only one minimum at $z_2^0=z^0_1$.
Since
\begin{eqnarray}
f\left(z^0_1\right)<0\quad \wedge \quad f\left(z^0_1-d\left(\boldsymbol{z}_1,\boldsymbol{z}_2\right)\right)<0\\
\Rightarrow f\left(z_2^0\right)<0, \quad\forall\,
z_2^0\in\left[z^0_1,z^0_1-d\left(\boldsymbol{z}_1,\boldsymbol{z}_2\right)\right],
\end{eqnarray} we have shown that $z_1$ and $z_2$ have spacelike separation over the whole integration interval.
This is obvious, since $f\left(z_2^0\right)$ becomes zero at
$z_2^0=z^0_1\pm\left\vert\boldsymbol{z}_1-\boldsymbol{z}_2\right\vert$,
which is outside the integration interval. In addition we find
the following:
\begin{eqnarray}
\left\vert\boldsymbol{z}_1-\boldsymbol{z}_2\right\vert&>&\left\vert d\left(\boldsymbol{z}_1,\boldsymbol{z}_2\right)\right\vert;\nonumber\\
\sqrt{\left(\boldsymbol{z}_1-\tilde{\boldsymbol{a}}\right)^2+\tau^2}\sqrt{\left(\boldsymbol{z}_2-\tilde{\boldsymbol{a}}\right)^2+\tau^2}&>&
\tau^2+\left(\boldsymbol{z}_1-\tilde{\boldsymbol{a}}\right)\cdot\left(\boldsymbol{z}_2-\tilde{\boldsymbol{a}}\right)\nonumber\\\Rightarrow
\left(\boldsymbol{z}_1-\tilde{\boldsymbol{a}}\right)^2\left(\boldsymbol{z}_2-\tilde{\boldsymbol{a}}\right)^2\left(1-\mathrm{cos}^2\left(\boldsymbol{z}_1,\boldsymbol{z}_2\right)\right)&>&-\tau^2\left(\boldsymbol{z}_1-\boldsymbol{z}_2\right)^2.
\end{eqnarray}
\\
In the considered integration interval, where $z_1$ and $z_2$ are
always separated by a spacelike distance, it follows
from~(\ref{eq:mc}) and~(\ref{eq:scalarfield}) that
\begin{eqnarray}
&&\hat{\mathcal
{L}}_{\mathrm{int}}\left(\Lambda\,z_2\right)\hat{\mathcal
{L}}_{\mathrm{int}}\left(\Lambda\,z_1\right)=\hat{U}\left(\Lambda\right)
\hat{\mathcal {L}}_{\mathrm{int}}\left(z_2\right)\hat{\mathcal
{L}}_{\mathrm{int}}\left(z_1\right)\hat{U}\left(\Lambda\right)^{-1}
\nonumber\\&&=\hat{U}\left(\Lambda\right) \hat{\mathcal
{L}}_{\mathrm{int}}\left(z_1\right)\hat{\mathcal
{L}}_{\mathrm{int}}\left(z_2\right)\hat{U}\left(\Lambda\right)^{-1}.\label{eq:cancelout}
\end{eqnarray} Therefore, the last 2 integrals
in~(\ref{eq:2ndo1}) cancel each other.
\\
Thus, for the remaining terms in~(\ref{eq:2ndo1}) we find the
usual second order contribution as
\begin{eqnarray}
&&\int_{\mathbb{R}^4}\mathrm{d}^4z_1
\int_{\mathbb{R}^3}\mathrm{d}^3z_2
\left[\int_{-\infty}^{z^0_1}\mathrm{d}z_2^0:\hat{\mathcal
{L}}_{\mathrm{int}}\left(\Lambda\,z_1\right)::\hat{\mathcal
{L}}_{\mathrm{int}}\left(\Lambda\,z_2\right):\right.\nonumber\\&&\quad+\left.\int_{z^0_1}^{\infty}\mathrm{d}z_2^0:\hat{\mathcal
{L}}_{\mathrm{int}}\left(\Lambda\,z_2\right)::\hat{\mathcal
{L}}_{\mathrm{int}}\left(\Lambda\,z_1\right)\right]\nonumber\\&&=
\int_{\mathbb{R}^4}\mathrm{d}^4z_1
\int_{\mathbb{R}^4}\mathrm{d}^4z_2
\left[\theta\left(z^0_1-z^0_2\right):\hat{\mathcal
{L}}_{\mathrm{int}}\left(\Lambda\,z_1\right)::\hat{\mathcal
{L}}_{\mathrm{int}}\left(\Lambda\,z_2\right):\right.\nonumber\\&&\quad+\left.\theta\left(z^0_2-z^0_1\right):\hat{\mathcal
{L}}_{\mathrm{int}}\left(\Lambda\,z_2\right)::\hat{\mathcal
{L}}_{\mathrm{int}}\left(\Lambda\,z_1\right):
\right]\nonumber\\&&= \int_{\mathbb{R}^4}\mathrm{d}^4z_1
\int_{\mathbb{R}^4}\mathrm{d}^4z_2
\left[\theta\left(z^0_1-z^0_2\right):\hat{\mathcal
{L}}_{\mathrm{int}}\left(z_1\right)::\hat{\mathcal
{L}}_{\mathrm{int}}\left(z_2\right):\right.\nonumber\\&&\quad+\left.\theta\left(z^0_2-z^0_1\right):\hat{\mathcal
{L}}_{\mathrm{int}}\left(z_2\right)::\hat{\mathcal
{L}}_{\mathrm{int}}\left(z_1\right):\right],
\end{eqnarray}
where we have used that
$\theta\left(z^0\right)=\theta\left(\Lambda\left(v\right)^{0}_{~\lambda}z^{\lambda}\right)$
is Lorentz invariant. A timelike component of a four-vector does
not change the sign under a continuous Lorentz transformation of
$\mathcal{L}_+^{\uparrow}$.$\quad\bullet$
\section{Third Order } $\circ\quad$ For the third order contribution
to the S operator in~(\ref{eq:De}) we have
\begin{eqnarray}
&&2^3\int_{-\infty}^{\infty}\mathrm{d}s_1\,
k_{\mu}\int_{-\infty}^{\infty}\mathrm{d}s_2\,
k_{\nu}\,\int_{-\infty}^{\infty}\mathrm{d}s_3\,
k_{\lambda}\nonumber\\&&\quad\times\int_{\mathbb{R}^4}\mathrm{d}^4x_1\,\delta\left(x_{1\lambda}\,x_1^{\lambda}-\tau^2\right)
\theta\left(x_1^0\right)x_1^{\mu}\int_{\mathbb{R}^4}\mathrm{d}^4x_2\,
\delta\left(x_{2\lambda}\,x_2^{\lambda}-\tau^2\right)\theta\left(x_2^0\right)x_2^{\nu}\nonumber\\&&\quad
\times\int_{\mathbb{R}^4}\mathrm{d}^4x_3\,\delta\left(x_{3\lambda}\,x_3^{\lambda}-\tau^2\right)
\theta\left(x_3^0\right)x_3^{\lambda}\nonumber\\&&\quad
\times\mathcal{S}\left[:\hat{\mathcal{L}}_{\mathrm{int}}\left(x_1+s_1\,k\right)::\hat{\mathcal{L}}_{\mathrm{int}}
\left(x_2+s_2\,k\right)::\hat{\mathcal{L}}_{\mathrm{int}}\left(x_3+s_3\,k\right):\right]
\nonumber\end{eqnarray}\\\begin{eqnarray}&&=
2^3\int_{-\infty}^{\infty}\mathrm{d}s_1\,
k_{\mu}\int_{-\infty}^{\infty}\mathrm{d}s_2\,
k_{\nu}\,\int_{-\infty}^{\infty}\mathrm{d}s_3\,
k_{\lambda}\nonumber\\&&\quad\times\int_{\mathbb{R}^4}\mathrm{d}^4x_1\,\delta\left(x_{1\lambda}\,x_1^{\lambda}-\tau^2\right)
\theta\left(x_1^0\right)x_1^{\mu}\int_{\mathbb{R}^4}\mathrm{d}^4x_2\,\delta\left(x_{2\lambda}\,x_2^{\lambda}-\tau^2\right)
\theta\left(x_2^0\right)x_2^{\nu}\nonumber\\&&\quad\times\int_{\mathbb{R}^4}
\mathrm{d}^4x_3\,\delta\left(x_{3\lambda}\,x_3^{\lambda}-\tau^2\right)\theta\left(x_3^0\right)x_3^{\lambda}\nonumber\\&&\quad
\times
\left[\theta\left(s_1-s_2\right)\theta\left(s_2-s_3\right)\right.\nonumber\\
&&\qquad\times \left.:\hat{\mathcal{L}}_{\mathrm{int}}\left(x_1+s_1\,k+a\right)::
\hat{\mathcal{L}}_{\mathrm{int}}\left(x_2+s_2\,k+a\right)::\hat{\mathcal{L}}_{\mathrm{int}}\left(x_3+s_3\,k+a\right):
\right.\nonumber\\&&\quad+
\left.\theta\left(s_1-s_3\right)\theta\left(s_3-s_2\right)\right.\nonumber\\
&&\qquad\times \left.
:\hat{\mathcal
{L}}_{\mathrm{int}}\left(x_1+s_1\,k+a\right)::\hat{\mathcal{L}}_{\mathrm{int}}\left(x_3+s_3\,k+a\right)::
\hat{\mathcal{L}}_{\mathrm{int}}\left(x_2+s_2\,k+a\right):\right.\nonumber\\&&\quad+\left.
\theta\left(s_1-s_2\right)\theta\left(s_3-s_1\right)\right.\nonumber\\
&&\qquad\times \left.:\hat{\mathcal{L}}_{\mathrm{int}}\left(x_3+s_3\,k+a\right)::
\hat{\mathcal {L}}_{\mathrm{int}}\left(x_1+s_1\,k+a\right)
::\hat{\mathcal {L}}_{\mathrm{int}}\left(x_2+s_2\,k+a\right):
\right.\nonumber\\&&\quad+\left.
\theta\left(s_2-s_1\right)\theta\left(s_1-s_3\right)\right.\nonumber\\
&&\qquad\times \left.:\hat{\mathcal{L}}_{\mathrm{int}}\left(x_2+s_2\,k+a\right)::
\hat{\mathcal{L}}_{\mathrm{int}}\left(x_1+s_1\,k+a\right)
::\hat{\mathcal{L}}_{\mathrm{int}}\left(x_3+s_3\,k+a\right):\right.\nonumber\\&&\quad+\left.
\theta\left(s_2-s_3\right)\theta\left(s_3-s_1\right)\right.\nonumber\\
&&\qquad\times \left.
:\hat{\mathcal {L}}_{\mathrm{int}}\left(x_2+s_2\,k+a\right)::
\hat{\mathcal{L}}_{\mathrm{int}}\left(x_3+s_3\,k+a\right)::\hat{\mathcal{L}}_{\mathrm{int}}\left(x_1+s_1\,k+a\right)
: \right.\nonumber\\&&\quad+\left. \theta\left(s_2-s_1\right)
\theta\left(s_3-s_2\right)\right.\nonumber\\
&&\qquad\times \left.:\hat{\mathcal{L}}_{\mathrm{int}}\left(x_3+s_3\,k+a\right)
::\hat{\mathcal
{L}}_{\mathrm{int}}\left(x_2+s_2\,k+a\right)::\hat{\mathcal
{L}}_{\mathrm{int}}\left(x_1+s_1\,k+a\right): \right].\nonumber\\
\end{eqnarray}\\
Again making use of Lorentz invariance and introducing new
variables $z_1,z_2,z_3$ as before in~(\ref{eq:z}), we have (ignoring
the $\boldsymbol{z}$-integrations)
\begin{eqnarray}&&
\int_{-\infty}^{\infty}\mathrm{d}z^0_1\left\lbrace
\int_{-\infty}^{\infty}\mathrm{d}z_2^0\,
\theta\left(z^0_1-z_2^0-d\left(\boldsymbol{z}_1,\boldsymbol{z}_2\right)\right)\right.\nonumber\\
&&\quad\times\left.\left[
\int_{-\infty}^{\infty}\mathrm{d}z_3^0\left( \theta\left(z_2^0-z_3^0-d\left(\boldsymbol{z}_2,\boldsymbol{z}_3\right)\right):\hat{\mathcal {L}}_{\mathrm{int}}\left(\Lambda\,z_1\right)::\hat{\mathcal {L}}_{\mathrm{int}}\left(\Lambda\,z_2\right)::\hat{\mathcal {L}}_{\mathrm{int}}\left(\Lambda\,z_3\right):\right.\right.\right.\nonumber\\
&&\quad+\left.\left.\left.
\theta\left(z^0_1-z_3^0-d\left(\boldsymbol{z}_1,\boldsymbol{z}_3\right)\right)\theta\left(z_3^0-z_2^0+d
\left(\boldsymbol{z}_2,\boldsymbol{z}_3\right)\right):\hat{\mathcal {L}}_{\mathrm{int}}\left(\Lambda\,z_1\right)::\hat{\mathcal {L}}_{\mathrm{int}}\left(\Lambda\,z_3\right)::\hat{\mathcal {L}}_{\mathrm{int}}\left(\Lambda\,z_2\right):\right.\right.\right.\nonumber\\
&&\quad+\left.\left.\left.
\theta\left(z_3^0-z^0_1+d\left(\boldsymbol{z}_1,\boldsymbol{z}_3\right)\right):\hat{\mathcal{L}}_{\mathrm{int}}
\left(\Lambda\,z_3\right)::\hat{\mathcal{L}}_{\mathrm{int}}\left(\Lambda\,z_1\right)::\hat{\mathcal{L}}_{\mathrm{int}}\left(\Lambda\,z_2\right):
\right)
\right]\right.\nonumber\\
&&\quad+\left.\int_{-\infty}^{\infty}\mathrm{d}z_2^0\,
\theta\left(z_2^0-z^0_1+d\left(\boldsymbol{z}_1,\boldsymbol{z}_2\right)\right)\right.\nonumber\\&&\quad\times\left.\left[
\int_{-\infty}^{\infty}\mathrm{d}z_3^0\left(
\theta\left(z^0_1-z_3^0-d\left(\boldsymbol{z}_1,\boldsymbol{z}_3\right)\right):\hat{\mathcal
{L}}_{\mathrm{int}}\left(z_2\right)::\hat{\mathcal{L}}_{\mathrm{int}}\left(\Lambda\,z_1\right)::
\hat{\mathcal{L}}_{\mathrm{int}}\left(\Lambda\,z_3\right):\right.\right.\right.\nonumber\\
&&\quad+\left.\left.\left. \theta\left(z_2^0-z_3^0-d\left(\boldsymbol{z}_2,\boldsymbol{z}_3\right)\right)
\theta\left(z_3^0-z^0_1+d\left(\boldsymbol{z}_1,\boldsymbol{z}_3\right)\right):
\hat{\mathcal{L}}_{\mathrm{int}}\left(\Lambda\,z_2\right)::\hat{\mathcal{L}}_{\mathrm{int}}\left(\Lambda\,z_3\right)::\hat{\mathcal{L}}_{\mathrm{int}}\left(\Lambda\,z_1\right):\right.\right.\right.\nonumber\\
&&\quad+\left.\left.\left.\theta\left(z_3^0-z_2^0+d\left(\boldsymbol{z}_2,\boldsymbol{z}_3\right)\right):
\hat{\mathcal{L}}_{\mathrm{int}}\left(\Lambda\,z_3\right)::\hat{\mathcal{L}}_{\mathrm{int}}\left(\Lambda\,z_2\right)::
\hat{\mathcal{L}}_{\mathrm{int}}\left(\Lambda\,z_1\right):
\right)\right]
\right\rbrace\nonumber\end{eqnarray}\\\begin{eqnarray}&&=
\int_{-\infty}^{\infty}\mathrm{d}z^0_1\left\lbrace
\int_{-\infty}^{z^0_1-d\left(\boldsymbol{z}_1,\boldsymbol{z}_2\right)}\mathrm{d}z_2^0
\left[\int_{-\infty}^{z_2^0-d\left(\boldsymbol{z}_2,\boldsymbol{z}_3\right)}
\mathrm{d}z_3^0:\hat{\mathcal{L}}_{\mathrm{int}}\left(\Lambda\,z_1\right)::
\hat{\mathcal{L}}_{\mathrm{int}}\left(\Lambda\,z_2\right)::
\hat{\mathcal{L}}_{\mathrm{int}}\left(\Lambda\,z_3\right):\right.\right.\nonumber\\
&&\quad+\left.\left.
\int_{z_2^0-d\left(\boldsymbol{z}_2,\boldsymbol{z}_3\right)}^{z^0_1
-d\left(\boldsymbol{z}_1,\boldsymbol{z}_3\right)}\mathrm{d}z_3^0:
\hat{\mathcal{L}}_{\mathrm{int}}\left(\Lambda\,z_1\right)::\hat{\mathcal{L}}_{\mathrm{int}}\left(\Lambda\,z_3\right)::
\hat{\mathcal{L}}_{\mathrm{int}}\left(\Lambda\,z_2\right):
\right.\right.\nonumber\\
&&\quad+\left.\left.\int_{z^0_1-d\left(\boldsymbol{z}_1,\boldsymbol{z}_3\right)}^{\infty}\mathrm{d}z_3^0:
\hat{\mathcal{L}}_{\mathrm{int}}\left(\Lambda\,z_3\right)::\hat{\mathcal{L}}_{\mathrm{int}}\left(\Lambda\,z_1\right)::
\hat{\mathcal{L}}_{\mathrm{int}}\left(\Lambda\,z_2\right):
\right]\right.\nonumber\\&&\quad+\left.
\int_{z^0_1-d\left(\boldsymbol{z}_1,\boldsymbol{z}_2\right)}^{\infty}\mathrm{d}z_2^0
\left[
\int_{-\infty}^{z^0_1-d\left(\boldsymbol{z}_1,\boldsymbol{z}_3\right)}\mathrm{d}z_3^0:
\hat{\mathcal{L}}_{\mathrm{int}}\left(\Lambda\,z_2\right)::
\hat{\mathcal{L}}_{\mathrm{int}}\left(\Lambda\,z_1\right)::\hat{\mathcal{L}}_{\mathrm{int}}\left(\Lambda\,z_3\right):\right.\right.\nonumber\\
&&\quad+\left.\left.
\int_{z^0_1-d\left(\boldsymbol{z}_1,\boldsymbol{z}_3\right)}^{z_2^0
-d\left(\boldsymbol{z}_2,\boldsymbol{z}_3\right)}\mathrm{d}z_3^0:\hat{\mathcal{L}}_{\mathrm{int}}\left(\Lambda\,z_2\right)::\hat{\mathcal{L}}_{\mathrm{int}}\left(\Lambda\,z_3\right)::\hat{\mathcal{L}}_{\mathrm{int}}\left(\Lambda\,z_1\right):
\right.\right.\nonumber\\
&&\quad+\left.\left.\int_{z_2^0-d\left(\boldsymbol{z}_2,\boldsymbol{z}_3\right)}^{\infty}\mathrm{d}z_3^0:\hat{\mathcal{L}}_{\mathrm{int}}\left(\Lambda\,z_3\right)::\hat{\mathcal{L}}_{\mathrm{int}}\left(\Lambda\,z_2\right)::\hat{\mathcal{L}}_{\mathrm{int}}\left(\Lambda\,z_1\right):\right]\right\rbrace.\nonumber\\
\label{eq:3rdo}\end{eqnarray} The first expression in the square
brackets can be written as
\begin{eqnarray}&&
\int_{-\infty}^{z_2^0}\mathrm{d}z_3^0:\hat{\mathcal{L}}_{\mathrm{int}}\left(\Lambda\,z_1\right)::
\hat{\mathcal{L}}_{\mathrm{int}}\left(\Lambda\,z_2\right)::\hat{\mathcal{L}}_{\mathrm{int}}\left(\Lambda\,z_3\right):\nonumber\\&&+
\int_{z_2^0}^{z^0_1}\mathrm{d}z_3^0:\hat{\mathcal{L}}_{\mathrm{int}}\left(\Lambda\,z_1\right)::
\hat{\mathcal{L}}_{\mathrm{int}}\left(\Lambda\,z_3\right)::\hat{\mathcal{L}}_{\mathrm{int}}\left(\Lambda\,z_2\right):
\nonumber\\&&+\int_{z^0_1}^{\infty}\mathrm{d}z_3^0:
\hat{\mathcal{L}}_{\mathrm{int}}\left(\Lambda\,z_3\right)::\hat{\mathcal{L}}_{\mathrm{int}}\left(\Lambda\,z_1\right)::
\hat{\mathcal{L}}_{\mathrm{int}}\left(\Lambda\,z_2\right):
\nonumber\\&&+\int_{z_2^0}^{z_2^0-d\left(\boldsymbol{z}_2,\boldsymbol{z}_3\right)}\mathrm{d}z_3^0:
\hat{\mathcal{L}}_{\mathrm{int}}\left(\Lambda\,z_1\right)::
\hat{\mathcal{L}}_{\mathrm{int}}\left(\Lambda\,z_2\right)::
\hat{\mathcal{L}}_{\mathrm{int}}\left(\Lambda\,z_3\right):\nonumber\\&&+
\int_{z_2^0-d\left(\boldsymbol{z}_2,\boldsymbol{z}_3\right)}^{z_2^0
}\mathrm{d}z_3^0:\hat{\mathcal{L}}_{\mathrm{int}}\left(\Lambda\,z_1\right)::\hat{\mathcal{L}}_{\mathrm{int}}\left(\Lambda\,z_3\right)::
\hat{\mathcal{L}}_{\mathrm{int}}\left(\Lambda\,z_2\right):\nonumber\\&&+
\int_{z^0_1}^{z^0_1
-d\left(\boldsymbol{z}_1,\boldsymbol{z}_3\right)}\mathrm{d}z_3^0:\hat{\mathcal{L}}_{\mathrm{int}}\left(\Lambda\,z_1\right)::\hat{\mathcal{L}}_{\mathrm{int}}\left(\Lambda\,z_3\right)::\hat{\mathcal{L}}_{\mathrm{int}}\left(\Lambda\,z_2\right):\nonumber\\&&+\int_{z^0_1-d\left(\boldsymbol{z}_1,\boldsymbol{z}_3\right)}^{z^0_1}\mathrm{d}z_3^0:
\hat{\mathcal{L}}_{\mathrm{int}}\left(\Lambda\,z_3\right)::\hat{\mathcal{L}}_{\mathrm{int}}\left(\Lambda\,z_1\right)::\hat{\mathcal{L}}_{\mathrm{int}}\left(\Lambda\,z_2\right):
\nonumber.\\
\end{eqnarray}
The last 4 terms cancel with the same argument as
in~(\ref{eq:cancelout2})--(\ref{eq:cancelout}). The same holds for the second square
 bracket in~(\ref{eq:3rdo}).
The $z^0_2$-integrals of~(\ref{eq:3rdo}) can again be split into 2 terms
\begin{eqnarray}
\int_{-\infty}^{z^0_1-d\left(\boldsymbol{z}_1,\boldsymbol{z}_2\right)}\mathrm{d}z_2^0=\int_{-\infty}^{z^0_1}\mathrm{d}z_2^0+\int_{z^0_1}^{z^0_1-d\left(\boldsymbol{z}_1,\boldsymbol{z}_2\right)}\mathrm{d}z_2^0.
\end{eqnarray} The first terms give the usual instant-form contributions since $\int_{\mathbb{R}^4}\mathrm{d}^4z\,\theta\left(z^0\right)$ is Lorentz invariant. The second terms are integrations over intervals, where
$z_1,z_2$ are spacelike separated, thus the corresponding Lagrangian densities commute~(cf.~Section~\ref{sec:mc}). These terms are (leaving the $z_1$ and $z_2$-integrations away)
\begin{eqnarray}&& \int_{-\infty}^{z_2^0}\mathrm{d}z_3^0:\hat{\mathcal{L}}_{\mathrm{int}}\left(\Lambda\,z_1\right)::
\hat{\mathcal{L}}_{\mathrm{int}}\left(\Lambda\,z_2\right)::\hat{\mathcal
{L}}_{\mathrm{int}}\left(z_3\right):\nonumber\\&&\quad+
\int_{z_2^0}^{z^0_1}\mathrm{d}z_3^0:\hat{\mathcal{L}}_{\mathrm{int}}\left(\Lambda\,z_1\right)::
\hat{\mathcal{L}}_{\mathrm{int}}\left(\Lambda\,z_3\right)::\hat{\mathcal{L}}_{\mathrm{int}}\left(\Lambda\,z_2\right):
\nonumber\\&&\quad+\int_{z^0_1}^{\infty}\mathrm{d}z_3^0:\hat{\mathcal{L}}_{\mathrm{int}}\left(\Lambda\,z_3\right)::
\hat{\mathcal{L}}_{\mathrm{int}}\left(\Lambda\,z_1\right)::\hat{\mathcal{L}}_{\mathrm{int}}\left(\Lambda\,z_2\right):\nonumber\\&&\quad+
\int_{z^0_1}^{-\infty}\mathrm{d}z_3^0:\hat{\mathcal{L}}_{\mathrm{int}}\left(\Lambda\,z_2\right)::
\hat{\mathcal{L}}_{\mathrm{int}}\left(\Lambda\,z_1\right)::\hat{\mathcal
{L}}_{\mathrm{int}}\left(\Lambda\,z_3\right):\nonumber\\&&\quad+
\int_{z_2^0}^{z^0_1}\mathrm{d}z_3^0:\hat{\mathcal{L}}_{\mathrm{int}}\left(\Lambda\,z_2\right)::
\hat{\mathcal{L}}_{\mathrm{int}}\left(\Lambda\,z_3\right)::\hat{\mathcal{L}}_{\mathrm{int}}\left(\Lambda\,z_1\right):\nonumber\\&&\quad+
\int_{\infty}^{z_2^0}\mathrm{d}z_3^0:\hat{\mathcal{L}}_{\mathrm{int}}\left(\Lambda\,z_3\right)::
\hat{\mathcal{L}}_{\mathrm{int}}\left(\Lambda\,z_2\right)::\hat{\mathcal{L}}_{\mathrm{int}}\left(\Lambda\,z_1\right):
\nonumber\\&&=
 \int_{z^0_1}^{z_2^0}\mathrm{d}z_3^0:\hat{\mathcal{L}}_{\mathrm{int}}\left(\Lambda\,z_1\right)::
 \hat{\mathcal{L}}_{\mathrm{int}}\left(\Lambda\,z_2\right)::\hat{\mathcal{L}}_{\mathrm{int}}\left(\Lambda\,z_3\right):\nonumber\\&&\quad+
 \int_{z^0_1}^{z_2^0}\mathrm{d}z_3^0:\hat{\mathcal{L}}_{\mathrm{int}}\left(\Lambda\,z_3\right)::
 \hat{\mathcal{L}}_{\mathrm{int}}\left(\Lambda\,z_1\right)::\hat{\mathcal{L}}_{\mathrm{int}}\left(\Lambda\,z_2\right):
 \nonumber\\&&\quad+
 \int_{z_2^0}^{z^0_1}\mathrm{d}z_3^0\left[:\hat{\mathcal{L}}_{\mathrm{int}}\left(\Lambda\,z_1\right)::
 \hat{\mathcal{L}}_{\mathrm{int}}\left(\Lambda\,z_3\right)::\hat{\mathcal{L}}_{\mathrm{int}}\left(\Lambda\,z_2\right):\right.\nonumber\\&&\qquad+\left.
 :\hat{\mathcal{L}}_{\mathrm{int}}\left(\Lambda\,z_2\right)::\hat{\mathcal{L}}_{\mathrm{int}}\left(\Lambda\,z_3\right)::
 \hat{\mathcal
 {L}}_{\mathrm{int}}\left(z_1\right):\right]\nonumber.\\
\end{eqnarray}
$z_1$ and $z_2$ are spacelike separated and the integration
limits of the $z^0_3$-integration are $z^0_1$ and $z^0_2$.
Therefore $z_3$ is spacelike separated with either $z_1$ or $z_2$
(or both $z_1$ and $z_2$). In either case, the last contributions
cancel out by performing the appropriate commutations. Then the 3
remaining terms are just the usual time ordering as in instant
form. Thus, the third order perturbation theory is the same as in
instant-form quantum field theory. Consequently, we have shown
that also the third order contributions for the S operator are
equivalent to the usual time-ordered perturbation
theory.$\quad\bullet$\\
Similarly, it can be shown by complete induction that this is true
for all orders of the perturbation
series.\\

%
%
%
%
 \end{appendix}
 \backmatter
%
%
%
%
%
%
%
%
%
%

\clearpage
\addcontentsline{toc}{chapter}{Bibliography}

\clearpage \pagebreak \clearpage\thispagestyle{empty}
\subsubsection*{Acknowledgements}
A fulfilling activity, be it a study, a profession or a hobby, may
be very hard to find, sometimes even harder to realize. I am in
the lucky situation having found a study I can be creative with.
Many people have supported me in different ways to pursue my
studies. They made it possible to accomplish this diploma thesis.
First of all, I would like to thank my supervisor Wolfgang
Schweiger for all his help and patience and for introducing me to
such an interesting field of research. I want to thank William H.
Klink for the constructive talks and his helpful ideas. I also
want to thank Robert J. Weibel for proof-reading this thesis.
Furthermore, I will always be grateful for the love and support I
have received from my parents and from my family. I would also
like to mention my friends and colleges. Thank you for being
always there for me! Finally, I thank Justine. I
dedicate this thesis to her.

\end{document}